\documentclass[10pt,twocolumn,superscriptaddress,aps,prd,preprintnumbers,amsmath,amssymb,nofootinbib]{revtex4-2}
\usepackage{iftex,ifpdf}
\ifpdf
  \usepackage{graphicx,xcolor}
\else
  \ifXeTeX
    \usepackage{graphicx,xcolor}
  \else
    \usepackage[dvipdfmx]{graphicx,xcolor}
    \PassOptionsToPackage{dvipdfmx}{hyperref}
  \fi
\fi
\graphicspath{{./}{figures/}{PRD/}{PRD/figures/}}
\usepackage{bm,booktabs,dcolumn,mathtools,placeins}
\usepackage[colorlinks=true,linkcolor=blue,citecolor=blue,urlcolor=blue]{hyperref}
\numberwithin{equation}{section}
\renewcommand{\theequation}{\arabic{section}.\arabic{equation}}
\newcommand{\Mp}{M_{\rm Pl}}
\newcommand{\cF}{\mathcal F}
\newcommand{\cG}{\mathcal G}
\newcommand{\cL}{\mathcal L}
\newcommand{\ah}{\hat\alpha}
\newcommand{\bh}{\hat\beta}

\allowdisplaybreaks
\begin{document}

\preprint{YITP-26-161, WUCG-26-13}

\title{Quasinormal modes of dyonic axion black holes with nonlinear electrodynamics}
\author{Jialun Zhao}
\email{zhao@asagi.waseda.jp}
\affiliation{Department of Physics, Waseda University,
3-4-1 Okubo, Shinjuku, Tokyo 169-8555, Japan}
\author{Antonio De Felice}
\email{antonio.defelice@yukawa.kyoto-u.ac.jp}
\affiliation{Center for Gravitational Physics and Quantum Information,
Yukawa Institute for Theoretical Physics, Kyoto University, Kyoto 606-8502, Japan}
\author{Shinji Tsujikawa}
\email{tsujikawa@waseda.jp}
\affiliation{Department of Physics, Waseda University,
3-4-1 Okubo, Shinjuku, Tokyo 169-8555, Japan}

\begin{abstract}
We study quasinormal modes (QNMs) of static charged black holes with a massless axion and nonlinear electrodynamics (NED).
Without NED corrections, we compute the complex QNM frequencies of five oscillatory mode families in the system of five dynamical degrees of freedom.
These families are continued from the fundamental Reissner--Nordstr\"om (RN) modes. Axion coupling generically breaks gravitational and electromagnetic parity isospectrality. At representative nonzero couplings, all five frequencies are distinct even on purely electric RN backgrounds without axion hair.
Adding magnetic charge sources axion hair, couples all five perturbations, and modifies their oscillation frequencies and damping rates.
Additional purely imaginary growing modes occur in some strong-coupling benchmarks.
The known electromagnetic--scalar spectral degeneracy at the charge-to-mass ratio $Q/M=\sqrt{7}/4$ for angular multipole number $\ell=2$ in the decoupled RN limit explains the linear frequency splitting induced by a small axion coupling. 
We include two quartic NED interactions obtained by squaring the two standard electromagnetic invariants.
With the axion decoupled and equal quartic coefficients, electromagnetic duality makes the first-order NED frequency shift of each individual gravitational or electromagnetic branch independent of electric--magnetic charge composition at fixed $Q/M$.
For unequal coefficients, duality instead constrains the sum of the first-order NED frequency shifts within each isolated RN parity pair.
Depending on the coefficients, NED increases or decreases the damping rates by about one percent in representative backgrounds, while the quartic terms remain small compared with the local Maxwell energy density throughout the exterior.
The QNM frequencies and NED shifts obtained with the spectral method agree with independent integration-and-matching results at representative parameter values.
\end{abstract}

\maketitle

\section{Introduction}
\label{sec:intro}

The ringdown of a black hole (BH) is characterized by quasinormal modes (QNMs), whose complex frequencies encode the oscillation frequencies and damping rates of linear perturbations. Gravitational-wave observations have opened a new avenue for testing BH dynamics in the strong-gravity regime~\cite{LIGOScientific:2016aoc}. The theoretical framework of BH QNMs and its extensions to systems with additional fields and interactions are reviewed in Refs.~\cite{Kokkotas:1999bd,Nollert:1999ji,Berti:2009kk,Konoplya:2011qq,Pani:2013pma}.

BH QNMs are determined within linear perturbation theory by solving the field equations linearized about a BH background, with ingoing boundary conditions at the event horizon and outgoing boundary conditions at infinity. The foundations of this framework were established by the Regge--Wheeler and Zerilli equations for Schwarzschild BHs~\cite{Regge:1957td,Zerilli:1970se,Zerilli:1970wzz}. Chandrasekhar and collaborators investigated the Schwarzschild QNM spectrum and developed transformations relating the two parity sectors~\cite{Chandrasekhar:1975zza,Chandrasekhar:1975nkd,Chandrasekhar:1985kt}. Perturbations of charged BHs were studied by Moncrief, Zerilli, and Chandrasekhar~\cite{Moncrief:1974gw,Moncrief:1974ng,Moncrief:1975sb,Zerilli:1974ai,Chandrasekhar:1979iz}.

Several complementary methods have been developed to compute QNM frequencies. Direct numerical integration was used in early calculations~\cite{Gunter:1980,Kokkotas:1988fm}; its matrix-valued extension treats coupled perturbations through a determinant condition on the unwanted asymptotic amplitudes or on a matching matrix~\cite{Pani:2013pma,Macedo:2016wgh}. Semiclassical Wentzel--Kramers--Brillouin (WKB) approximations~\cite{Schutz:1985km,Iyer:1986np}, including improvements based on Pad\'e approximants~\cite{Matyjasek:2017psv}, provide efficient estimates when a suitable potential description is available. Leaver's continued-fraction method~\cite{Leaver:1985ax,Leaver:1986gd,Leaver:1990zz} and its applications to charged and near-extremal BHs~\cite{Berti:2003zu,Onozawa:1995vu,Andersson:1996xw} provide accurate frequency-domain benchmarks.

Spectral and pseudospectral methods instead approximate the regularized perturbations in a finite basis and reduce the boundary-value problem to an algebraic eigenvalue problem, as in Chebyshev collocation~\cite{Jansen:2017oag} and Bernstein-polynomial implementations~\cite{Fortuna:2020obg}. First-order formulations also permit asymptotic analysis and spectral discretization without reduction to a Schr\"odinger-like equation~\cite{Langlois:2021xzq}. 
Chung \textit{et al.} developed a spectral approach to metric perturbations that directly solves the coupled linearized Einstein equations on Schwarzschild and Kerr backgrounds~\cite{Chung:2023zdq,Chung:2023wkd}. The Kerr implementation is known as METRICS. Horizon-penetrating hyperboloidal coordinates provide a formulation in which the event horizon and future null infinity are included in a compactified computational domain~\cite{Ansorg:2016ztf,Ripley:2022ypi}.

QNM frequencies can also be extracted from time-domain waveforms~\cite{Gundlach:1993tp,Berti:2007dg} or computed using the asymptotic iteration method~\cite{Cho:2009cj}. Since these methods have different convergence properties and domains of applicability, comparisons between independent approaches provide useful checks of numerical accuracy, particularly for coupled perturbations. The QNM spectrum is determined by the full perturbation operator and therefore depends on both the background geometry and the couplings among the perturbed fields. Interactions that leave a background solution unchanged can still modify the linearized equations and hence shift its QNM frequencies.

Charged BHs provide a useful setting for examining the relation between the background geometry and the QNM spectrum. In Einstein--Maxwell theory, the QNM spectrum of a Reissner--Nordstr\"om (RN) BH depends on its mass and total charge magnitude, but not on the relative electric and magnetic contributions. This degeneracy follows from electromagnetic duality, which can be formulated directly at the level of the coupled perturbation equations~\cite{Pereniguez:2023wxf,DeFelice:2023rra}. The gravitational and electromagnetic (EM) QNM families also exhibit axial--polar isospectrality: within each family, the two parity sectors have identical spectra. Additional matter interactions can break these spectral degeneracies. Beyond Einstein--Maxwell theory, static, spherically symmetric hairy BH solutions have been constructed in $U(1)$ gauge-invariant scalar--vector--tensor (SVT) theories~\cite{Heisenberg:2018SVThairy}. Their odd-parity perturbations and the associated no-ghost and gradient-stability conditions have also been studied~\cite{Heisenberg:2018SVTperturbations}.

Axions provide a particle-physics motivation for such interactions. The Peccei--Quinn mechanism addresses the strong CP problem by dynamically suppressing charge-conjugation and parity (CP) violation in quantum chromodynamics (QCD) through a pseudoscalar field, the axion~\cite{Peccei:1977hh,Weinberg:1977ma,Wilczek:1977pj}. Invisible axion models implement this mechanism through heavy-quark or extended Higgs sectors~\cite{Kim:1979if,Shifman:1979if,Dine:1981rt,Zhitnitsky:1980tq}, while coherent oscillations of the axion field can contribute to the dark matter abundance~\cite{Preskill:1982cy,Abbott:1982af,Dine:1982ah}. More general axion-like particles (ALPs) arise in string compactifications~\cite{Witten:1984dg,Svrcek:2006yi,Arvanitaki:2009fg,Acharya:2010zx,Cicoli:2012sz}. Their masses and photon couplings need not obey the relation characteristic of QCD axion models. We therefore study a generic ALP in the massless limit, treating its photon coupling as an independent parameter.

Early exact solutions describing gravitating dyons and dyonic BHs were obtained by Kasuya and collaborators~\cite{Kasuya:1981tq,Kasuya:1981ef}. In axion theories, the coupling to the electromagnetic Chern--Pontryagin invariant sources axion hair in the presence of both electric and magnetic fields. Static dyonic BHs with axion hair have been constructed in several settings~\cite{Lee:1991jw,Filippini:2019cqk,Fernandes:2019kmh}. Their QNMs can distinguish electric--magnetic charge configurations that would yield the same RN geometry in Einstein--Maxwell theory~\cite{DeFelice:2024axionQNM}. Even on a purely electric RN background, for which a regular massless axion is constant, axion and electromagnetic perturbations remain coupled in the odd-parity sector. 
The coupled system can exhibit avoided crossings in the QNM spectrum and resonant enhancement of the excitation factors~\cite{Boskovic:2018lkj,Takahashi:2025resonance,Takahashi:2026ringdown}. These effects must be established through the coupled eigenvalue and excitation problems; a sign change in a frequency correction alone is insufficient to demonstrate a resonance.
At sufficiently strong coupling, the same axion--EM interaction
can also trigger mode instabilities~\cite{Boskovic:2018lkj}.

Nonlinear electrodynamics (NED) provides another theoretically motivated extension of Maxwell theory. Euler--Heisenberg theory describes local nonlinear electromagnetic interactions induced by charged-particle loops~\cite{Heisenberg:1936nmg,Schwinger:1951nm,Dunne:2004nc}, while Born--Infeld theory offers a distinct nonlinear generalization~\cite{Born:1934gh}. When coupled to gravity, NED admits charged BH solutions whose geometries can differ substantially from that of RN. Solutions with regular centers include those constructed by Ay\'on-Beato and Garc\'ia~\cite{Ayon-Beato:1998hmi,Ayon-Beato:1999kuh,Ayon-Beato:2000mjt}, as well as electrically charged regular configurations~\cite{Dymnikova:2004zc}. More general constructions and their properties have also been investigated~\cite{Ansoldi:2008jw,Balart:2014cga,Balart:2014jia,Fan:2016hvf,Rodrigues:2018bdc,Maeda:2021jdc}. Other notable examples include BH solutions in Euler--Heisenberg and Born--Infeld theories, with and without a cosmological constant~\cite{Yajima:2000kw,Fernando:2003tz,Cai:2004eh,Dey:2004yt,Kruglov:2017mpj}.

Regularity of the background does not guarantee perturbative stability. For the NED BHs with regular centers analyzed in Refs.~\cite{DeFelice:2024seu,DeFelice:2025NED}, an angular gradient instability develops near the center. This conclusion applies to the electromagnetic and scalar extensions considered there and does not automatically extend to arbitrary matter Lagrangians. Conversely, positive-power corrections such as the Euler--Heisenberg-type $\mathcal F^2$ term need not eliminate the curvature singularity at $r=0$, although singular BH branches can satisfy local no-ghost and propagation conditions~\cite{Chen:2025aom}. The exterior configurations studied here in the weak-NED regime satisfy sufficient local conditions of this kind. These exterior conditions do not, however, establish global mode stability. Moreover, a truncated low-energy action cannot determine the ultraviolet resolution of the central singularity.

Low-energy effective field theory (EFT) provides a direct motivation for combining axion hair with nonlinear electromagnetic corrections. In a theory containing an axion--photon interaction, integrating out heavy charged matter also generates quartic field-strength operators. Their Wilson coefficients encode the masses, charges, spins, and multiplicities of the heavy particles~\cite{Heisenberg:1936nmg,Dunne:2004nc,Preucil:2017wen}, which may belong to the visible or a dark sector. Such corrections can therefore carry information about heavy states while remaining subdominant to the Maxwell term.

In a dyonic background with axion hair, NED simultaneously modifies the electric field, the axion source, the metric, and the coupled perturbation equations. The resulting QNM shifts cannot generally be obtained by simply adding separate axion-hair and NED corrections to the RN spectrum. Building on previous NED QNM analyses without axion hair~\cite{Nomura:2020tpc,Nomura:2021efi}, we investigate this interplay in the full five-field system. Our analysis isolates the response to the retained quartic operators; interpreting this response as a prediction of a specific EFT additionally requires control over the omitted operators.

We first follow five quadrupolar ($\ell=2$) QNM families from the fundamental RN modes in the minimal axion theory, extending the two-family dyonic analysis of Ref.~\cite{DeFelice:2024axionQNM}. Even without NED, axion perturbations couple to the odd-parity gravitoelectromagnetic sector on a purely electric RN background, while the even-parity sector retains its Einstein--Maxwell spectrum. This coupling lifts the axial--polar frequency degeneracy despite the absence of background axion hair. A separate search identifies additional growing modes at selected strong-coupling benchmarks. To interpret the small-coupling splitting, we use an EM--scalar Darboux degeneracy of the decoupled RN system, previously derived in Ref.~\cite{Rout:2026RNdegeneracy}.

We then determine the additional NED shifts, distinguishing the full correction from its magnetic modification at fixed charge-to-mass ratio. With the axion decoupled, equal quartic coefficients make the theory invariant under electromagnetic duality through first NED order. Consequently, each individual gravitational or EM branch has a first-order frequency shift independent of charge composition. For unequal coefficients, the rotated theory contains an additional parity-odd interaction; duality then constrains the sum of the shifts within each isolated RN parity pair. Calculating the individual shifts in the degenerate subspace generally requires a coupled treatment~\cite{McManus:2019qnm}. We examine how finite axion coupling removes the equal-coefficient cancellation, then determine the individual finite-NED shifts for separately varied and unequal coefficients. A derivative expansion provides a local approximation to these directly computed frequencies.

We compute the spectra by compactified Chebyshev
collocation~\cite{Jansen:2017oag} of five coupled radial equations
obtained by eliminating the nondynamical fields from the
quadratic action.
Factoring out the QNM boundary phases gives a quadratic
eigenvalue problem retaining the first-derivative couplings.
All figures use these spectral results.
Independent integration-and-matching calculations and tests
of radial resolution, background accuracy, and parameter-step
dependence support representative frequencies and shifts.

The paper is organized as follows. Section~\ref{sec:model} introduces the theory, its Wilson-coefficient interpretation, the background solutions, and the normalization conventions. Section~\ref{sec:pert} develops the five-field numerical formulation. Section~\ref{sec:electric} establishes the minimal-theory spectrum and magnetic response, and Sec.~\ref{sec:ned} presents the NED frequency shifts and their physical interpretation. We conclude in Sec.~\ref{concludesec}.
Appendix~\ref{app:EFT} discusses the EFT interpretation, and Appendix~\ref{app:reduction} gives the coupled perturbation equations. Appendix~\ref{app:numerics} collects the supporting numerical checks. Appendices~\ref{app:response} and~\ref{app:duality} present the frequency-response formulas and the duality argument, respectively.

We use natural units, $\hbar=c=1$, and adopt the metric signature $(-,+,+,+)$ throughout.

\section{Theory, backgrounds, and control of the NED expansion}
\label{sec:model}

The interplay between axion hair and electromagnetic nonlinearity
is described by Einstein gravity coupled to a canonical pseudoscalar
$\phi$ and a $U(1)$ gauge potential $A_\mu$.
Let $g_{\mu\nu}$ denote the spacetime metric and
$g_{\rm M}\equiv\det(g_{\mu\nu})$ its determinant.
The electromagnetic field strength is
$F_{\mu\nu}=\partial_\mu A_\nu-\partial_\nu A_\mu$.
Greek indices label spacetime components and are contracted
with the metric.
We define the electromagnetic invariants and Hodge-dual convention by
\begin{align}
\cF&=-\frac14F_{\mu\nu}F^{\mu\nu},&
\cG&=-\frac14F_{\mu\nu}\widetilde F^{\mu\nu},\\
\widetilde F^{\mu\nu}&=-\frac{\varepsilon^{\mu\nu\rho\sigma}}{2\sqrt{-g_{\rm M}}}F_{\rho\sigma},&
\varepsilon^{tr\theta\varphi}&=+1.\label{eq:dual}
\end{align}
The tensor density $\varepsilon^{\mu\nu\rho\sigma}$ is the
totally antisymmetric symbol in the oriented coordinates
$(t,r,\theta,\varphi)$ used in Eq.~\eqref{eq:metric}.

We retain the axion--photon interaction and the two parity-even
quartic invariants without derivatives of the field strength.
The action is
\begin{align}
S=\int {\rm d}^4 x\sqrt{-g_{\rm M}}\biggl[&\frac{\Mp^2}{2}R
-\frac12\nabla_\mu\phi\nabla^\mu\phi+\cL\biggr],\label{eq:action}\\
\cL={}&\cF+\alpha\cF^2+\beta\cG^2+g\phi\cG\,.
\label{eq:model}
\end{align}
Here $R$ is the Ricci scalar, $\nabla_\mu$ is the metric-compatible
covariant derivative, and $\Mp^2=(8\pi G_N)^{-1}$ defines the
reduced Planck mass in terms of Newton's constant $G_N$.
The constant $g$ denotes the axion--photon coupling, while
$\alpha$ and $\beta$ control the quartic NED interactions
$\cF^2$ and $\cG^2$, respectively.
The mass dimensions are $[\phi]=[A_\mu]=1$, $[g]=-1$, and $[\alpha]=[\beta]=-4$. We set the pseudoscalar potential to zero throughout the background and QNM calculations. 
We work in the massless ALP limit. Applying our results to a massive axion requires the effects of its potential on the background, the perturbation equations, and the QNM boundary conditions to be negligible.

The coefficients $\alpha$ and $\beta$ are Wilson coefficients of the retained quartic electromagnetic operators. For example, integrating out heavy Dirac particles with masses $m_i$, coupled to $A_\mu$ with charges $Q_i e_D$, yields the leading spinor Euler--Heisenberg coefficients~\cite{Heisenberg:1936nmg,Dunne:2004nc,Preucil:2017wen}
\begin{equation}
\alpha=\sum_i\frac{8\alpha_D^2Q_i^4}{45m_i^4},\qquad
\beta=\sum_i\frac{14\alpha_D^2Q_i^4}{45m_i^4},
\label{eq:EH}
\end{equation}
where $e_D$ is the gauge coupling, $Q_i$ is the dimensionless charge of species $i$, and $\alpha_D=e_D^2/(4\pi)$. The microscopic charges $Q_i$ are distinct from the geometric BH charge $Q$ defined below. This matching gives $\beta/\alpha=7/4$, while $\alpha=\beta$ corresponds to the leading quartic expansion of Born--Infeld theory. These ratios characterize the retained quartic interactions without specifying a complete nonlinear theory. Appendix~\ref{app:EFT} gives the one-loop interaction, its multiplicity dependence, and the additional requirements for an EFT interpretation. In our numerical analysis, we vary $\alpha$ and $\beta$ independently to isolate their respective effects. The signed variations used for numerical differentiation are computational probes and are not required to satisfy the heavy-Dirac matching relations.

\subsection{Static dyonic equations}

We consider static, spherically symmetric dyonic backgrounds of the form
\begin{align}
{\rm d}s^2&=-Nh\,{\rm d}t^2+\frac{{\rm d}r^2}{h}+r^2{\rm d}\Omega_2^2,\label{eq:metric}\\
A_\mu {\rm d}x^\mu&=A_0(r){\rm d}t-q_M\cos\theta\,{\rm d}\varphi,\qquad\phi=\phi(r).
\end{align}
Here $N(r)$ and $h(r)$ are metric functions, $r$ is the areal radius, and ${\rm d}\Omega_2^2={\rm d}\theta^2+\sin^2\theta\,{\rm d}\varphi^2$ is the line element on the unit two-sphere. The function $A_0(r)$ is the electric potential. The signed magnetic charge $q_M$ is constant, as required by the Bianchi identity $\nabla_\mu\widetilde F^{\mu\nu}=0$, and is normalized by $\int_{S^2}F=4\pi q_M$, where 
$F=(1/2)F_{\mu\nu}\,{\rm d}x^\mu\wedge{\rm d}x^\nu$ is the electromagnetic field-strength two-form and $S^2$ is a sphere at fixed $t$ and $r$. Primes and overdots denote derivatives with respect to $r$ and $t$, respectively. The signed radial electric and magnetic fields measured in a local orthonormal frame are $E=A_0'/\sqrt N$ and $B=q_M/r^2$, giving $\cF=(E^2-B^2)/2$ and $\cG=-EB$. A comma denotes differentiation with respect to the indicated invariant, for example $\cL_{,\cF}=\partial\cL/\partial\cF$.

Varying the action with respect to $A_\nu$ gives the gauge-field equation
\begin{equation}
\nabla_\mu\left(\cL_{,\cF}F^{\mu\nu}
+\cL_{,\cG}\widetilde F^{\mu\nu}\right)=0.
\label{eq:gauge_eom}
\end{equation}
On this background, its $t$ component reduces to $(r^2\cL_{,\cF}E-q_M\cL_{,\cG})'=0$. Integrating once yields
\begin{equation}
r^2\cL_{,\cF}E-q_M\cL_{,\cG}=q_E,
\label{eq:gauss_first_integral}
\end{equation}
where the integration constant $q_E$ is the electric displacement charge, with conserved displacement flux $4\pi q_E$. Substituting $\cL_{,\cF}=1+2\alpha\cF$ and $\cL_{,\cG}=2\beta\cG+g\phi$, together with the above expressions for the invariants, gives the algebraic electric-field equation
\begin{equation}
E\bigl[1+\alpha E^2+(2\beta-\alpha)B^2\bigr]
=\frac{q_E+gq_M\phi}{r^2}.
\label{eq:gauss}
\end{equation}
We select the electric-field branch continuously connected to the $\alpha=\beta=0$ solution.

The independent Einstein and scalar equations become
\begin{align}
\frac{N'}{N}&=\frac{r\phi'^2}{\Mp^2},\label{eq:Nbg}\\
h'&=\frac{1-h}{r}-\frac{r\rho_{\rm EM}}{\Mp^2}
-\frac{rh\phi'^2}{2\Mp^2},\label{eq:hbg}\\
(r^2h\sqrt N\,\phi')'&=gq_M\sqrt N E,\label{eq:phibg}
\end{align}
where the electromagnetic energy density is
\begin{align}
\rho_{\rm EM}={}&\frac{E^2+B^2}{2}+\beta E^2B^2\nonumber\\
&+\frac\alpha4(3E^4-2E^2B^2-B^4).
\label{eq:rho}
\end{align}
The topological axion interaction has no explicit stress-energy contribution. It affects the metric through the electric field and the scalar profile.

Regularity at a nonextremal event horizon of areal radius $r_h$ gives
\begin{align}
h(r_h)&=0,&
h'_h&=\frac{1-r_h^2\rho_{{\rm EM},h}/\Mp^2}{r_h}>0,\\
\phi'_h&=\frac{gq_ME_h}{r_h^2h'_h}.\label{eq:horizon}
\end{align}
A subscript $h$ denotes evaluation at $r=r_h$. We impose asymptotic flatness, fix the time normalization by $N(\infty)=1$, and choose the asymptotic axion value $\phi(\infty)=0$. The geometric mass $M$ is extracted from the asymptotic metric and is related to the Arnowitt--Deser--Misner (ADM) mass $M_{\rm ADM}$ by
\begin{equation}
h(r)=1-\frac{2M}{r}+O(r^{-2}),\qquad M=G_N M_{\rm ADM}.
\end{equation}

We define the total flux magnitude $q_T\geq0$, the geometric total charge $Q$, the signed magnetic fraction $\eta$, and the dimensionless couplings and frequency by
\begin{gather}
q_T^2=q_E^2+q_M^2,\qquad Q=\frac{q_T}{\sqrt2\Mp},\\
\eta=\frac{q_M}{q_T},\qquad \bar g=g\Mp,\nonumber\\
\ah=\frac{\alpha\Mp^2}{r_h^2},\qquad
\bh=\frac{\beta\Mp^2}{r_h^2},\nonumber\\
\Omega=M\omega.\label{eq:params}
\end{gather}
Here $\bar g$ is the dimensionless axion coupling, while $\ah$ and $\bh$ are the NED coefficients normalized by the horizon radius. We adopt the Fourier convention $e^{-i\omega t}$ and report the dimensionless QNM frequency $\Omega=M\omega$, with damping rate $\Gamma=-\operatorname{Im}\Omega$. These conventions give $h=1-2M/r+Q^2/r^2$ in the RN limit. In the numerical calculations, we set $r_h=\Mp=1$ and adjust $q_T$ along each parameter sequence to maintain the prescribed $Q/M$. Holding $q_T/r_h$ fixed instead would generally change $Q/M$ and thus define a different frequency response.

We normalize the asymptotic time coordinate before constructing the perturbation operator. The subscript ``raw'' denotes quantities before this normalization. For a shooting solution with $C_\infty=N_{\rm raw}(\infty)>0$, we apply the rescaling
\begin{align}
N&=\frac{N_{\rm raw}}{C_\infty},&
t&=\sqrt{C_\infty}\,t_{\rm raw},\nonumber\\
A_0&=\frac{A_{0,{\rm raw}}}{\sqrt{C_\infty}},&
\omega&=\frac{\omega_{\rm raw}}{\sqrt{C_\infty}}.
\label{eq:asymptotic_clock_normalization}
\end{align}
The local electric field $E=A_0'/\sqrt N$, the charges, and the ADM mass are invariant under this rescaling. The perturbation operator and all reported frequencies use the normalized time coordinate. Appendix~\ref{app:reduction} gives the corresponding rescaling of the auxiliary fields.

Both numerical implementations use these conventions.
The corresponding mass-normalized NED coefficients are
\begin{align}
\alpha_M&=\frac{\alpha\Mp^2}{M^2}=\frac{\ah}{(M/r_h)^2},\\
\beta_M&=\frac{\beta\Mp^2}{M^2}=\frac{\bh}{(M/r_h)^2}.
\label{eq:masscouplings}
\end{align}

Our NED-response sequences vary $\ah$ and $\bh$ at fixed $Q/M$, $\bar g$, and $\eta$. Along a magnetic-composition sequence at fixed $\ah,\bh$, the ratio $M/r_h$ generally changes, so $\alpha_M,\beta_M$ need not remain constant. Keeping the latter fixed would require adjusting the hatted coefficients as the magnetic composition varies. Expressing frequencies as $M\omega$ does not make these two parameter prescriptions equivalent.

\subsection{Amplitude and local stability}

We characterize the strength of the retained NED corrections by
\begin{equation}
\epsilon_{\rm NED}(r)=\max(|\alpha|,|\beta|)(E^2+B^2).
\label{eq:eps}
\end{equation}
This measure avoids cancellations between the electric and magnetic contributions that can make $\cF$ small. Since $\cF^2+\cG^2=(E^2+B^2)^2/4$, the magnitude of the quartic Lagrangian term $\alpha\cF^2+\beta\cG^2$ is bounded by $\epsilon_{\rm NED}/2$ times the local Maxwell energy density $(E^2+B^2)/2$. We impose the exterior amplitude criterion $\epsilon_{\rm NED}^{\max}\equiv\sup_{r\geq r_h}\epsilon_{\rm NED}(r)<0.05$. The conditions $|\alpha\cF|\ll1$ and $|\beta\cG|\ll1$ are insufficient by themselves. For example, on a purely electric background, $\cG=0$ leaves $\beta$ unconstrained, even when $\beta E^2$ produces large corrections to the perturbation equations.

Applying the general dyonic stability conditions of Ref.~\cite{Taniguchi:2025dyonic} (Sec.~IV, in particular Eq.~(4.67)) to the model~\eqref{eq:model}, we obtain the following sufficient local no-ghost and propagation conditions:
\begin{align}
a&=1+\alpha(E^2-B^2)>0,\nonumber\\
d&=1+3\alpha E^2+(2\beta-\alpha)B^2>0,\nonumber\\
c_{\rm em}&=1+(\alpha-2\beta)E^2-3\alpha B^2>0.\label{eq:adc}
\end{align}
Here $a,d,c_{\rm em}$ are local constitutive coefficients. Let $c_\Omega^2$ denote a squared angular propagation speed measured in a local orthonormal frame. In the high-frequency limit, the scalar and two gravitational modes have $c_\Omega^2=1$, while all five modes have unit squared radial propagation speeds. Defining $b=2(\beta-\alpha)EB$ and $\tau_\Omega=c_{\rm em}d+a^2+b^2$, the remaining two angular speeds satisfy
\begin{equation}
ad z^2-\tau_\Omega z+ac_{\rm em}=0,\qquad z=c_\Omega^2.
\label{eq:speeds}
\end{equation}
For positive $a,d,c_{\rm em}$, both roots are real and positive because $\tau_\Omega\geq2a\sqrt{c_{\rm em}d}$. Moreover, $|a-1|\leq\epsilon_{\rm NED}$ and $|c_{\rm em}-1|,|d-1|\leq3\epsilon_{\rm NED}$, so our amplitude criterion guarantees $a>0.95$ and $c_{\rm em},d>0.85$ throughout the exterior. These are sufficient local high-frequency conditions; they do not prove global mode stability for arbitrary multipoles.
Explicit growing modes at strong axion coupling are discussed
in Sec.~\ref{sec:magnetic}.

The QNM boundary-value problem is posed on the exterior region, where the amplitude criterion is imposed. Small retained NED amplitudes do not constrain omitted derivative, curvature, or additional axion operators. Appendix~\ref{app:EFT} discusses the heavy-particle scale hierarchies and the additional conditions required for a controlled EFT interpretation.

\section{Five coupled perturbations and numerical construction}
\label{sec:pert}

The linear perturbations are expanded in spherical harmonics $Y_{\ell m}(\theta,\varphi)$, where $\ell$ and $m$ denote the angular multipole and azimuthal numbers, respectively. We define 
\begin{equation}
L=\ell(\ell+1)\,,
\end{equation}
and set $m=0$ without loss of generality by spherical symmetry. We restrict the radiative problem to $\ell\geq2$ and include both metric parities, the gauge field, and the pseudoscalar. Magnetic charge mixes the parity sectors. After eliminating the nondynamical fields, we use the five 
dynamical perturbations
\begin{equation}
\bm y=(V,\Psi,\delta\phi,\chi,\delta A)^T.
\label{eq:dynamical_fields}
\end{equation}
For a single harmonic $Y_\ell\equiv Y_{\ell0}$, the scalar-field
and odd-parity electromagnetic perturbations are defined by
\begin{align}
\delta\phi_{\rm full}&=\delta\phi(t,r)Y_\ell,\nonumber\\
\delta A_\varphi&=-\delta A(t,r)\sin\theta\,\partial_\theta Y_\ell.
\label{eq:scalar_gauge_harmonics}
\end{align}
The even-parity electromagnetic perturbations are written as
$\delta A_t=\delta A_0(t,r)Y_\ell$ and
$\delta A_r=\delta A_1(t,r)Y_\ell$.
The remaining three reduced fields are
\begin{align}
V={}&\delta A_0'-\dot{\delta A}_1\nonumber\\
&+\frac{s_2H_0+s_3H_2+Ls_4\delta A+s_6\delta\phi}{2s_1},\nonumber\\
\Psi={}&rH_2-Lh_1,\nonumber\\
\chi={}&r\dot{\mathcal W}-r\mathcal Q'+2\mathcal Q
+\frac{p_2}{2p_1}\delta A.
\label{eq:main_auxiliary_definitions}
\end{align}
Here $H_0,H_2,h_1$ and $\mathcal Q,\mathcal W$ denote the even-
and odd-parity metric perturbations, respectively. 
The functions $s_i(r)$ and $p_i(r)$ are background-dependent coefficients of the unreduced quadratic action. Their explicit expressions are given in Eqs.~\eqref{eq:coeff_s} and~\eqref{eq:coeff_p} of Appendix~\ref{app:reduction}, which also specifies the harmonic conventions and the constraint elimination. Thus $V$ carries the electric gauge degree of freedom, $\Psi$ and $\chi$ carry the two gravitational degrees of freedom, and $\delta\phi$ and $\delta A$ carry the pseudoscalar and odd electromagnetic degrees of freedom. These assignments refer to the electric limit; for nonzero axion coupling and magnetic charge, the five amplitudes form a coupled system. A superscript $T$ denotes matrix transpose.

The reduced quadratic action takes the form
\begin{align}
S^{(2)}=\int {\rm d}t\,{\rm d}r\bigl[&
\dot{\bm y}^T K\dot{\bm y}
+\bm y'^T G\bm y'\nonumber\\
&+2\bm y^T C\bm y'
+\bm y^T M_0\bm y\bigr].
\label{eq:redaction}
\end{align}
Here $K$, $G$, and $M_0$ are the real symmetric $5\times5$
kinetic, gradient, and potential matrices, respectively,
while the real matrix $C$ encodes the couplings linear in radial
derivatives. All four matrices depend on the background and
$L=\ell(\ell+1)$ and are independent of frequency.
On the background field equations, the reduced matrices obey
the exact finite-$L$ relation $G=-Nh^2K$.
We verify this identity by extracting $K$ and $G$ independently
from the quadratic action.

For perturbations proportional to $e^{-i\omega t}$, variation
of Eq.~\eqref{eq:redaction} gives
\begin{equation}
-(G\bm y')'+(C-C^T)\bm y'
+(M_0-C^{T\prime})\bm y+\omega^2K\bm y=0.
\label{eq:euler}
\end{equation}
Our numerical formulation retains the first-derivative couplings
$C-C^T$ explicitly and solves the coupled system in this field basis.
Appendix~\ref{app:reduction} presents the quadratic coefficients
obtained from Eq.~\eqref{eq:action}, the elimination of the
nondynamical perturbations, and the five coupled radial equations.

The identity $G=-Nh^2K$ implies that all five modes have unit
squared radial propagation speeds in a local orthonormal frame
throughout the regular exterior. We introduce the tortoise
coordinate $r_*$ and rescaled amplitudes $\bm\psi$ by
\begin{align}
r_*'&=\frac1{h\sqrt N},\nonumber\\
\bm y&=\operatorname{diag}
\left(r^{-2},\frac{r}{r-r_h},r^{-1},1,1\right)\bm\psi.
\label{eq:tortoise_fields}
\end{align}
Near a regular nonextremal horizon, the leading equations
in these variables reduce to
\begin{equation}
\partial_*^2\bm\psi+\omega^2\bm\psi\simeq0,
\qquad r\to r_h,
\label{eq:horizon_wave_equation}
\end{equation}
where $\partial_*=\partial/\partial r_*$.
With the Fourier convention $e^{-i\omega t}$, all five components
therefore share the ingoing phase $e^{-i\omega r_*}$ at the future
horizon. Asymptotic flatness and the massless nature of the fields
give the outgoing phase $e^{i\omega r_*}$ at infinity.
The QNM boundary conditions are thus
\begin{equation}
\begin{aligned}
\bm\psi&\sim e^{-i\omega r_*}\qquad(r\to r_h),\\
\bm\psi&\sim e^{i\omega r_*}\qquad(r\to\infty).
\end{aligned}
\label{eq:bc}
\end{equation}
These phases are determined by the radial characteristics;
the angular propagation speeds need not all be unity.
At complex frequency, the outgoing branch is understood by
analytic continuation of the outgoing wave solution.

Using the compactified coordinate $x=r_h/r$, we factor out
the boundary phases as $\bm\psi=e^{i\omega s(r)}\bm u(x)$, with
\begin{align}
s(r)&=r+2M\log(r/r_h)-a_H\log(1-r_h/r),\nonumber\\
a_H&=\frac1{h'_h\sqrt{N_h}}.
\label{eq:phase}
\end{align}
Since $s\simeq-r_*$ near the horizon and $s\simeq r_*$
at infinity, this factor incorporates both QNM boundary phases
and leaves $\bm u$ regular at the endpoints.
The compactified domain includes infinity, so no finite outer
boundary is introduced.

Applying Chebyshev--Gauss collocation with $N_{\rm c}$ nodes
to the five coupled radial equations gives the quadratic
eigenvalue problem
\begin{equation}
[\mathbb A_0+\omega\mathbb A_1+\omega^2\mathbb A_2]\bm u=0.
\label{eq:QEP}
\end{equation}
Here the frequency-independent matrices $\mathbb A_k$ ($k=0,1,2$)
represent the discretized radial operators, and $\bm u$ collects
the nodal values of the five regular amplitudes.
We linearize this quadratic system into an equivalent generalized
eigenvalue problem. We assess frequency convergence by varying
$N_{\rm c}$ and track the convergent mode families through
neighboring points in parameter space.
The five plotted families are oscillatory modes continued
from the fundamental RN modes.
Purely imaginary roots are examined in a separate unfiltered
search, as discussed in Sec.~\ref{sec:magnetic}.

The plotted spectra use $N_{\rm c}=36$, except for
Figs.~\ref{fig:V_09} and~\ref{fig:V_11}, which use $N_{\rm c}=48$.
Selected configurations are repeated with 28, 48, and 60 nodes
and tighter background integration tolerances.
The first- and second-order NED response coefficients are obtained
using centered finite differences with fourth-order accuracy.
Each stencil point uses a separately computed background, with
the total charge retuned to maintain the prescribed $Q/M$.
We assess parameter-step sensitivity separately from frequency
convergence.

The background solver evolves the mass function $m=r(1-h)/2$,
$\phi$, ${\rm d}\phi/{\rm d}x$, and $\log N$ to $x=0$,
where the geometric mass is obtained directly as $M=m|_{x=0}$.
A near-horizon series supplies the initial conditions.
Shooting determines the horizon scalar value and total charge
simultaneously by imposing $\phi(\infty)=0$ and the prescribed
$Q/M$. Among the production backgrounds tested, the largest
residual in either shooting condition is below $1.5\times10^{-13}$.

For 25 spectral benchmark configurations, including the
strong-coupling electric modes, the largest absolute changes
in $\Omega=M\omega$ between 36 and 48 nodes and between 48 and
60 nodes are $3.1\times10^{-9}$ and $6.0\times10^{-11}$,
respectively. Tightening the background tolerance from
$2\times10^{-11}$ to $4\times10^{-13}$ changes the checked
frequencies by at most $1.1\times10^{-11}$.
We use these differences as empirical measures of numerical
sensitivity. For purely electric backgrounds with
$\alpha=\beta=0$, the odd-sector frequencies are also checked
using the independent three-field potential formulation and
complex-contour ODE integration.

The spectral solver is implemented in Python using NumPy and SciPy
in double precision. The background equations and local
quadratic-action matrices are implemented separately, and the
spectral operator is constructed directly from them.
For purely electric backgrounds, we solve the even two-field
and odd three-field blocks separately. This identifies the
degenerate RN parity partners within their respective sectors
without assuming simple eigenvalues of the full five-field system.

To validate these spectral results, we also use an independent
two-sided numerical integration-and-matching method.
The radial equations are integrated outward from a near-horizon
point with ingoing boundary conditions and inward from a large
radius with outgoing boundary conditions.
The horizon integration uses a fourth-order expansion initialized
at $1.001r_h$, while the outer integration uses an outgoing
inverse-radius expansion initialized at $50r_h$, $55r_h$, or $60r_h$.
The QNM frequencies are determined by matching the two sets
of solutions at $8r_h$.
The outgoing expansion is retained through order 14, increased
to order 16 for the sensitive strong-coupling odd-gravitational
benchmarks. The working precision is 45 decimal digits, with
accuracy and precision goals of 22 and 19, respectively.
Because the expansion at infinity is asymptotic, its truncation
order and initialization radius are varied independently.
The two implementations agree on the tested five-family
splittings and NED shifts along the $\alpha$-only, $\beta$-only,
and equal-coefficient directions within the measured numerical
variations.

Appendix~\ref{app:numerics} compares the production radial
operators with an independent symbolic reference and gives
the RN parity-pair tests and frequency-response convergence checks.

\section{Breaking of isospectrality without nonlinear electrodynamics}
\label{sec:electric}

This section follows five oscillatory QNM families, resolves
five distinct frequencies at representative parameters, and establishes the breaking of
RN axial--polar isospectrality, extending the two-family analysis
of Ref.~\cite{DeFelice:2024axionQNM}.
We set $\hat\alpha=\hat\beta=0$ throughout, including
Sec.~\ref{sec:magnetic}, to isolate the axion-induced effects.
Five dynamical degrees of freedom do not by themselves imply
five distinct frequencies: each eigenmode can involve several
perturbation components.
We identify the families by continuous tracking from the
fundamental purely electric RN modes.
The additional growing modes in Sec.~\ref{sec:magnetic} show
that these five families do not exhaust the spectrum.

At $\bar g=0$, the massless scalar decouples, while the gravitational and EM fundamental 
modes each form an isospectral axial--polar pair. The five families therefore 
correspond to only three distinct frequencies at the reference charge considered here. 
Establishing the breaking of isospectrality requires resolving the frequency splitting within 
each RN parity pair after the axion coupling is introduced. 
This parity degeneracy is distinct from the independence of the RN spectrum from 
the electric--magnetic charge composition at fixed $Q/M$.

\subsection{Electric RN background and axion-induced splitting}
\label{sec:electric_splitting}

For $q_M=0$ and $\alpha=\beta=0$, Eq.~\eqref{eq:phibg}
and horizon regularity give $\phi'=0$.
The asymptotic condition $\phi(\infty)=0$ then fixes
$\phi(r)=0$.
With $N(\infty)=1$, Eq.~\eqref{eq:Nbg} gives $N=1$,
and Eqs.~\eqref{eq:gauss}, \eqref{eq:hbg}, and~\eqref{eq:rho}
yield the electric RN solution,
\begin{equation}
N=1,\qquad h=f=1-\frac{2M}{r}+\frac{Q^2}{r^2}.
\label{eq:electric_RN_background}
\end{equation}
The even-parity two-field system coincides with that of Einstein--Maxwell theory. By contrast, the odd-parity three-field system retains the quadratic interaction $g\,\delta\phi\,\delta\cG$, despite the vanishing background axion. This interaction couples the scalar and EM perturbations, while the background electric charge couples the EM and gravitational perturbations. The odd-parity frequencies can therefore differ from their even-parity counterparts even in the absence of background scalar hair and NED corrections.

To display this coupling explicitly, we introduce
$\bm Z=(Z_{\rm grav},Z_{\rm EM},Z_\phi)^T$ by rescaling
the odd-parity subset $(\chi,\delta A,\delta\phi)$ of
Eq.~\eqref{eq:dynamical_fields}:
\begin{equation}
\begin{aligned}
Z_{\rm grav}&=\Mp\sqrt{\frac{L}{2(L-2)}}\,\chi,\\
Z_{\rm EM}&=\sqrt{L}\,\delta A,\qquad
Z_\phi=r\,\delta\phi.
\end{aligned}
\label{eq:RN_odd_canonical_fields}
\end{equation}
These variables describe the same three degrees of freedom
in a canonically normalized basis. They satisfy
\begin{equation}
\frac{{\rm d}^2\bm Z}{{\rm d}r_*^2}
+\left(\omega^2I_3-\mathsf V_{\rm RN}\right)\bm Z=0,
\label{eq:RN_odd_master}
\end{equation}
where ${\rm d}r_*/{\rm d}r=f^{-1}$ and $I_3$ is the
$3\times3$ identity matrix. In this gravitational, EM, and scalar basis,
\begin{align}
\mathsf V_{\rm RN}
&=\frac{f}{r^2}\left[
L I_3+\frac{\mathsf U}{r}
+\frac{Q^2}{r^2}\operatorname{diag}(4,4,-2)
\right],\nonumber\\
\mathsf U
&=\begin{pmatrix}
-6M & 2Q\sqrt{L-2} & 0\\
2Q\sqrt{L-2} & 0 & \sqrt{2L}\,\bar g Q\\
0 & \sqrt{2L}\,\bar g Q & 2M
\end{pmatrix}.
\label{eq:RNmatrix}
\end{align}
Here $\operatorname{diag}(4,4,-2)$ denotes the diagonal
$3\times3$ matrix with entries $4$, $4$, and $-2$ in the
gravitational, EM, and scalar order, respectively.
The factor $\sqrt{2}$ in the scalar--EM coupling follows from
$q_E=\sqrt{2}\Mp Q$ and $\bar g=g\Mp$ in our normalization.
After accounting for coupling normalizations and field-sign conventions,
this system agrees with the established pure-electric
Einstein--Maxwell--axion (EMA) perturbation equations~
\cite{Boskovic:2018lkj,Takahashi:2025resonance,Takahashi:2026ringdown}.

\begin{table*}[t]
\centering
\caption{Frequencies $\Omega=M\omega$ of the five oscillatory QNM
families continued from the fundamental RN modes in the minimal
axion model at $\ell=2$, $Q/M=0.6$, and
$\hat\alpha=\hat\beta=0$. The RN reference
($\bar g=0$, $\eta=0$) has two isospectral axial--polar pairs,
one gravitational and one EM, together with a distinct scalar
frequency, giving three distinct values for the five families.
At $\bar g=5$, all five frequencies are distinct for both
$\eta=0$ and $\eta=0.4$. In the purely electric case
($\eta=0$), the background remains RN and the even-parity
frequencies are unchanged, whereas the axion coupling shifts
the odd-parity frequencies. At $\eta=0.4$, the row labels
identify families continued from the purely electric limit. 
The tabulated frequencies are computed with $N_{\rm c}=36$
and rounded to eight decimal places; convergence is checked
using $N_{\rm c}=48$ and $60$.}
\label{tab:minimal_spectrum}
\small
\begin{tabular}{|l|c|c|c|}
\hline
Family & RN: $\bar g=0$, $\eta=0$ & $\bar g=5$, $\eta=0$ & $\bar g=5$, $\eta=0.4$\\
\hline
even grav. & $0.38621747-0.08981367i$ & $0.38621747-0.08981367i$ & $0.38769583-0.09142218i$\\
even EM & $0.51201092-0.09801663i$ & $0.51201092-0.09801663i$ & $0.49560332-0.09876332i$\\
odd grav. & $0.38621747-0.08981367i$ & $0.32447691-0.09057384i$ & $0.37013222-0.08951900i$\\
odd EM & $0.51201092-0.09801663i$ & $0.41007348-0.09123271i$ & $0.41368659-0.09231075i$\\
scalar & $0.51738763-0.09832464i$ & $0.65334365-0.10464778i$ & $0.61097442-0.10570224i$\\
\hline
\end{tabular}
\end{table*}
\begin{figure*}[t]
\centering
\includegraphics[width=\textwidth]{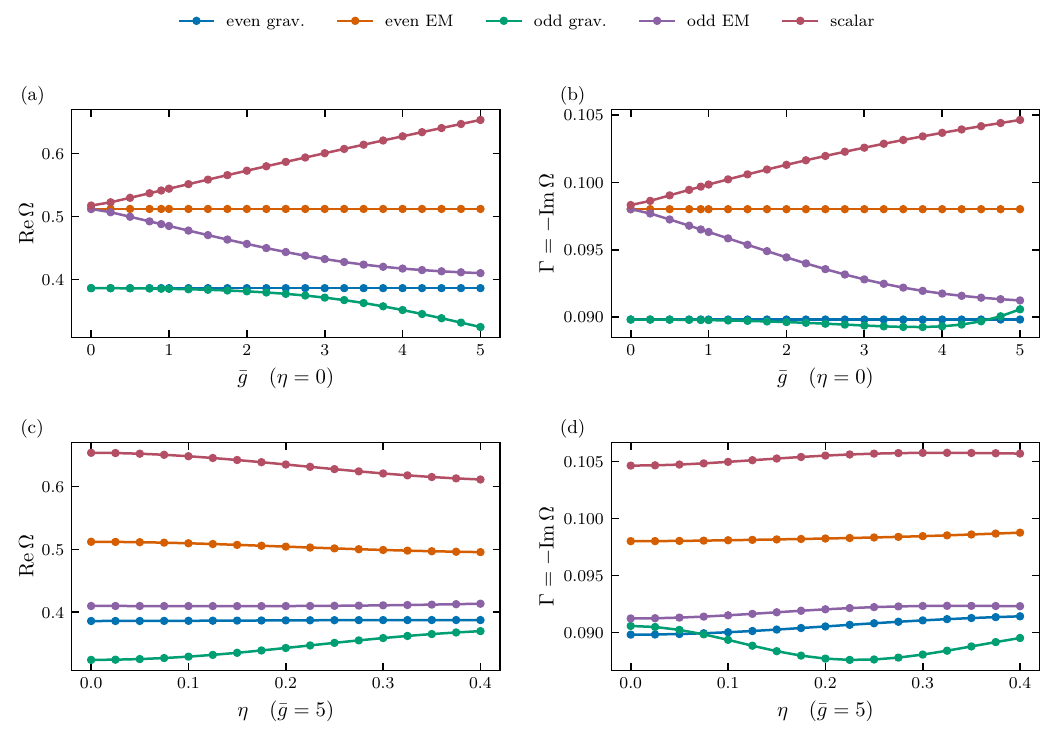}
\caption{Fundamental $\ell=2$ QNM frequencies $\Omega=M\omega$ at
$\hat\alpha=\hat\beta=0$ and $Q/M=0.6$, shown through their
real parts and damping rates $\Gamma=-\operatorname{Im}\Omega$.
Top panels: dependence on the axion coupling $\bar g$ in the
purely electric case ($\eta=0$). At $\bar g=0$, the gravitational
and EM axial--polar pairs are each degenerate. The axion coupling
lifts both degeneracies by shifting the odd-parity frequencies,
while the even-parity frequencies remain unchanged. The background
metric remains RN with $\phi=0$ throughout this sequence.
Bottom panels: dependence on the magnetic fraction $\eta$ at
$\bar g=5$, tracing five distinct frequency branches into the
dyonic regime. At nonzero $\eta$, the labels ``even'' and ``odd''
refer to the parity of the corresponding purely electric limits.
Each curve represents an eigenfrequency branch of the full
coupled perturbation problem. Lines connect independently
computed points.}
\label{fig:MIN_01}
\end{figure*}

Table~\ref{tab:minimal_spectrum} and Fig.~\ref{fig:MIN_01}
show the five fundamental QNM families. At $Q/M=0.6$, the
RN reference column of Table~\ref{tab:minimal_spectrum}
corresponds to $\bar g=0$ and $\eta=0$. The even- and
odd-gravitational modes share the frequency
$0.38621747-0.08981367i$, while the even- and odd-EM modes
share $0.51201092-0.09801663i$. These pairwise equalities
express RN axial--polar isospectrality. The decoupled scalar
perturbation $\delta\phi$ has the fundamental frequency
$\Omega=0.51738763-0.09832464i$, distinct from the gravitational
and EM frequencies. Thus, the five families correspond to
three distinct complex frequencies in the RN reference limit.

Turning on the axion coupling lifts this parity degeneracy.
Table~\ref{tab:minimal_spectrum} lists the resulting spectra
for $(\bar g,\eta)=(5,0)$ and $(5,0.4)$, each of which contains
five distinct complex frequencies. In the purely electric case
$(\bar g,\eta)=(5,0)$, the background metric remains RN and
the two even-parity frequencies retain their reference values.
The axion coupling nevertheless modifies the odd-parity
operator. The odd-gravitational frequency shifts to
$\Omega_{{\rm grav},o}=0.32447691-0.09057384i$, breaking its degeneracy
with the even-gravitational mode. Likewise, the odd-EM
frequency shifts to
$\Omega_{{\rm EM},o}=0.41007348-0.09123271i$, while the
even-EM frequency remains at
$\Omega_{{\rm EM},e}=0.51201092-0.09801663i$.
Together with the scalar branch, these modes yield five
distinct fundamental frequencies. Thus, the purely electric
system already exhibits the breaking of RN axial--polar
isospectrality at $\hat\alpha=\hat\beta=0$, without NED
corrections or background scalar hair.

The breaking of RN axial--polar isospectrality originates from
axion-induced mixing in the odd-parity sector. On the purely
electric background, the axion perturbation couples to the
angular curl of the EM perturbation, which has odd parity.
The background electric charge also couples the odd-parity
EM and gravitational perturbations. Through these interactions,
the axion modifies the eigenfrequencies of the entire odd
three-field system, while the even-parity operator remains
unchanged. The splitting therefore arises from the coupled
perturbation equations despite the unchanged background geometry.

Starting from a simple isolated QNM of the odd-parity block
in the $\bar g=0$ reference system, turning on a small nonzero
axion coupling produces a frequency shift that generically
begins at $O(\bar g^2)$. Analyticity and the symmetry
$(\bar g,\delta\phi)\to(-\bar g,-\delta\phi)$ imply that the
frequency expansion contains only even powers of $\bar g$.
The isolation condition applies within the odd-parity block;
degeneracy with an even-parity mode does not invalidate the
argument because the two parity sectors remain decoupled
on the purely electric background. In the uncharged limit
$Q=0$, the background reduces to Schwarzschild and both the
gravitational--EM and scalar--EM mixing terms vanish.
The gravitational and EM parity pairs therefore remain
isospectral.

\subsection{Five coupled dyonic families}

We next increase the magnetic fraction $\eta$ at fixed $Q/M=0.6$
and $\bar g=5$, keeping $\hat\alpha=\hat\beta=0$.
Magnetic charge sources a background axion profile, changes
the geometry and electric field, and couples the two perturbation
blocks that were separated in the purely electric case.
The resulting modes generally involve gravitational, EM, and
axion perturbations. 
The labels \emph{even} and \emph{odd} are inherited from the
corresponding purely electric modes. At nonzero magnetic
charge, the even- and odd-parity variables are coupled
and must be solved together.

The bottom panels of Fig.~\ref{fig:MIN_01} follow all five
frequency branches from $\eta=0$ into this coupled regime.
At $\eta=0.4$, Table~\ref{tab:minimal_spectrum} gives five
distinct complex frequencies, extending the two-family
analysis of Ref.~\cite{DeFelice:2024axionQNM} to all five
fundamental branches.
Their magnetic frequency shifts are quantified in
Sec.~\ref{sec:magnetic}.
This result concerns the sampled parameters; special limits
or accidental crossings can still produce degeneracies.

For the three parameter sets in Table~\ref{tab:minimal_spectrum},
we test convergence using the collocation method of
Sec.~\ref{sec:pert} at $N_{\rm c}=36,48,60$.
Writing the frequency of family $j$ as $\Omega_j^{(N_{\rm c})}$,
the largest changes over all five families are below
$9.2\times10^{-12}$ between 36 and 48 nodes and
$1.4\times10^{-11}$ between 48 and 60 nodes.
At fixed $N_{\rm c}=36$, tightening the background-integration
tolerance changes the frequencies by less than $3.4\times10^{-12}$.

We compare the numerical variations above with the frequency
differences between distinct families at fixed physical parameters.
For the two $\bar g=5$ cases with $\eta=0$ and $\eta=0.4$,
we define the smallest separation in the complex-frequency
plane by
\begin{equation}
d_{\min}=\min_{j<k}|\Omega_j-\Omega_k|,
\label{eq:minimal_mode_gap}
\end{equation}
where $j,k$ label the five QNM families in
Table~\ref{tab:minimal_spectrum}.
At $\eta=0$, the closest pair consists of the
even-gravitational and odd-EM modes, with
$d_{\min}\simeq2.39\times10^{-2}$.
At $\eta=0.4$, the closest pair consists of the
families continued from the even- and odd-gravitational RN modes, with
$d_{\min}\simeq1.77\times10^{-2}$.
Both minimum separations greatly exceed the measured numerical
variations, supporting the identification of five distinct
frequencies at these benchmarks.
We next discuss an exact EM--scalar degeneracy in the
electric RN spectrum at $\bar g=0$.

\subsection{An exact RN electromagnetic--scalar degeneracy}
\label{sec:RN_Darboux}

The five-family spectra above were evaluated at $Q/M=0.6$.
We now vary $Q/M$ in the purely electric RN reference system
($\eta=0$), keeping $\bar g=\hat\alpha=\hat\beta=0$, to identify
the charge ratio at which the odd-EM and odd-scalar modes
have identical complex QNM frequencies.
In this limit, the scalar perturbation decouples from the
gravitational--EM system, so the two families obey separate
radial equations. Nevertheless, their frequencies can coincide
at a special charge, with equal oscillation frequencies and
damping rates. This degeneracy occurs within the odd-parity
sector and is distinct from the axial--polar degeneracy
discussed in Sec.~\ref{sec:electric_splitting}.
Identifying this special charge also helps explain how the
EM and scalar frequencies respond when a small nonzero
axion coupling is introduced and the two modes become coupled.

Rout~\cite{Rout:2026RNdegeneracy} derived this degeneracy
in the decoupled RN sector of dynamical Chern--Simons gravity.
Equation~(67) of that work gives
$Q_c/M=2\sqrt{L+1}/(L+2)$, with $L=\ell(\ell+1)$,
which reduces to $\sqrt7/4$ for $\ell=2$.
The overlap concerns the common Einstein--Maxwell plus free
massless-scalar limit; the finite-coupling interactions differ.
We retain the explicit $\ell=2$ Darboux map below to interpret
the mixing in our axion--photon theory.

To recover this coincidence, we use the potential matrix
$\mathsf V_{\rm RN}$ in Eq.~\eqref{eq:RNmatrix} at $\bar g=0$.
We denote by $V_E$ the EM eigenpotential obtained by
diagonalizing its gravitational--EM $2\times2$ block, and by
$V_\phi$ its third diagonal entry, which governs the decoupled
massless-scalar perturbation.
For $\ell=2$ ($L=6$), we determine the special charge $Q_\star$
by requiring $V_E$ and $V_\phi$ to admit the Darboux
representation in Eq.~\eqref{eq:Darboux} below.
Matching the coefficients of inverse powers of $r$ gives
$\sqrt{9M^2+16Q_\star^2}=4M$.
The resulting charge and corresponding outer and inner
horizon radii are
\begin{equation}
\frac{Q_\star}{M}=\frac{\sqrt7}{4},\qquad
r_+=\frac{7M}{4},\qquad r_-=\frac{M}{4}.
\label{eq:RN_special_charge}
\end{equation}
At this charge, the two potentials can be expressed as
\begin{align}
V_E&=W^2+\frac{{\rm d}W}{{\rm d}r_*}
-\frac{144}{M^2},\nonumber\\
V_\phi&=W^2-\frac{{\rm d}W}{{\rm d}r_*}
-\frac{144}{M^2},
\label{eq:Darboux}
\end{align}
where a common function $W$ is given by 
\begin{equation}
W=\frac{12}{M}+\frac{M}{4r^2}
-\frac{Q_\star^2}{r^3}.
\end{equation}
Here $\psi_E$ and $\psi_\phi$ denote the individual EM and
scalar radial master functions, respectively.
The function $\psi_E$ is a linear combination of
$Z_{\rm grav}$ and $Z_{\rm EM}$ obtained by diagonalizing
the gravitational--EM block.
The two radial solutions are connected by the first-order
Darboux transformation
\begin{equation}
\psi_E=\left(\frac{{\rm d}}{{\rm d}r_*}+W\right)\psi_\phi.
\label{eq:Darboux_map}
\end{equation}
The function $W$ is called the superpotential.
Since $W\to12/M$ at both the horizon and infinity, this
transformation preserves the ingoing and outgoing character
of the QNM solutions.
For $\omega\neq\pm12i/M$, the inverse transformation also
exists, establishing equality of the EM and scalar spectra
away from these exceptional factorization frequencies.

For the common fundamental mode, we obtain numerically
\begin{equation}
M\omega_E^{(0)}=M\omega_\phi^{(0)}
=0.52618687-0.09850443i,
\end{equation}
where the superscript $(0)$ denotes the $\bar g=0$ reference
system and the displayed value is rounded.
The equality of both the oscillation frequencies and damping
rates follows analytically from the Darboux relation and
the preservation of the QNM boundary conditions.

Turning on a small nonzero $\bar g$ couples the EM and scalar
modes. At a fixed charge ratio, suppose that these uncoupled
fundamental modes are simple and isolated within the odd-parity
sector, with a nonzero complex-frequency gap,
$\omega_E^{(0)}\ne\omega_\phi^{(0)}$.
When the axion-induced mixing is weak compared with the
magnitude of this complex-frequency difference, nondegenerate
perturbation theory gives a mixing contribution proportional
to $\bar g^2/(\omega_E^{(0)}-\omega_\phi^{(0)})$.
The opposite trends of the odd-EM and scalar real frequencies
and damping rates shown in panels (a) and (b) of
Fig.~\ref{fig:MIN_01} at $Q/M=0.6$ are consistent with
this mixing mechanism.
Coupling to the remaining odd-parity mode also contributes,
so the two-mode estimate alone does not determine the full
complex shifts.

At $Q=Q_\star$, the uncoupled frequency gap vanishes,
and the preceding nondegenerate expansion is no longer valid.
The EM and scalar modes must instead be treated together
using degenerate perturbation theory. The coupling within
this two-mode subspace is linear in $\bar g$ and produces
an $O(\bar g)$ frequency splitting provided the product of
the projected off-diagonal couplings is nonzero, as verified
numerically for the fundamental pair.
This linear splitting at the degenerate point is distinct
from the quadratic corrections obtained at fixed nonzero
frequency gap. The degenerate case therefore lies outside
the isolated-mode assumption used in
Sec.~\ref{sec:electric_splitting}.

\subsection{Magnetic response of the five families}
\label{sec:magnetic}

We now study how magnetic charge modifies the five oscillatory
QNM families at nonzero axion coupling, retaining
$\hat\alpha=\hat\beta=0$.
Varying the magnetic fraction $\eta$ at fixed $Q/M$ and
$\bar g$ changes the axion profile, geometry, electric field,
and coupled perturbation operator.
For each family $j$, let $\Omega_j(\eta)$ denote its frequency
and $\Omega_j(0)$ its purely electric reference at the same
$Q/M$ and $\bar g$.
The family labels are assigned by continuation from the
corresponding modes in the axion-decoupled electric RN limit
at each fixed charge ratio.
We define the magnetic frequency shift and, for $\eta\ne0$,
its normalized magnitude by
\begin{equation}
\Delta_\eta\Omega_j=\Omega_j(\eta)-\Omega_j(0),\qquad
\mathcal R_j=\frac{|\Delta_\eta\Omega_j|}{\eta^2}.
\label{eq:R}
\end{equation}
The normalization reflects the symmetry
$(q_M,\phi)\to(-q_M,-\phi)$.
For an isolated branch analytic at $\eta=0$, its frequency
is even in $\eta$, so $\Delta_\eta\Omega_j=O(\eta^2)$.
Degenerate reference modes require a coupled perturbative
treatment.
At finite $\eta$, $\mathcal R_j$ also includes higher-order
magnetic corrections.

We first consider the $Q/M=0.6$, $\bar g=5$ sequence
in the bottom panels of Fig.~\ref{fig:MIN_01}.
Panel (c) shows that all five real frequencies vary with
$\eta$, including the two families of electric even-parity
origin whose frequencies at $\eta=0$ are independent of
$\bar g$.
Increasing $\eta$ from $0$ to $0.4$ raises the real part of
the odd-gravitational frequency and lowers that of the
scalar frequency.
The frequencies in Table~\ref{tab:minimal_spectrum} give
$\mathcal R_j\simeq0.285$ and $0.265$ for the
odd-gravitational and scalar families at $\eta=0.4$,
respectively, the two largest responses at this point.
Panel (d) shows the corresponding damping rates,
$\Gamma_j=-\operatorname{Im}\Omega_j$.
The odd-gravitational damping rate first decreases and then
increases, crossing the even-gravitational rate, while
remaining separate from the odd-EM rate over the plotted range.
The real frequencies remain distinct at this crossing.
Equality of the damping rates therefore does not imply
the complex-frequency degeneracy established at the
electric Darboux point.

To examine how the magnetic response depends on the axion
coupling, we next fix $Q/M=0.6$ and $\eta=0.1$ and compare
$\bar g=0.1,0.9,3,10$.
The even-EM family has the largest $\mathcal R_j$ at
$\bar g=0.1,0.9$, whereas the scalar family has the largest
at $\bar g=3,10$.
The odd-gravitational response also increases substantially
over these sampled couplings.
These changes describe the relative magnitudes of the
frequency shifts; they do not by themselves establish
a resonance or an exchange of mode content.

We next use the largest sampled coupling, $\bar g=10$,
to test the charge dependence of these enhanced shifts
and search for additional growing modes.
Keeping $\eta=0.1$, we compare $Q/M=0.4,0.6,0.7,0.8$.
The families are identified separately at each charge as
defined above, without continuous branch tracking in $Q/M$.
Across these benchmarks, the odd-EM response increases
from $\mathcal R_j\simeq0.0250$ at $Q/M=0.4$ to $8.65$
at $Q/M=0.8$.
The scalar response varies nonmonotonically, with
$\mathcal R_j\simeq0.772,2.89,0.460,0.954$, respectively.
The odd-gravitational response is largest at $Q/M=0.7$
among the four benchmarks and is smaller at $Q/M=0.8$.
The five complex QNM frequencies at $Q/M=0.8$ and their
resolution tests are given in Table~\ref{tab:spectral_resolution}
(Appendix~\ref{app:numerics}).

At this stronger coupling, the oscillatory families alone
do not determine the stability of the backgrounds.
An $\ell=2$ search including purely imaginary frequencies
at $\bar g=10$ and $\eta=0.1$ also finds
$\Omega\simeq0.06720i,0.13343i,0.17560i$ at
$Q/M=0.6,0.7,0.8$, respectively
(Table~\ref{tab:growing_modes} in Appendix~\ref{app:numerics}).
With the time dependence $e^{-i\omega t}$, these positive
imaginary frequencies imply exponential growth, so all three
backgrounds are mode unstable.
The responses quoted above at these points thus describe
oscillatory modes coexisting with an instability.
The growing-mode frequencies converge with radial resolution
and are independently reproduced by a bound-state calculation
whose solutions decay spatially toward both the horizon
and infinity (Appendix~\ref{app:growing_modes}).
This finite-frequency instability is not excluded by the
local high-frequency conditions in Sec.~\ref{sec:model};
strong-coupling axionic instabilities are already known
in the electric EMA system~\cite{Boskovic:2018lkj}.
These unstable benchmarks are separate from the lower-coupling
parameter sequences shown in Figs.~\ref{fig:MIN_01}--\ref{fig:V_11}.

\section{Nonlinear-electrodynamic frequency responses}
\label{sec:ned}

We now determine the additional frequency shifts due to the two
quartic interactions. We separate background and perturbative
effects on purely electric BHs, study charge composition and
the RN parity-pair relation, and compute individual shifts
at finite NED coefficients.

\subsection{Reference spectra and frequency shifts}
\label{sec:ned_shifts}

For each chosen $(Q/M,\bar g,\eta)$, we compare the NED
frequencies with those at $\hat\alpha=\hat\beta=0$
and the same parameters.
In this section the two arguments of $\Omega_j(\hat\alpha,
\hat\beta)$ are the horizon-normalized NED coefficients defined
in Eq.~\eqref{eq:params}.
The index $j$ identifies a QNM family, and each difference
compares continuously connected eigenmodes.
Let $\Omega_{j,0}=\Omega_j(0,0)$ and
$\Gamma_{j,0}=-\operatorname{Im}\Omega_{j,0}$ denote the reference
frequency and damping rate of family $j$.
We define the full NED shift, the percentage changes in
the real frequency and damping rate, and the magnetic
modification of the NED shift as
\begin{align}
\Delta\Omega_j&=\Omega_j(\ah,\bh)-\Omega_j(0,0),\label{eq:ned_shift}\\
\delta_{R,j}&=100\frac{\operatorname{Re}\Delta\Omega_j}{\operatorname{Re}\Omega_{j,0}},\qquad
\delta_{\Gamma,j}=100\frac{-\operatorname{Im}\Delta\Omega_j}{\Gamma_{j,0}},\label{eq:percent}\\
B_j(\eta)&=\Delta\Omega_j(\eta)-\Delta\Omega_j(0).\label{eq:B}
\end{align}
The damping-rate shift is
$\Delta\Gamma_j=\Gamma_j-\Gamma_{j,0}
=-\operatorname{Im}\Delta\Omega_j$.
In Eq.~\eqref{eq:B}, the displayed argument is the magnetic
fraction $\eta$, while the NED arguments are suppressed.
The quantity $B_j$ compares the NED shift at $\eta$ with
that at the electric endpoint $\eta=0$, keeping
$\hat\alpha$, $\hat\beta$, $Q/M$, and $\bar g$ fixed.
It therefore isolates the magnetic dependence of the
NED correction; the magnetic shift already present in
the minimal theory is included in the reference at each $\eta$.
We form $B_j$ from complex shifts rather than percentages,
whose normalization generally depends on the family and $\eta$.
For a damped mode, $\Gamma_j=-\operatorname{Im}\Omega_j>0$,
so $\delta_{\Gamma,j}<0$ means that NED reduces the damping
rate and slows the decay.

To vary the NED coefficients at a fixed ratio, we write
$(\hat\alpha,\hat\beta)=\lambda(a_\alpha,a_\beta)$,
where the dimensionless parameter $\lambda$ sets the scale
of the two hatted coefficients and $(a_\alpha,a_\beta)$
fixes their ratio.
We use $(1,0),(0,1),(1,1)$ to separate the two interactions
and test the equal-coefficient case, and $(4/7,1)$ for
the unequal heavy-Dirac matching ratio.
At $\lambda=0$, all sequences reduce to the minimal axion theory.
At each parameter point, we recompute the background and
ADM mass and adjust the total charge to keep $Q/M$ fixed.
When varying $\eta$, we hold $\lambda$ and the chosen
coefficient ratio fixed, so both $\hat\alpha$ and $\hat\beta$
remain constant.
The corresponding coefficients normalized by the ADM mass
can nevertheless vary because $M/r_h$ need not remain
constant, as explained in Sec.~\ref{sec:model}.

\subsection{Purely electric backgrounds and NED shifts}
\label{sec:electric_ned}

\begin{figure*}[t]
\centering
\includegraphics[width=\textwidth]{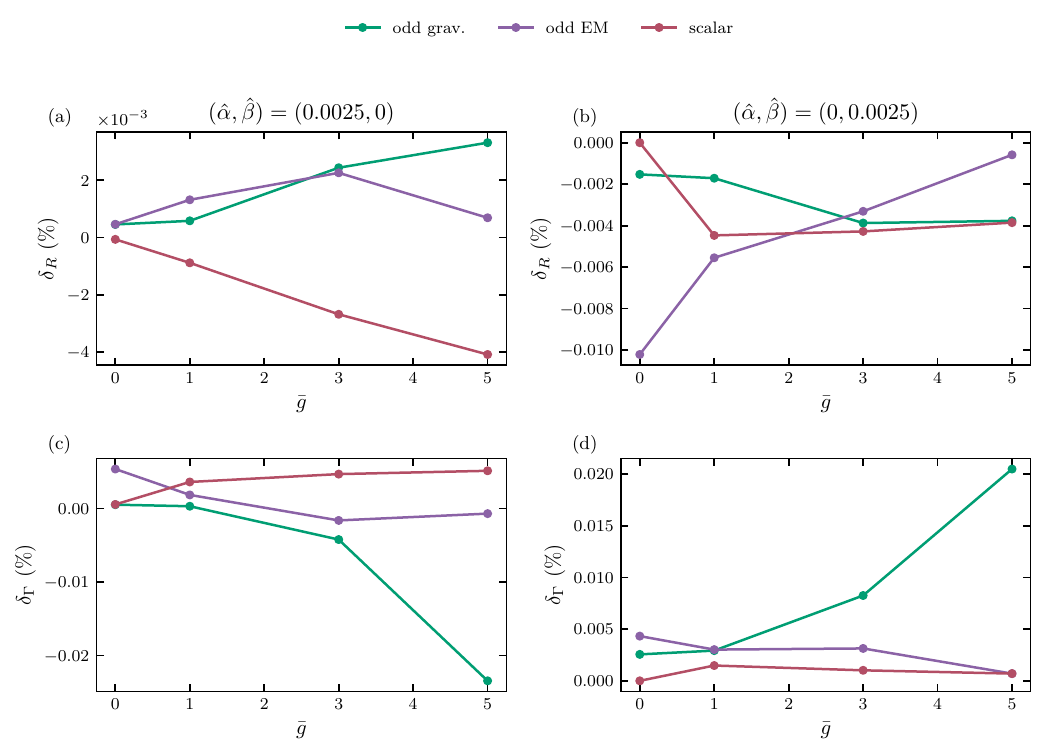}
\caption{NED-induced percentage shifts of the odd-gravitational,
odd-EM, and scalar $\ell=2$ QNMs for purely electric
BHs ($\eta=0$) at $Q/M=0.6$.
The horizontal axis is the axion coupling $\bar g$
in all panels.
Panels (a,c) use $(\hat\alpha,\hat\beta)=(0.0025,0)$,
and panels (b,d) use $(\hat\alpha,\hat\beta)=(0,0.0025)$.
The upper and lower panels show the real-frequency and
damping-rate shifts, respectively, defined in
Eq.~\eqref{eq:percent}.
Each shift is measured relative to the corresponding mode
in the minimal axion theory ($\hat\alpha=\hat\beta=0$)
at the same $\bar g$.
Negative damping-rate shifts indicate slower decay;
all plotted modes remain damped.}\label{fig:IV_02}
\end{figure*}

We first consider purely electric BHs ($\eta=0$).
With $q_M=0$, Eq.~\eqref{eq:phibg} and horizon regularity
give $\phi'=0$; Eq.~\eqref{eq:Nbg} then gives $N'=0$.
The asymptotic conditions $\phi(\infty)=0$ and $N(\infty)=1$
fix $\phi=0$ and $N=1$.
For $B=0$, the electric-field equation~\eqref{eq:gauss}
and the metric equation~\eqref{eq:hbg}, using the energy
density in Eq.~\eqref{eq:rho}, reduce to
\begin{align}
r^2E(1+\alpha E^2)&=q_E,\nonumber\\
m'&=\frac{r^2}{\Mp^2}
\left(\frac{E^2}{4}+\frac{3\alpha E^4}{8}\right).
\end{align}
Here $m=r(1-h)/2$ is the mass function.
Of the two NED coefficients, only $\alpha$ modifies this
purely electric background.
The $\beta$ and axion interactions nevertheless contribute to
the quadratic perturbation action through
\begin{equation}
S_{\beta,g}^{(2)}=\int {\rm d}^4 x\sqrt{-\bar g_{\rm M}}
\left[\beta(\delta\cG)^2+g\delta\phi\,\delta\cG\right].
\end{equation}
Here $\bar g_{\mu\nu}$ is the background metric and
$\bar g_{\rm M}=\det(\bar g_{\mu\nu})$.
The quantities $\delta\phi$ and $\delta\cG$ denote the
first-order perturbations of the axion field and the
electromagnetic invariant $\cG$, respectively.
Although $\bar{\cG}=0$ on the purely electric background,
$\delta\cG$ can be nonzero.
The angular curl of an even-parity electromagnetic harmonic
vanishes, giving $\delta\cG_{\rm even}=0$, whereas odd-parity
electromagnetic perturbations can contribute to $\delta\cG$.
The perturbation equations therefore separate into an
even-parity two-field block depending only on $\alpha$
and an odd-parity three-field block depending on
$\alpha,\beta,g$.
In particular, varying $\beta$ can shift the odd-parity
frequencies while leaving the electric background and
even-parity frequencies unchanged.

Nomura and Yoshida~\cite{Nomura:2021efi} computed QNMs
for magnetic BHs and argued in Appendix A4 that their
first-order NED shifts also apply to electric BHs through
an electric--magnetic duality mapping.
Our independently computed electric even-parity modes
provide a numerical test of this prediction.
We compare the fundamental even-gravitational and even-EM
modes at $\ell=2$, $\bar g=0$, and $Q/M=0.4,0.6,0.8$.
The comparison uses their dimensionless coefficients
$\bar\alpha_{\rm NY}=\alpha q_E^2/M^4$ and
$\bar\beta_{\rm NY}=\beta q_E^2/M^4$, together with the
negative-real-frequency branch $\Omega_{\rm NY}=-\Omega^*$,
where the asterisk denotes complex conjugation.
In these conventions, our derivatives of $\Omega_{\rm NY}$
with respect to $\bar\alpha_{\rm NY}$ at zero NED coefficients
agree with the magnetic type-II coefficients in their
Table I within the rounding precision of the published values.
At $Q/M=0.6$ and $\bar g=0$, the odd-gravitational
and odd-EM first-order responses to both coefficients
also agree with their magnetic type-I results under
the same conversion.
These benchmarks concern the axion-decoupled theory
to first NED order; the finite-axion results below
are computed from the coupled axion system.

We next quantify finite NED shifts for purely electric BHs.
At $\ell=2$, $Q/M=0.8$, $\hat\alpha=0.1$, and
$\hat\beta=0$, the even-EM real frequency is about $0.87\%$
lower and its damping rate about $0.28\%$ higher than
their values at $\hat\alpha=\hat\beta=0$, with the same
$Q/M$ and $\bar g$.
Both even-parity frequencies remain independent of $\bar g$,
because the axion interaction enters only the odd-parity
block on this background.

Figure~\ref{fig:IV_02} shows the odd-sector NED shifts
at fixed $Q/M=0.6$ and $\eta=0$.
We vary $\bar g=0,1,3,5$ for the two fixed choices
$(\hat\alpha,\hat\beta)=(0.0025,0)$ and $(0,0.0025)$.
Each shift is measured relative to the minimal axion
spectrum at the same $\bar g$.
For $(\hat\alpha,\hat\beta)=(0.0025,0)$, the
odd-gravitational and odd-EM damping corrections
in panel (c) are positive at $\bar g=0,1$ and negative
at $\bar g=3,5$.
For $(\hat\alpha,\hat\beta)=(0,0.0025)$, both corrections
in panel (d) remain positive at all sampled couplings,
and the odd-gravitational correction increases with
$\bar g$.
The scalar damping corrections are nonnegative in both
NED cases.
The vanishing correction at $\bar g=0$ in panel (d)
follows from scalar decoupling and the $\beta$-independence
of the purely electric background.
Negative damping corrections indicate slower decay
relative to the minimal axion model; the full damping
rates remain positive.

\subsection{Charge composition and electromagnetic duality}
\label{sec:ned_composition}

\begin{figure*}[t]
\centering
\includegraphics[width=\textwidth]{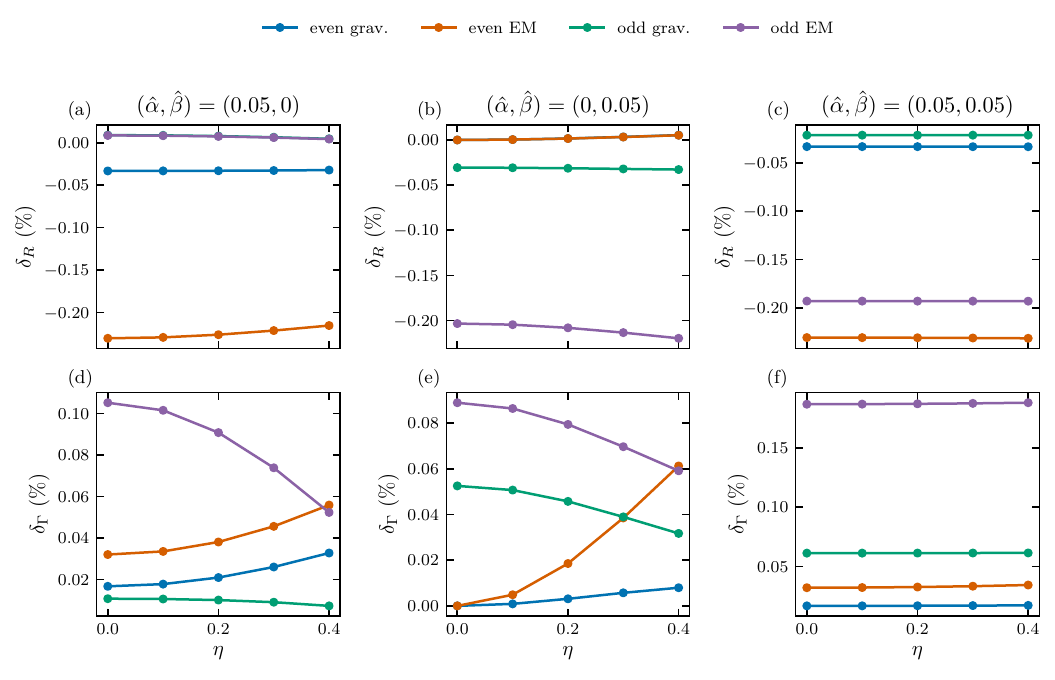}
\caption{NED-induced percentage shifts of the $\ell=2$ gravitational
and EM QNMs as functions of the magnetic fraction $\eta$,
at fixed $Q/M=0.6$ and $\bar g=0$.
From left to right, the columns correspond to
$(\hat\alpha,\hat\beta)=(0.05,0),(0,0.05),(0.05,0.05)$.
The upper and lower panels show the real-frequency and
damping-rate shifts, respectively, defined in
Eq.~\eqref{eq:percent}. Each shift is measured relative 
to the NED-free spectrum at the same $\eta$.
Family labels identify modes continued from the purely
electric limit at the same NED coefficients.
For $(\hat\alpha,\hat\beta)=(0.05,0.05)$, each branch's
shift is independent of $\eta$ through first NED order
[Eq.~\eqref{eq:duality_individual}]; residual dependence
at finite coefficients is allowed at higher order.
The full NED shifts remain nonzero.}
\label{fig:V_02}
\end{figure*}

Here we determine how electric--magnetic charge composition
modifies the NED frequency shifts.
We first vary $\eta$ at fixed $Q/M$, $\bar g$, and NED
coefficients, computing the full individual shifts
$\Delta\Omega_j(\eta)$ in Eq.~\eqref{eq:ned_shift}
and their magnetic modifications $B_j(\eta)$ in
Eq.~\eqref{eq:B}.
The duality map in Appendix~\ref{app:duality} explains two
different results at $\bar g=0$.
For $\hat\alpha=\hat\beta=\lambda$, it relates different
charge compositions within the same theory through first
order in $\lambda$, making each individual branch's linear
NED shift independent of $\eta$
[Eq.~\eqref{eq:duality_individual} below].
For unequal coefficients, it instead constrains the
gravitational and EM pair sums $B_J$ through
Eq.~\eqref{eq:dualitylaw}.
The response formulas are given in Appendix~\ref{app:response}.
We also vary $\lambda$ at fixed $\eta$ to separate linear
from higher-order magnetic dependence and test the loss
of the cancellation at nonzero axion coupling.

Figure~\ref{fig:V_02} compares the full NED shifts at
$Q/M=0.6$ and $\bar g=0$ for
$(\hat\alpha,\hat\beta)=(0.05,0),(0,0.05),(0.05,0.05)$,
with each coefficient choice held fixed as $\eta$ varies.
The gravitational and EM families are continued from $\eta=0$
at the same NED coefficients.
The plotted percentages refer to the NED-free spectrum
at each $\eta$, as defined in Eq.~\eqref{eq:percent}.

For $(\hat\alpha,\hat\beta)=(0.05,0)$, the
even-gravitational and even-EM families already have
nonzero NED shifts at $\eta=0$.
For $(\hat\alpha,\hat\beta)=(0,0.05)$, the shifts of
these families vanish at $\eta=0$ and become nonzero
when magnetic charge is introduced.
In both single-coefficient cases, the even-EM damping
correction increases with $\eta$ over the plotted range,
whereas the odd-EM correction decreases.
For equal coefficients, $(\hat\alpha,\hat\beta)=(0.05,0.05)$,
the frequency and damping shifts of each individual branch
are nearly constant as $\eta$ varies.
This is the finite-coefficient manifestation of the stronger
branchwise relation~\eqref{eq:duality_individual}, rather
than only a cancellation between parity partners.
The full shifts remain nonzero; their residual magnetic
dependence can begin at second NED order.

\begin{figure*}[t]
\centering
\includegraphics[width=\textwidth]{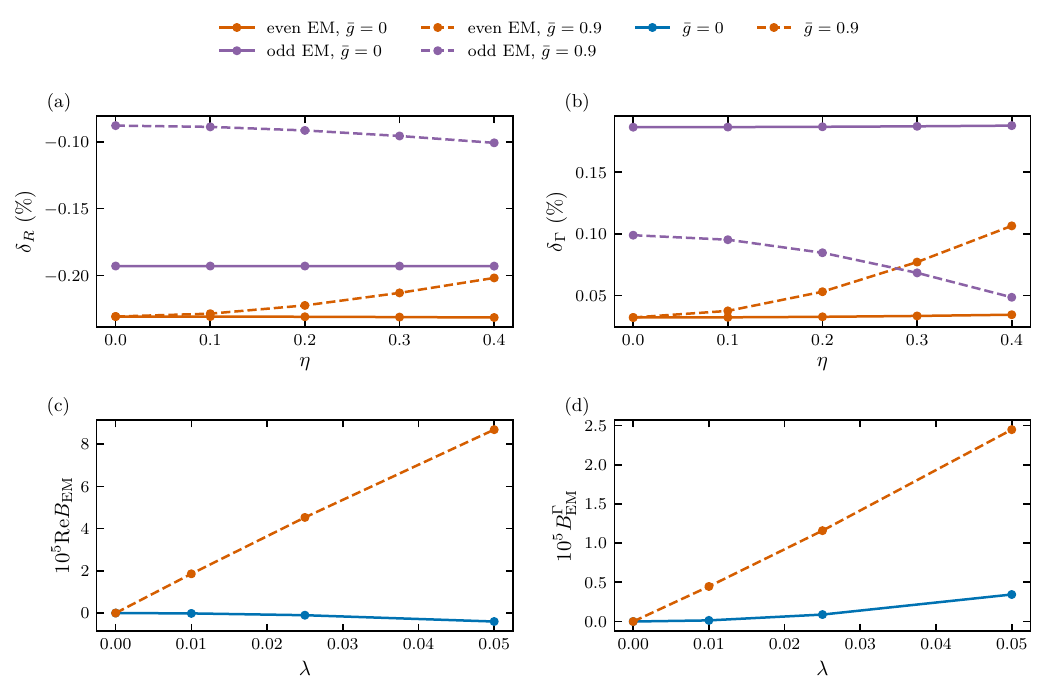}
\caption{Full NED shifts and their magnetic modification for
equal coefficients $(\hat\alpha,\hat\beta)=(\lambda,\lambda)$
at $\ell=2$ and $Q/M=0.6$.
Upper panels: real-frequency and damping percentages of the
two EM families as functions of $\eta$ at $\lambda=0.05$.
Lower panels: $\operatorname{Re}B_{\rm EM}$ and
$B_{\rm EM}^{\Gamma}\equiv-\operatorname{Im}B_{\rm EM}$,
both scaled by $10^5$,
as functions of $\lambda$ at $\eta=0.4$.
Solid and dashed lines denote $\bar g=0$ and $0.9$,
respectively.
At $\bar g=0$, Eq.~\eqref{eq:duality_individual} predicts
first-order magnetic cancellation for each individual
branch, and hence for their sum.
The full NED shifts remain nonzero.}\label{fig:V_05}
\end{figure*}

To interpret the magnetic dependence of the pair sums,
we retain $\bar g=0$ and use the purely electric RN
reference $\eta=\hat\alpha=\hat\beta=0$.
At this reference, the even- and odd-gravitational modes
share one complex QNM frequency, while the even- and
odd-EM modes share a different complex frequency.
These define the gravitational and EM RN parity pairs,
respectively.

We expand each pair's frequency sum to first order in
$\hat\alpha$ and $\hat\beta$ at fixed $Q/M$.
Electromagnetic duality then determines its magnetic
dependence from derivatives evaluated at this electric reference.
For $J=\mathrm{grav},\mathrm{EM}$, we define
\begin{align}
\Sigma_J(\eta)&=\Omega_{J,e}(\eta)+\Omega_{J,o}(\eta),
\nonumber\\
B_J(\eta)&=B_{J,e}(\eta)+B_{J,o}(\eta).
\label{eq:RN_pair_sums}
\end{align}
Here $\Omega_{J,e}$ and $\Omega_{J,o}$ are the corresponding
members of $\Omega_j(\hat\alpha,\hat\beta)$ in
Eq.~\eqref{eq:ned_shift}; their NED arguments are implicit.
The indices $e,o$ specify parity at $\eta=0$ and label the
continued families at nonzero $\eta$, rather than individual
perturbation fields.
Each $B_{J,e}$ or $B_{J,o}$ is defined by Eq.~\eqref{eq:B}.
Thus $B_J$ is the sum of the pair's two NED shifts
at $\eta$, minus their sum at $\eta=0$, with the same
NED coefficients, $Q/M$, and $\bar g$.

At $\bar g=0$, the NED-free reference frequency is independent
of $\eta$ and common to both members of each RN pair.
Writing it as $\Omega_{J,0}$, with
$\Gamma_{J,0}=-\operatorname{Im}\Omega_{J,0}$,
Eqs.~\eqref{eq:percent} and~\eqref{eq:RN_pair_sums} give
\begin{align}
100\frac{\operatorname{Re}B_J(\eta)}{\operatorname{Re}\Omega_{J,0}}
&=\sum_{s=e,o}[\delta_{R,J,s}(\eta)-\delta_{R,J,s}(0)],\\
100\frac{-\operatorname{Im}B_J(\eta)}{\Gamma_{J,0}}
&=\sum_{s=e,o}[\delta_{\Gamma,J,s}(\eta)-\delta_{\Gamma,J,s}(0)].
\end{align}
Thus, to recover $B_J$ from Fig.~\ref{fig:V_02}, we add
the two curves of a pair, subtract their summed value
at $\eta=0$, and undo the percentage normalization.
This conversion is exact at the plotted finite NED
coefficients; the duality prediction below is first order.

The two electric-reference response coefficients are
\begin{align}
T_{J,\alpha}=\left.
\frac{\partial\Sigma_J}{\partial\hat\alpha}\right|_0,\qquad
T_{J,\beta}=\left.
\frac{\partial\Sigma_J}{\partial\hat\beta}\right|_0.
\label{eq:RN_pair_derivatives}
\end{align}
The subscript $0$ indicates evaluation at
$\eta=\bar g=\hat\alpha=\hat\beta=0$ for the chosen $Q/M$.
Each partial derivative varies one hatted NED coefficient
while keeping the other coefficient zero and
$\eta=\bar g=0$.
The derivatives include the charge retuning and
background-dependent ADM-mass normalization at fixed $Q/M$.

We assume that the common frequency of each RN pair is
isolated from the remaining spectrum and semisimple,
with two independent eigenmodes and no generalized
eigenmodes, and that its projected frequency-derivative
matrix $N_J$ is invertible.
Equations~\eqref{eq:RN_projected_matrices}
and~\eqref{eq:RN_pair_response_trace} in
Appendix~\ref{app:response} define this matrix and give
the summed frequency derivatives.
For both RN pairs at $Q/M=0.6$ and $\ell=2$,
the two diagonal entries of $N_J$ in
Eq.~\eqref{eq:RN_parity_frequency_projection} remain
nonzero at $N_{\rm c}=36,48,60$, supporting its invertibility.
Appendix~\ref{app:operator_checks} summarizes these numerical checks.

At $\bar g=0$, combining Eqs.~\eqref{eq:duality_pair_shift}
and~\eqref{eq:duality_pair_magnetic_difference} in
Appendix~\ref{app:duality}, we obtain the magnetic
modification of the summed NED frequency shift:
\begin{align}
B_J={}&-4\eta^2(1-\eta^2)(\ah-\bh)(T_{J,\alpha}-T_{J,\beta})
\nonumber\\*
&+O(\hat\alpha^2,\hat\alpha\hat\beta,\hat\beta^2).
\label{eq:dualitylaw}
\end{align}
The duality relation~\eqref{eq:dualitylaw} is restricted
to $\bar g=0$.
Its remainder contains terms quadratic or higher order
in the NED coefficients at this fixed axion coupling.
The factor $4\eta^2(1-\eta^2)=\sin^2(2\vartheta)$
arises from the electromagnetic duality rotation,
whose angle $\vartheta$ satisfies $\eta=\sin\vartheta$.

For equal coefficients, a stronger statement holds.
We define the linear response along
$(\hat\alpha,\hat\beta)=(\lambda,\lambda)$ by
$\kappa_j^{\rm eq}(\eta)=
\left.\partial_\lambda\Omega_j(\eta;\lambda,\lambda)
\right|_{\lambda=0}$ at fixed $Q/M$ and $\bar g=0$.
For each gravitational or EM branch continued from its
electric-reference mode, the same-theory duality map gives
\begin{equation}
\begin{aligned}
\kappa_j^{\rm eq}(\eta)&=\kappa_j^{\rm eq}(0),\\
B_j(\eta;\lambda,\lambda)&=O(\lambda^2).
\end{aligned}
\label{eq:duality_individual}
\end{equation}
Appendix~\ref{app:duality_equal} derives this result without
taking a trace over the RN pair and checks it using the
independent integration-and-matching data.
It implies the equal-coefficient pair-sum cancellation in
Eq.~\eqref{eq:dualitylaw}, but also rules out compensating
first-order magnetic shifts of the individual partners.
The nonzero values of $\kappa_j^{\rm eq}(0)$ still allow
nonzero full first-order NED shifts.
Nor does Eq.~\eqref{eq:duality_individual} require the two
parity partners to have equal NED shifts.
The quartic truncation is self-dual only through first
NED order, so no exact finite-$\lambda$ composition
independence is asserted.

For $(\hat\alpha,\hat\beta)=(\lambda,0)$ and $(0,\lambda)$
with the same small $\lambda$, the first-order contribution
to $B_J$ changes by an overall minus sign:
its real and imaginary parts each reverse sign
without changing their magnitudes.
This sign reversal concerns $B_J$, rather than the individual
full NED shifts plotted as percentages in Fig.~\ref{fig:V_02}.

At $\bar g=0$ and fixed $Q/M$, $\hat\alpha$, and $\hat\beta$,
Eq.~\eqref{eq:dualitylaw} shows that the part of $B_J$
linear in the NED coefficients is proportional to
$\eta^2(1-\eta^2)$.
When this linear contribution is nonzero, its value
at $\eta=0.4$ is therefore $3.5$ times its value
at $\eta=0.2$.
The first-order coefficients of $B_J$ extracted from
the numerical frequencies are consistent with this ratio.

We next test the equal-coefficient cancellation at nonzero
axion coupling.
Figure~\ref{fig:V_05} compares $\bar g=0$ and $0.9$
at fixed $Q/M=0.6$, with
$\hat\alpha=\hat\beta=\lambda$ in all panels.
At $\bar g=0.9$ and $\eta=0.4$, $B_{\rm EM}$ contains
a nonzero term linear in $\lambda$, demonstrating the
loss of the first-order magnetic cancellation at this point.
At $\bar g=0.9$, the EM pair denotes the two families
continued from the even- and odd-EM RN modes;
their frequencies need not coincide.
We still define $B_{\rm EM}$ by
Eq.~\eqref{eq:RN_pair_sums} and compute it directly
from the individual frequency shifts in
Eqs.~\eqref{eq:ned_shift} and~\eqref{eq:B}.
Neither zero-axion duality prediction,
Eq.~\eqref{eq:dualitylaw} or~\eqref{eq:duality_individual},
is used for the $\bar g=0.9$ results.

The upper panels fix $\lambda=0.05$ and vary $\eta$,
plotting $\delta_{R,j}$ and $\delta_{\Gamma,j}$ for
the two EM families, as in Fig.~\ref{fig:V_02}.
The lower panels instead fix $\eta=0.4$ and vary
$\lambda$, plotting $\operatorname{Re}B_{\rm EM}$ and
$-\operatorname{Im}B_{\rm EM}$ scaled by $10^5$.
These lower-panel quantities are absolute dimensionless
frequency differences, rather than percentages.
Each point subtracts the pair's summed NED shift at
$\eta=0$ from that at $\eta=0.4$, with the same
$\lambda$ and $\bar g$.
At $\bar g=0.9$, the reference frequencies and damping
rates depend on both family and $\eta$.
To construct $B_{\rm EM}$ from percentages, each shift
must therefore first be converted to $\Delta\Omega_j$
using its own reference values in Eq.~\eqref{eq:percent};
the common RN normalization used above no longer applies.

The suppressed $\bar g=0$ response in the lower panels
is consistent with $B_{\rm EM}=O(\lambda^2)$ for equal
coefficients. Equation~\eqref{eq:duality_individual}
predicts the same first-order cancellation separately
for each branch in the upper panels.
At nonzero axion coupling, the duality rotation of
$g\phi\cG$ also generates a $\phi\cF$ interaction,
which is absent from Eq.~\eqref{eq:model}.
The transformed theory therefore has a different set
of interactions, so equal NED coefficients no longer
guarantee the first-order magnetic cancellation.

To test whether unequal NED coefficients could restore
the cancellation at this parameter point, we compute
the linear responses of $B_{\rm EM}$ to $\hat\alpha$
and $\hat\beta$ separately at $\hat\alpha=\hat\beta=0$.
We obtain these responses from differences of the individual
mode-frequency derivatives at $\eta=0.4$ and $0$,
using the frequency-difference formulas in
Eq.~\eqref{eq:frequency_differences} of
Appendix~\ref{app:response}.
Canceling the real component would require
$\hat\beta/\hat\alpha\simeq-3.58$, whereas canceling
the damping component, $-\operatorname{Im}B_{\rm EM}$,
would require $\hat\beta/\hat\alpha\simeq0.287$.
These different ratios rule out any nonzero real choice
of $(\hat\alpha,\hat\beta)$ that cancels both components
of $B_{\rm EM}$ to first order.
They also exclude simultaneous cancellation of the
EM- and gravitational-pair magnetic modifications
at this point.

\subsection{Individual frequency shifts at finite NED coefficients}
\label{sec:ned_expansion}

\begin{figure*}[t]
\centering
\includegraphics[width=\textwidth]{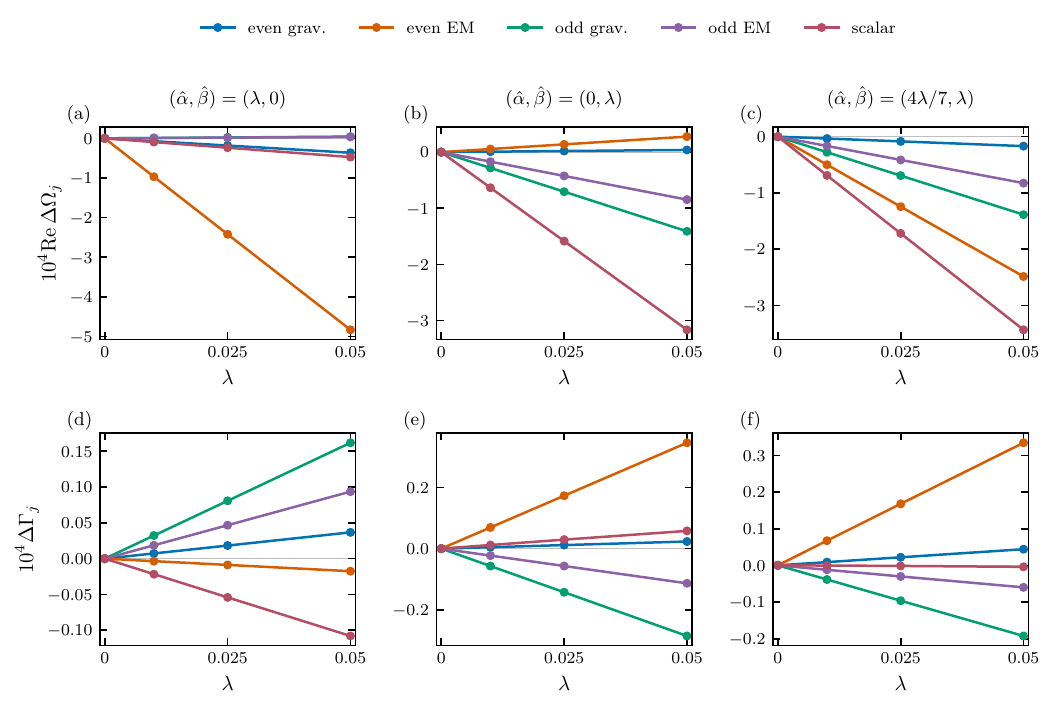}
\caption{NED-induced shifts of the five fundamental $\ell=2$
QNM families at fixed $Q/M=0.6$, $\bar g=5$, and $\eta=0.4$.
From left to right, the columns correspond to
$(\hat\alpha,\hat\beta)=(\lambda,0),(0,\lambda),
(4\lambda/7,\lambda)$, with $\lambda$ on the horizontal axis.
The upper panels show the real-frequency shifts
$10^4\operatorname{Re}\Delta\Omega_j$, and the lower panels
show the damping-rate shifts $10^4\Delta\Gamma_j$,
where $\Delta\Gamma_j=-\operatorname{Im}\Delta\Omega_j$.
Positive damping-rate shifts indicate faster decay.
All shifts are dimensionless and are measured relative
to the corresponding mode at $\hat\alpha=\hat\beta=0$,
with the same $Q/M$, $\bar g$, and $\eta$, as defined
in Eq.~\eqref{eq:ned_shift}.
Points are obtained by solving the full coupled QNM
eigenvalue problem at each finite-coefficient point
using $N_{\rm c}=48$ radial collocation points;
lines connect the computed points.}
\label{fig:V_09}
\end{figure*}

The preceding subsection focused on charge composition
and the magnetic modification $B_J$ of a pair's summed
NED shift.
Here we vary the NED coefficients at fixed charge
composition and determine the five individual frequencies
$\Omega_j(\hat\alpha,\hat\beta)$ and their full shifts
$\Delta\Omega_j$ in Eq.~\eqref{eq:ned_shift}.
The reference is the spectrum at
$\hat\alpha=\hat\beta=0$, with the same $Q/M$, $\bar g$,
and $\eta$; no electric-endpoint subtraction is made here.

We set $\ell=2$, $Q/M=0.6$, $\bar g=5$, and $\eta=0.4$,
and compare three coefficient sequences:
$(\hat\alpha,\hat\beta)=(\lambda,0),(0,\lambda),
(4\lambda/7,\lambda)$, with
$\lambda=0,0.01,0.025,0.05$.
The first two sequences separate the two quartic interactions.
The third has two nonzero, unequal coefficients with
$\hat\beta/\hat\alpha=7/4$, the heavy-Dirac matching ratio
introduced in Sec.~\ref{sec:model}.
We use this ratio within the retained quartic model.
We solve the coupled five-field QNM problem at each
finite-coefficient point, using the background and mass
normalization prescription of Sec.~\ref{sec:ned_shifts}.
Each family is continued from its frequency at
$\hat\alpha=\hat\beta=0$ along the chosen sequence.

Figure~\ref{fig:V_09} shows the shifts in the real frequencies
and damping rates of all five families.
Over the plotted range, the largest real-frequency shift
in magnitude occurs in the even-EM family for $\hat\beta=0$
and in the scalar family for $\hat\alpha=0$.
The odd-gravitational and odd-EM damping corrections are
positive for the sampled $\hat\alpha>0$, $\hat\beta=0$
points and negative for $\hat\alpha=0$, $\hat\beta>0$.
The relative strengths of the two NED interactions can
therefore determine whether these modes decay faster or
more slowly than their NED-free counterparts.

In the purely electric limit ($\eta=0$), the even-parity
frequencies are independent of $\hat\beta$, as established
in Sec.~\ref{sec:electric_ned}.
On the dyonic background considered here ($\eta=0.4$),
however, the $\beta\cG^2$ interaction contributes to both
the background equations and the full coupled perturbation
operator.
The two QNM families continued from the electric
even-parity modes therefore also depend on $\hat\beta$.
For the even-EM family, the damping-rate shift is
$\Delta\Gamma_{{\rm EM},e}\simeq-1.75\times10^{-6}$
at $(\hat\alpha,\hat\beta)=(0.05,0)$, whereas at
$(0,0.05)$ it is $+3.47\times10^{-5}$.
The former reduces the damping rate, while the latter
increases it.

\begin{figure*}[t]
\centering
\includegraphics[width=\textwidth]{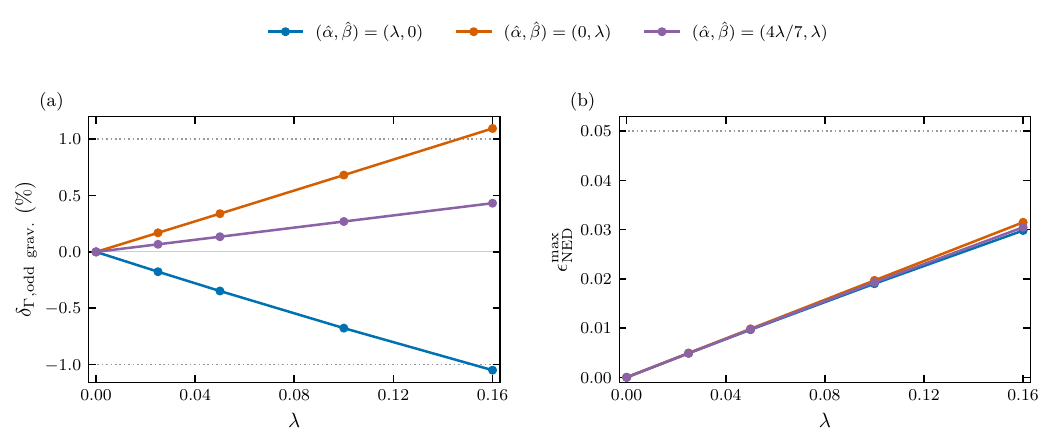}
\caption{NED-induced damping-rate changes and exterior NED amplitudes
at fixed $\ell=2$, $Q/M=0.6$, $\bar g=5$, and $\eta=0.1$.
The horizontal axis is $\lambda$ for the three sequences
$(\hat\alpha,\hat\beta)=(\lambda,0),(0,\lambda),
(4\lambda/7,\lambda)$, extending to $\lambda=0.16$.
Left: the odd-gravitational damping-rate change
$\delta_{\Gamma,j}$ in percent, defined in
Eq.~\eqref{eq:percent}, relative to the corresponding
NED-free mode at the same $Q/M$, $\bar g$, and $\eta$.
Right: $\epsilon_{\rm NED}^{\max}$, the maximum over
the BH exterior of the NED amplitude defined in
Eq.~\eqref{eq:eps}, evaluated on each background.
Dotted horizontal lines indicate damping-rate changes
of $\pm1\%$ in the left panel and the adopted NED
amplitude bound of $0.05$ in the right panel.
Points use directly computed QNM frequencies with
$N_{\rm c}=48$ radial collocation points and backgrounds
recomputed at each choice of NED coefficients;
lines connect the computed points.}
\label{fig:V_11}
\end{figure*}

When both interactions are present along
$(\hat\alpha,\hat\beta)=(4\lambda/7,\lambda)$,
all five families have nonzero real and imaginary
frequency shifts at each of the sampled values
$\lambda=0.01,0.025,0.05$.
For example, at $\lambda=0.05$, the even-EM shift is
$\Delta\Omega_{{\rm EM},e}\simeq
-2.48537\times10^{-4}-3.34318\times10^{-5}i$.
Its real frequency therefore decreases.
The negative imaginary shift gives
$\Delta\Gamma_{{\rm EM},e}
=-\operatorname{Im}\Delta\Omega_{{\rm EM},e}
\simeq+3.34318\times10^{-5}$,
so the damping rate increases and the mode decays faster.

For these fixed parameters, the reference damping rates
in the $\bar g=5$, $\eta=0.4$ column of
Table~\ref{tab:minimal_spectrum} range from approximately
$0.0895$ to $0.1057$.
All corrections in Fig.~\ref{fig:V_09} satisfy
$|\Delta\Gamma_j|<3.5\times10^{-5}$ and are therefore
much smaller than the reference rates.
Hence $\Gamma_j=\Gamma_{j,0}+\Delta\Gamma_j>0$
at every plotted point, so all five sampled families
remain damped.

As a local approximation to these directly computed shifts,
we also expand each complex frequency about
$\hat\alpha=\hat\beta=0$.
The derivatives and the two-coefficient Taylor expansion
are given in Appendix~\ref{app:response},
Eqs.~\eqref{eq:K} and~\eqref{eq:Taylor}.
For the even-EM example at $\lambda=0.05$ along
$(\hat\alpha,\hat\beta)=(4\lambda/7,\lambda)$,
the absolute difference between the predicted and
directly computed complex shifts decreases from
approximately $7.8\times10^{-7}$ at first order to
$3.2\times10^{-9}$ at second order.
If the second-order term proportional to
$\hat\alpha\hat\beta$ is omitted, the difference is
$3.4\times10^{-7}$.
Thus, the mixed term substantially improves the
prediction for this unequal-coefficient example.

We quantify the strength of the retained NED interactions
using $\epsilon_{\rm NED}^{\max}$, the maximum of
$\epsilon_{\rm NED}(r)$ in Eq.~\eqref{eq:eps}
over the BH exterior.
Every background in Fig.~\ref{fig:V_09} satisfies
$\epsilon_{\rm NED}^{\max}<3.8\times10^{-3}$,
so the quartic electromagnetic terms remain small
compared with the local Maxwell energy density
throughout the exterior.

We next assess whether individual damping-rate shifts
can reach the percent level within the adopted NED
amplitude criterion. 
For Fig.~\ref{fig:V_11}, we fix $Q/M=0.6$, $\bar g=5$,
and $\eta=0.1$, and vary $\lambda$ up to $0.16$
along the same three coefficient sequences.
Compared with Fig.~\ref{fig:V_09}, this extends the
coefficient range beyond $\lambda=0.05$ and uses
a smaller magnetic fraction, $\eta=0.1$ instead of $0.4$.
We focus on the odd-gravitational family because
its damping rate has the largest fractional changes
among the five families at these sampled parameters.

The sampled nonzero coefficients are
$\lambda=0.025,0.05,0.1,0.16$.
The left panel shows the odd-gravitational damping-rate
change in percent, $\delta_{\Gamma,j}$ defined in
Eq.~\eqref{eq:percent}, relative to the NED-free mode
at the same parameters.
The right panel shows $\epsilon_{\rm NED}^{\max}$
evaluated on each corresponding background.

At $\lambda=0.16$, the odd-gravitational damping-rate
changes are approximately $-1.05\%$, $+1.10\%$,
and $+0.43\%$ for
$(\hat\alpha,\hat\beta)=(\lambda,0),(0,\lambda),
(4\lambda/7,\lambda)$, respectively.
The damping-rate changes of the other four families
are smaller than $0.23\%$ of their respective NED-free
rates throughout these sampled sequences.
All backgrounds in Fig.~\ref{fig:V_11} satisfy
$\epsilon_{\rm NED}^{\max}<0.032$, below the adopted
bound of $0.05$.
Even the largest odd-gravitational damping decrease
is only about one percent of its NED-free value,
so this mode remains damped at every plotted point.
These examples show that percent-level damping changes
are possible while the retained quartic interactions
remain small throughout the exterior.

The sign of the correction also depends on the magnetic
fraction.
At $\lambda=0.05$ along
$(\hat\alpha,\hat\beta)=(4\lambda/7,\lambda)$,
the odd-gravitational damping rate increases at
$\eta=0.1$ in Fig.~\ref{fig:V_11} but decreases at
$\eta=0.4$ in Fig.~\ref{fig:V_09}.
Here $Q/M$, $\bar g$, and both NED coefficients are
the same in the two calculations.
Changing $\eta$ modifies both the background and
the coupled perturbation operator, allowing the
damping correction to reverse sign.

Interpreting the model as a controlled EFT additionally
requires the scale hierarchies and suppression of
omitted operators discussed in Appendix~\ref{app:EFT}.
Whether the resulting damping-rate shifts can be measured
depends on how strongly the relevant modes are excited
and on the precision of the ringdown observations.

\section{Conclusions}
\label{concludesec}

We have studied the QNMs of static charged BHs with a massless
axion and two quartic electromagnetic interactions.
We first derived the reduced quadratic action and coupled
linear equations for five dynamical degrees of freedom,
with a regular elimination of the nondynamical fields on
the stated nonextremal exterior branch.
The studied backgrounds satisfy sufficient local high-frequency
conditions for the absence of ghosts and gradient instabilities
throughout the BH exterior; these conditions alone do not
establish global mode stability.

Our spectral calculations use the asymptotic clock and
the ADM mass of each background, with the charge retuned
to keep $Q/M$ fixed.
Spectral-resolution tests, background-tolerance tests,
and independent integration-and-matching calculations
support the sampled fundamental frequencies and NED shifts.

The first main result is the computation of the complex QNM frequencies of five oscillatory quadrupolar mode families
continued from the fundamental RN modes, already without NED,
extending the two-family dyonic analysis of
Ref.~\cite{DeFelice:2024axionQNM}.
A purely electric BH has no background axion hair and
retains the RN geometry, but the axion--photon coupling
mixes the scalar and odd-parity gravitoelectromagnetic
perturbations.
At representative nonzero couplings, the odd frequencies split
from the unchanged even frequencies, yielding five distinct
complex frequencies and breaking both gravitational and EM
axial--polar isospectrality.
Magnetic charge then sources axion hair and couples
all five perturbations in the axion-coupled theory.
The damping rates can vary nonmonotonically with magnetic charge
and cross while the corresponding real frequencies remain distinct.
Additional purely imaginary growing modes occur in the
$\bar g=10$, $\eta=0.1$ benchmarks at $Q/M=0.6,0.7,0.8$.
These backgrounds are therefore mode unstable despite
satisfying the local high-frequency conditions.
They are distinct from the lower-coupling backgrounds
used for the NED sequences below.

We have used the RN EM--scalar Darboux degeneracy
at $Q/M=\sqrt7/4$ for $\ell=2$, previously derived in
Ref.~\cite{Rout:2026RNdegeneracy}, to interpret the
small-coupling splitting in the axion--photon theory.
The transformation preserves the ingoing condition at the
horizon and the outgoing condition at infinity.
The complex EM and scalar QNM spectra therefore coincide,
except possibly at $\omega=\pm12i/M$, where the transformation
is not invertible.
This degeneracy occurs within the odd-parity sector and
is distinct from RN axial--polar isospectrality.
For the fundamental pair, a small axion coupling at this
special charge produces a splitting linear in $\bar g$.
At a fixed charge ratio where the uncoupled fundamental EM
and scalar modes are simple and isolated within the odd-parity
sector, with a nonzero complex-frequency gap, their frequency
shifts generically begin at $O(\bar g^2)$ when the axion-induced
mixing is weak compared with this gap.

We then included the quartic NED interactions
$\alpha\cF^2+\beta\cG^2$ and determined their additional
QNM frequency shifts.
For purely electric BHs, only the $\cF^2$ interaction
modifies the background, whereas the $\cG^2$ and axion
interactions affect the odd-parity perturbation equations
and their QNM frequencies.

With the axion decoupled and equal quartic coefficients
($\hat\alpha=\hat\beta=\lambda$), the duality-rotated
theory agrees with the original theory through first
order in $\lambda$.
Consequently, the first-order NED shift of each individual
gravitational or EM branch is independent of $\eta$ at
fixed $Q/M$, as expressed by
Eq.~\eqref{eq:duality_individual}.
The full NED shifts remain nonzero and can differ between
the two parity partners.
Independent integration-and-matching differences support
the branchwise cancellation.
For unequal coefficients, Eq.~\eqref{eq:dualitylaw}
retains its role as a constraint on the summed shifts of
each isolated semisimple RN parity pair.
The equal-coefficient result is restricted to first NED
order and permits higher-order magnetic dependence.
This cancellation is lost in the tested axion-coupled
configuration at $Q/M=0.6$, $\bar g=0.9$, and $\eta=0.4$.
Canceling the real-frequency and damping components of
the EM-pair magnetic modification requires different
ratios $\hat\beta/\hat\alpha$.
Thus no nonzero real choice of the two NED coefficients
makes both components vanish simultaneously to first
order at this point.

Direct calculations at finite NED coefficients determine
the individual complex-frequency shifts at fixed charge
composition.
At $Q/M=0.6$, $\bar g=5$, and $\eta=0.4$, positive
$\hat\alpha$ alone increases the odd-gravitational and
odd-EM damping rates, whereas positive $\hat\beta$ alone
decreases them over the sampled range.
With both NED coefficients nonzero at the heavy-Dirac
matching ratio $\hat\beta/\hat\alpha=7/4$, all five families
exhibit nonzero shifts in their real frequencies and
damping rates at the sampled points.
For the representative even-EM example, including the
mixed term proportional to $\hat\alpha\hat\beta$ in the
second-order Taylor expansion improves agreement with
the directly computed shift.

At $\eta=0.1$ with the same $Q/M$ and $\bar g$,
the odd-gravitational damping-rate changes reach about
$-1.05\%$ and $+1.10\%$ for
$(\hat\alpha,\hat\beta)=(0.16,0)$ and $(0,0.16)$,
respectively.
This family has the largest fractional damping changes
along the sampled sequences, while the other four
remain below $0.23\%$.
All five oscillatory families remain damped along these
NED sequences, and the retained quartic terms remain small
compared with the local Maxwell energy
density throughout the exterior.
A specific heavy-particle EFT interpretation also requires
the scale hierarchies and control of omitted operators
discussed in Appendix~\ref{app:EFT}.
Observability depends on mode excitation and the precision
of ringdown measurements.

Future work includes mapping instability thresholds and
the stable parameter domain across multipoles and overtones,
as well as extending the model to a finite axion mass.
An especially interesting direction is to construct
rotating BH backgrounds in the same axion--NED theory
and compute their coupled gravitational, electromagnetic,
and axion QNMs.
Rotation introduces spin dependence and coupling between
angular harmonics, requiring an extension of the
spherical reduction.
Following the modes from the static limit would clarify
how spin changes the five families, their splittings,
and their responses to the axion and NED interactions,
and would connect these results with rotating
BH remnants relevant to gravitational-wave ringdown.

\FloatBarrier

\appendix
\renewcommand{\theequation}{\Alph{section}.\arabic{equation}}

\section{Heavy-particle matching and EFT interpretation}
\label{app:EFT}

For a canonically normalized gauge field, integrating out
heavy Dirac particles of masses $m_i$ and charges $Q_i e_D$
gives the one-loop interaction
\cite{Heisenberg:1936nmg,Dunne:2004nc,Preucil:2017wen}
\begin{equation}
\Delta\mathcal L^{(4)}
=\sum_i\frac{(Q_i e_D)^4}{360\pi^2m_i^4}
\left(4\cF^2+7\cG^2\right).
\label{eq:EH_four_field}
\end{equation}
The Maxwell correction is absorbed into the renormalized
gauge coupling $e_D$.
With $\alpha_D=e_D^2/(4\pi)$, this reproduces
Eq.~\eqref{eq:EH}.
For $N_f$ identical species of mass $m_\star$ and
dimensionless charge $q_\star$, including multiplicities,
\begin{equation}
\alpha=\frac{8N_f\alpha_D^2q_\star^4}{45m_\star^4},
\qquad
\beta=\frac{14N_f\alpha_D^2q_\star^4}{45m_\star^4}.
\label{eq:EH_identical}
\end{equation}
Thus the heavy-Dirac ratio is $\beta/\alpha=7/4$;
other spins or interactions can change it
\cite{Preucil:2017wen}.
The matching applies to ordinary or dark $U(1)$ fields,
with $Q_i e_D$ interpreted as the coupling to the retained
canonically normalized field.
Additional light gauge fields would require their mixed
operators to be retained or consistently truncated.

A local heavy-particle description requires
\begin{equation}
|\omega|,\quad r_h^{-1},\quad |\mathcal K|^{1/4}\ll m_i,
\qquad
\mathcal K=R_{\mu\nu\rho\sigma}R^{\mu\nu\rho\sigma},
\label{eq:EFT_scales}
\end{equation}
where $\mathcal K$ is the Kretschmann scalar, together with
\begin{equation}
|Q_i e_D E|\ll m_i^2,\qquad
|Q_i e_D B|\ll m_i^2
\label{eq:EFT_fields}
\end{equation}
throughout the exterior.
Local gradients must also remain within the derivative
expansion in regular freely falling frames; a static
observer's divergent horizon blueshift alone is not an
EFT breakdown criterion.
In the absence of large threshold logarithms,
$\sum_i\alpha_DQ_i^2/(4\pi)\ll1$ is a conservative
loop-expansion condition.

On the RN reference fields, $E^2+B^2=q_T^2/r^4$ gives
\begin{equation}
\epsilon_{\rm NED}^{\max}
=2\max(|\ah|,|\bh|)\left(\frac{Q}{r_h}\right)^2.
\label{eq:EFT_charge_epsilon}
\end{equation}
For axion-hairy backgrounds, we instead evaluate
Eq.~\eqref{eq:eps} using the electric field from
Eq.~\eqref{eq:gauss}.
The frequency responses in Appendix~\ref{app:response}
include both direct operator corrections and the induced
background changes at fixed $Q/M$, $\bar g$, and $\eta$.

Heavy-particle matching in curved spacetime also generates
curvature and derivative operators, schematically
$RF_{\mu\nu}F^{\mu\nu}/m_i^2$ and
$(\nabla_\mu F^{\mu\nu})^2/m_i^2$
\cite{Drummond:1979pp}.
For a specific EFT these terms, and additional axion
operators, must be included or shown to be negligible
at the required accuracy in a consistent operator basis.
Our calculation isolates the two quartic electromagnetic
operators by setting the other coefficients to zero.
The exterior amplitude criterion therefore controls
the retained interactions, rather than the entire EFT
or the central singularity.

\section{Five coupled perturbation equations}
\label{app:reduction}

We construct the five coupled radial equations using the
mixed-parity conventions of
Refs.~\cite{DeFelice:2024axionQNM,DeFelice:2025NED,Taniguchi:2025dyonic}.
As in Sec.~\ref{sec:pert}, we take $m=0$, $\ell\geq2$,
and $L=\ell(\ell+1)$.
For a single harmonic $Y_\ell\equiv Y_{\ell0}$, the even-parity
metric perturbations $H_0,H_1,H_2,h_1$ and the odd-parity metric
perturbations  $\mathcal Q,\mathcal W$ are defined by
\begin{align}
\delta g_{tt}&=NhH_0Y_\ell,\qquad
\delta g_{tr}=H_1Y_\ell,\nonumber\\*
\delta g_{rr}&=h^{-1}H_2Y_\ell,\qquad
\delta g_{r\theta}=h_1\partial_\theta Y_\ell,
\label{eq:even_metric_harmonics}\\
\delta g_{t\varphi}&=-\mathcal Q\sin\theta\,
\partial_\theta Y_\ell,\nonumber\\*
\delta g_{r\varphi}&=-\mathcal W\sin\theta\,
\partial_\theta Y_\ell.
\label{eq:odd_metric_harmonics}
\end{align}
All perturbation amplitudes depend on $(t,r)$, and the symmetry $\delta g_{\mu\nu}=\delta g_{\nu\mu}$ is implicit. We fix the coordinate gauge by setting the purely angular metric perturbations and the even-parity $t\theta$ component to zero. The $U(1)$ gauge freedom is used to eliminate the even-parity angular electromagnetic component.

The electromagnetic and scalar harmonics are defined in
Eq.~\eqref{eq:scalar_gauge_harmonics} and its accompanying text.
The pseudoscalar perturbation belongs to the electric odd sector;
at nonzero magnetic charge, parity labels specify the harmonic
basis without implying decoupled sectors.

\subsection{Unreduced action and background coefficients}
\label{app:background_coefficients}
We use unit-normalized spherical harmonics. Expanding
Eq.~\eqref{eq:action} to second order in the perturbations
and integrating over the angular coordinates yields, for each mode,
\begin{equation}
S^{(2)}=\int {\rm d}t\,{\rm d}r\,
\bigl(\cL_1+\cL_2\bigr).
\label{eq:quadratic_action_before_reduction}
\end{equation}
The unreduced radial densities contain background-dependent coefficients
$a_i,b_i,c_i,d_i,s_i,u_i,p_i$. Up to integrations by parts, they are
expressed using
\begin{align}
\mathcal E_\delta=\delta A_0'-\dot{\delta A}_1,\qquad
\mathcal Z_\delta=r\dot{\mathcal W}-r\mathcal Q'+2\mathcal Q.
\label{eq:unreduced_combinations}
\end{align}
The two parts of the Lagrangian density are
\begin{align}
\cL_1={}&a_0H_0^2\nonumber\\*
&+H_0\bigl[a_1H_2'+La_2h_1'+(a_3+La_4)H_2\nonumber\\*
&+La_5h_1+La_6\delta A+a_7\delta\phi'+a_8\delta\phi\bigr]\nonumber\\*
&+b_1H_1^2+H_1\bigl(b_2\dot H_2+Lb_3\dot h_1+b_4\dot{\delta\phi}\bigr)\nonumber\\*
&+c_0H_2^2+LH_2(c_1h_1+c_2\delta A)\nonumber\\*
&+c_3H_2\delta\phi'+c_4H_2\delta\phi+Ld_0\dot h_1^{\,2}+Ld_1h_1^2\nonumber\\*
&+Lh_1(d_2\delta A_0+d_3\delta A'+d_4\delta\phi)+s_1\mathcal E_\delta^2\nonumber\\*
&+\bigl(s_2H_0+s_3H_2+Ls_4\delta A+s_6\delta\phi\bigr)\mathcal E_\delta\nonumber\\*
&+Ls_7\delta A_0^2+Ls_8\delta A_1^2\nonumber\\*
&+u_1\dot{\delta\phi}^{\,2}+u_2\delta\phi'^{\,2}+Lu_3\delta\phi^2\nonumber\\*
&+u_5\delta\phi\delta A+Lu_6\delta A_0\delta A,
\label{eq:unreduced_L1}\\
\cL_2=L\bigl[&p_1\mathcal Z_\delta^2+p_2\delta A\mathcal Z_\delta
+p_3\dot{\delta A}^{\,2}+p_4\delta A'^{\,2}\nonumber\\*
&+Lp_5\delta A^2+(Lp_6+p_7)\mathcal W^2+(Lp_8+p_9)\mathcal Q^2\nonumber\\*
&+p_{10}\mathcal Q\delta A_0+p_{11}\mathcal Qh_1\nonumber\\*
&+p_{12}\mathcal W\delta A_1+p_{13}\delta A\mathcal Q\bigr].
\label{eq:unreduced_L2}
\end{align}
These expressions specialize the mixed-parity action of Ref.~\cite{Taniguchi:2025dyonic} to Eq.~\eqref{eq:model}. The coefficient labels refer to the terms in Eqs.~\eqref{eq:unreduced_L1} and~\eqref{eq:unreduced_L2} and need not coincide with those of the general action in that reference.

All coefficients below are evaluated on the background equations, and all functions have the same radial argument $r$. In terms of $E,B$, the constitutive coefficients $a,d,c_{\rm em}$ in Eq.~\eqref{eq:adc}, $b=2(\beta-\alpha)EB$, and $\rho_{\rm EM}$ in Eq.~\eqref{eq:rho}, the metric coefficients are
\begin{align}
a_0&=\frac{r^2\sqrt N\,dE^2}{8},\qquad
a_1=-\frac{\Mp^2rh\sqrt N}{2},\nonumber\\*
a_2&=-\frac{a_1}{r},\qquad a_4=-\frac{\Mp^2\sqrt N}{4},\nonumber\\*
a_3&=-\frac{\sqrt N}{4}\bigl(2\Mp^2-2r^2\rho_{\rm EM}+r^2dE^2\bigr),\nonumber\\*
a_5&=\frac{\Mp^2\sqrt N(2h+rh')}{4r},\qquad
a_6=\frac{\sqrt N}{2}(aB+bE),\nonumber\\*
a_7&=\frac{r^2h\sqrt N\,\phi'}{2},\qquad a_8=0,
\label{eq:coeff_a}\\
b_1&=\frac{L\Mp^2}{4\sqrt N},\qquad b_2=\frac{\Mp^2r}{\sqrt N},\nonumber\\*
b_3&=-\frac{\Mp^2}{2\sqrt N},\qquad b_4=-\frac{r^2\phi'}{\sqrt N},
\label{eq:coeff_b}\\
c_0&=-\frac{a_3}{2},\qquad
c_1=-\frac{\Mp^2[rNh'+h(2N+rN')]}{4r\sqrt N},\nonumber\\*
c_2&=-a_6,\qquad c_3=a_7,\qquad c_4=0,
\label{eq:coeff_c}\\
d_0&=\frac{\Mp^2}{4\sqrt N},\qquad
d_1=\frac{h\sqrt N}{2r^2}(\Mp^2-r^2aB^2),\nonumber\\*
d_2&=-aE,\qquad d_3=-aBh\sqrt N,\nonumber\\*
d_4&=h\sqrt N\,\phi'.
\label{eq:coeff_d}
\end{align}
The electromagnetic coefficients entering the auxiliary-field construction below are
\begin{align}
s_1&=\frac{r^2d}{2\sqrt N},\qquad
s_2=A_0's_1,\qquad s_3=-s_2,\nonumber\\*
s_4&=b,\qquad s_6=-gq_M,\nonumber\\*
s_7&=\frac{a}{2h\sqrt N},\qquad
s_8=-\frac{ah\sqrt N}{2}.
\label{eq:coeff_s}
\end{align}
The odd metric and electromagnetic coefficients are
\begin{align}
p_1&=\frac{\Mp^2}{4r^2\sqrt N},\qquad
p_2=-\frac{aE}{r},\nonumber\\*
p_3&=s_7,\qquad p_4=s_8,\qquad
p_5=-\frac{\sqrt N\,c_{\rm em}}{2r^2},\nonumber\\*
p_6&=-\frac{\Mp^2h\sqrt N}{4r^2},\qquad p_7=d_1,\nonumber\\*
p_8&=\frac{\Mp^2}{4r^2h\sqrt N},\qquad
p_9=-\frac{\Mp^2-r^2aB^2}{2r^2h\sqrt N},\nonumber\\*
p_{10}&=\frac{aB}{h\sqrt N},\qquad p_{11}=-aEB,\nonumber\\*
p_{12}&=d_3,\qquad p_{13}=Bu_6.
\label{eq:coeff_p}
\end{align}
Finally, the scalar and scalar--electromagnetic coefficients are
\begin{align}
u_1&=\frac{r^2}{2h\sqrt N},\qquad
u_2=-\frac{r^2h\sqrt N}{2},\nonumber\\*
u_3&=-\frac{\sqrt N}{2},\qquad u_5=-LgA_0',\nonumber\\*
u_6&=(\cL_{,\cG})'=2\beta\cG'+g\phi'.
\label{eq:coeff_u}
\end{align}
In the massless model, the terms $\delta\phi'\mathcal E_\delta$,
$\delta\phi'\delta A$, and $(\delta\phi)^2$ without $L$ vanish.
Also, $a_8=c_4=0$ follows from $\cL_{,\phi}=g\cG$,
$\cL_{,\cG\phi}=g$, and $\cG=-EB$.
The background simplifications involve no division by $\phi'$,
so the hairless limit is regular.
We keep $\Mp$ explicit; numerically $\Mp=1$.

\subsection{Local constraint reduction}

We linearize the squared combinations in
Eq.~\eqref{eq:unreduced_combinations} with the auxiliary
fields $V,\chi$ defined in Eq.~\eqref{eq:main_auxiliary_definitions}.
Completing the squares is equivalent for $s_1p_1\neq0$.
The same definitions give $H_2=(\Psi+Lh_1)/r$.
The identity $a_1/r+a_2=0$ eliminates the $h_1'$ term from 
the $H_0$ constraint. The background equations further imply 
$a_0-s_2^2/(4s_1)=0$, so the auxiliary-field action contains no $H_0^2$ term.
Thus, $H_0$ acts as a Lagrange multiplier whose constraint
is algebraic in $h_1$.

The asymptotic time normalization in
Eq.~\eqref{eq:asymptotic_clock_normalization} rescales
the auxiliary fields according to
\begin{align}
V&=\frac{V_{\rm raw}}{\sqrt{C_\infty}},\qquad
\chi=\frac{\chi_{\rm raw}}{\sqrt{C_\infty}}.
\label{eq:auxiliary_clock_normalization}
\end{align}
The other three fields are unchanged. The first and fourth regularized fields, defined later in 
Eq.~(\ref{eq:five_fields}), inherit the same factors. All operators use the normalized basis.

To express the $H_0$ constraint compactly, we introduce the
background-dependent combinations
\begin{align}
P_\Psi&=-\frac{a_1}{r^2}
+\frac{a_3+La_4-s_2s_3/(2s_1)}{r},\\
P_h&=L(P_\Psi+a_5).
\end{align}
For $P_h\neq0$, the constraint can be solved algebraically for $h_1$:
\begin{align}
h_1=-\frac{1}{P_h}\biggl[&
s_2V+P_\Psi\Psi+\frac{a_1}{r}\Psi'
+a_7\delta\phi' \nonumber\\*
&+\left(a_8-\frac{s_2s_6}{2s_1}\right)\delta\phi
+L\left(a_6-\frac{s_2s_4}{2s_1}\right)\delta A\biggr].
\label{eq:h1_constraint}
\end{align}
The original $h_1$ equation reconstructs $H_0$ because $P_h\neq0$, so eliminating this constraint and its multiplier loses no field equation.
After integrating by parts, the pairs
$(\delta A_0,\mathcal Q)$ and $(\delta A_1,\mathcal W)$
enter the action algebraically.
For a real pair $\bm z$ with density
$\bm z^TH\bm z+\bm J^T\bm z$ and symmetric invertible
$2\times2$ matrix $H$, its equation gives
$\bm z=-H^{-1}\bm J/2$ and contributes
$-\bm J^TH^{-1}\bm J/4$ after elimination.

For the pair $\bm z_s=(\delta A_0,\mathcal Q)^T$,
the matrix is
\begin{equation}
H_s=L\begin{pmatrix}
s_7&p_{10}/2\\
p_{10}/2&Lp_8+p_9
\end{pmatrix},
\label{eq:constraint_spatial_matrix}
\end{equation}
and the source vector has components
\begin{align}
\bm J_s&=(J_{s,1},J_{s,2})^T,\nonumber\\*
J_{s,1}&=-2(s_1V)'+Ld_2h_1+Lu_6\delta A,\nonumber\\*
J_{s,2}&=2L[rp_1\chi'+(rp_1'+3p_1)\chi]\nonumber\\*
&\quad+Lp_{11}h_1+Lp_{13}\delta A.
\label{eq:constraint_source_vector}
\end{align}
For the pair $\bm z_t=(\delta A_1,\mathcal W)^T$,
the corresponding quantities are
\begin{align}
H_t&=L\begin{pmatrix}
s_8&p_{12}/2\\
p_{12}/2&Lp_6+p_7
\end{pmatrix},\nonumber\\*
\bm J_t&=(2s_1\dot V,-2Lrp_1\dot\chi)^T.
\label{eq:constraint_time_pair}
\end{align}

To obtain the reduced matrices explicitly, we define
\begin{align}
U_{\rm aux}&=s_3H_2+Ls_4\delta A+s_6\delta\phi,\nonumber\\*
\bm b_H&=(0,b_2/r,b_4,0,0)^T,\nonumber\\*
J_H&=\bm b_H^T\dot{\bm y}
=\frac{b_2}{r}\dot\Psi+b_4\dot{\delta\phi}.
\label{eq:reduction_sources}
\end{align}
The terms containing $H_1$ are $b_1H_1^2+H_1(J_H+2b_1\dot h_1)$, since $L(b_2/r+b_3)=2b_1$. Hence
\begin{equation}
H_1=-\dot h_1-\frac{J_H}{2b_1}.
\label{eq:H1_elimination}
\end{equation}
Its elimination cancels the original $Ld_0\dot h_1^2$ term because $Ld_0=b_1$. After eliminating both algebraic pairs, the remaining spatial and kinetic densities are
\begin{align}
\cL_{\rm sp}={}&c_0H_2^2+Lc_1H_2h_1+Lc_2H_2\delta A\nonumber\\*
&+c_3H_2\delta\phi'+c_4H_2\delta\phi+Ld_1h_1^2\nonumber\\*
&+Ld_3h_1\delta A'+Ld_4h_1\delta\phi\nonumber\\*
&-s_1V^2+VU_{\rm aux}-\frac{U_{\rm aux}^2}{4s_1}\nonumber\\*
&+u_2\delta\phi'^2+Lu_3\delta\phi^2+u_5\delta\phi\delta A\nonumber\\*
&+L\biggl[-p_1\chi^2+p_2\delta A\chi
-\frac{p_2^2}{4p_1}\delta A^2\nonumber\\*
&\qquad+Lp_5\delta A^2+p_4\delta A'^2\biggr]\nonumber\\*
&-\frac14\bm J_s^TH_s^{-1}\bm J_s,
\label{eq:reduced_spatial_density}\\
\cL_{\rm kin}={}&u_1\dot{\delta\phi}^{\,2}
+Lp_3\dot{\delta A}^{\,2}-\frac{J_H^2}{4b_1}\nonumber\\*
&-J_H\dot h_1-\frac14\bm J_t^TH_t^{-1}\bm J_t.
\label{eq:reduced_kinetic_density}
\end{align}
Here $h_1$ is given by Eq.~\eqref{eq:h1_constraint}, $H_2=(\Psi+Lh_1)/r$, and $\dot h_1$ is the time derivative of that expression. The reduction uses only the scalar divisions by $s_1,p_1,P_h,b_1$ and pointwise inverses of $H_s,H_t$; their nonvanishing conditions are established below.

After these substitutions, treating the fields and their derivatives as independent variables specifies the spatial matrices through the Hessians
\begin{align}
(M_0)_{IJ}&=\frac12\frac{\partial^2\cL_{\rm sp}}{\partial y_I\partial y_J},\nonumber\\*
C_{IJ}&=\frac12\frac{\partial^2\cL_{\rm sp}}{\partial y_I\partial y_J'},\nonumber\\*
G_{IJ}&=\frac12\frac{\partial^2\cL_{\rm sp}}{\partial y_I'\partial y_J'},
\label{eq:spatial_matrix_hessians}
\end{align}
where $I,J=1,\ldots,5$ are field indices. Similarly,
\begin{align}
(K_0)_{IJ}=\frac12\frac{\partial^2\cL_{\rm kin}}{\partial\dot y_I\partial\dot y_J},\qquad
(K_1)_{IJ}=\frac12\frac{\partial^2\cL_{\rm kin}}{\partial\dot y_I\partial\dot y_J'}.
\label{eq:kinetic_matrix_hessians}
\end{align}
After the $\dot h_1^{\,2}$ cancellation, no term quadratic in $\dot{\bm y}'$ remains. The factors of $1/2$ follow from the conventions $2\bm y^TC\bm y'$ and $2\dot{\bm y}^TK_1\dot{\bm y}'$. The coefficients of $\bm y'$ in $h_1$ are $hN\bm b_H/(2P_h)$, so the mixed kinetic matrix is symmetric:
\begin{align}
K_1&=-\frac{hN}{4P_h}\bm b_H\bm b_H^T=K_1^T,\nonumber\\*
K&=K_0-K_1'.
\label{eq:kinetic_integration_by_parts}
\end{align}
The symmetry of $K_1$ permits radial integration by parts, giving the action~\eqref{eq:redaction} with no derivatives higher than first order in the density. Substitution of the displayed coefficients and background equations then yields the exact identity $G=-Nh^2K$. Equations~\eqref{eq:reduced_spatial_density}--\eqref{eq:kinetic_integration_by_parts} provide a closed construction of every reduced matrix entry.

We examine the regularity of the constraint elimination on a smooth
nonextremal exterior branch satisfying $N,h>0$, $a>0$, and $d\neq0$,
where $a$ and $d$ are defined in Eq.~\eqref{eq:adc}.
The coefficient $P_h$ in the $h_1$ constraint can be expressed
in terms of the background combination
\begin{equation}
\Xi=L+rh'-2h+\frac{r^2h\phi'^2}{\Mp^2}.
\label{eq:constraint_Xi}
\end{equation}
Using the background equations, the coefficients entering the
algebraic elimination take the factorized forms
\begin{align}
s_1&=\frac{r^2d}{2\sqrt{N}},\qquad
b_1=\frac{L\Mp^2}{4\sqrt{N}},\nonumber\\*
P_h&=-\frac{L\Mp^2\sqrt{N}}{4r}\Xi.
\label{eq:constraint_pivots}
\end{align}
The two constraint matrices have determinants
\begin{align}
\det H_s&=\frac{L^2\Mp^2(L-2)a}{8r^2Nh^2},\\
\det H_t&=\frac{L^2\Mp^2(L-2)aNh^2}{8r^2}.
\label{eq:constraint_determinants}
\end{align}
Under the stated assumptions and for $\ell\geq2$, regularity
at finite $r>r_h$ therefore reduces to establishing that
$\Xi$ does not vanish.

Differentiating the electric Gauss law and using $B'=-2B/r$ gives
\[
\rho_{\rm EM}'=gEB\phi'-\frac{2a}{r}(E^2+B^2).
\]
Together with the scalar and metric equations, this relation gives the integrating-factor identity
\begin{align}
&\frac{{\rm d}}{{\rm d}r}\bigl[r\sqrt N\{\Xi-(L-2)\}\bigr]\nonumber\\*
&=\frac{r^2\sqrt N}{\Mp^2}[2a(E^2+B^2)+\phi'^2].
\label{eq:theory_Xi}
\end{align}
At a regular nonextremal horizon, $0<N_h<\infty$ and finite $\phi'_h$ give $\Xi_h=L+r_hh'_h$, with $h'_h>0$. The integrated quantity therefore starts at $r_h\sqrt{N_h}(2+r_hh'_h)>0$ and is nondecreasing for $a>0$. Hence $\Xi>L-2>0$ for every finite $r>r_h$ and $\ell\ge2$. In particular, $P_h$ cannot vanish there. The amplitude criterion~\eqref{eq:eps} implies $a>0.95$ and $d>0.85$, including the signed NED probes used for differentiation.

With $N_\infty=1$ and finite nonzero $d_h$, the rescaled pivots $s_1\sqrt N/r^2=d/2$, $b_1\sqrt N=L\Mp^2/4$, and $rP_h/\sqrt N=-L\Mp^2\Xi/4$ have finite nonzero endpoint limits, with $\Xi_\infty=L-2$. The endpoint powers of the raw constraint matrices are coordinate effects. The radial rescaling $D_r=\operatorname{diag}(1,r)$ gives the common scaled block
\begin{align}
\mathsf H_{\rm c}&=D_r\frac{h\sqrt N}{L}H_sD_r\nonumber\\*
&=-\frac{D_rH_tD_r}{Lh\sqrt N}\nonumber\\*
&=\begin{pmatrix}
 a/2&arB/2\\
 arB/2&\Mp^2(L-2)/4+ar^2B^2/2
\end{pmatrix},
\label{eq:scaled_constraint_block}\\
\det\mathsf H_{\rm c}&=\frac{a\Mp^2(L-2)}8.
\label{eq:scaled_constraints}
\end{align}
Since $H_t=-Nh^2H_s$, both rescalings give the same block, with finite nonsingular horizon and infinity limits. For the five-field gradient matrix $\widehat G=T^TGT$ defined in Eq.~\eqref{eq:transformed_action_matrices},
\begin{equation}
\det\widehat G=-\frac{\Mp^4r^2d^2\sqrt N h^7}
 {32(r-r_h)^2\Xi^2}.
\label{eq:gradient_determinant}
\end{equation}
Thus the radial normal form is well defined in the open exterior. Substitution of the regular nonextremal horizon expansions shows that every entry of $-\widehat G/h$ has a finite limit. This limit is nonsingular because
\begin{equation}
\det\!\left[\lim_{r\to r_h}\frac{-\widehat G}{h}\right]
=\frac{\Mp^4r_h^2d_h^2\sqrt{N_h}(h'_h)^2}{32\Xi_h^2}.
\label{eq:gradient_horizon}
\end{equation}
At infinity,
\begin{align}
&\lim_{r\to\infty}(-\widehat G)\nonumber\\*
&=\operatorname{diag}\!\biggl(\frac1{2L},\frac{\Mp^2}{L(L-2)},\frac12,
\frac{\Mp^2L}{4(L-2)},\frac L2\biggr).
\label{eq:gradient_infinity}
\end{align}
The QNM boundary conditions use the corresponding regular limiting expansions, rather than raw constraint inverses evaluated at $h=0$. These results justify the local reduction, without establishing global mode stability. Integrations by parts identify equivalent local field equations; QNM boundary conditions are imposed separately, without assuming vanishing endpoint fields.

The implementation evaluates the local matrices and their radial
derivatives directly, using a directional complex-step method.
The identity $G+Nh^2K=0$ holds at relative norm below
$5\times10^{-16}$ in the selected checks.
A symbolic comparison establishes equality of the archived and
independently reconstructed reduced Euler operators on the
background equations: the corresponding densities differ by a
total radial derivative after including the
radius-dependent auxiliary-field transformation and its derivatives.
The separate production-operator checks are given in
Appendix~\ref{app:operator_checks}.

\subsection{Explicit radial equations}
For each Fourier component, the same symbol denotes the radial field after removing $e^{-i\omega t}$. Here $r$ is the areal radius, distinct from the tortoise coordinate $r_*$, with $r_*'=1/(h\sqrt N)$. Setting $v=h\sqrt N$ and $\partial_*=\partial/\partial r_*=v\,{\rm d}/{\rm d}r$, we use
\begin{align}
\psi_1&=r^2V,& \psi_2&=\frac{r-r_h}{r}\Psi,&
\psi_3&=r\delta\phi,\nonumber\\*
\psi_4&=\chi,& \psi_5&=\delta A,&\bm y&=T\bm\psi,\label{eq:five_fields}
\end{align}
where $T$ is the diagonal matrix in Eq.~\eqref{eq:tortoise_fields}. No additional frequency-dependent redefinition is made. The action coefficients in this basis are
\begin{align}
\widehat G={}&T^TGT,\nonumber\\*
\widehat C={}&T^TCT+T'^TGT,\nonumber\\*
\widehat M={}&T^TM_0T+T^TCT'+T'^TC^TT+T'^TGT',
\label{eq:transformed_action_matrices}
\end{align}
with $\widehat K=T^TKT$ and $\widehat G=-v^2\widehat K$. Every prime in the coefficient formulas is a derivative with respect to $r$, including the derivative of the background fields. Applying Eq.~\eqref{eq:euler} in this basis and multiplying by $-v^2\widehat G^{-1}$ gives
\begin{equation}
\partial_*^2\bm\psi+\mathcal A\partial_*\bm\psi
 +(\omega^2I_5+\mathcal B)\bm\psi=0,
\label{eq:radial_matrix_equation}
\end{equation}
where $I_5$ is the $5\times5$ identity. The real, frequency-independent $5\times5$ matrices $\mathcal A(r)$ and $\mathcal B(r)$ multiply first tortoise derivatives and undifferentiated fields, respectively. Using $\partial_*^2=v^2{\rm d}^2/{\rm d}r^2+vv'{\rm d}/{\rm d}r$ yields
\begin{align}
\mathcal A&=v\widehat G^{-1}
\left(\widehat G'-\widehat C+\widehat C^T\right)-v'I_5,
\label{eq:radial_A}\\
\mathcal B&=-v^2\widehat G^{-1}
\left(\widehat M-\widehat C^{T\prime}\right).
\label{eq:radial_B}
\end{align}
These expressions determine the radial matrices without assuming a symmetric potential. Here $\mathcal A_{IJ}$ multiplies $\partial_*\psi_J$ and $\mathcal B_{IJ}$ multiplies $\psi_J$ in the $I$th equation, with $I,J=1,\ldots,5$ ordered as in Eq.~\eqref{eq:five_fields}. Thus $\mathcal A_{11},\mathcal B_{11}$ act on the first field, whereas $\mathcal A_{12},\mathcal B_{12}$ couple its equation to the second field. The component equations are
\begin{widetext}
\begin{subequations}\label{eq:five_radial_equations}
\begin{align}
0={}&\partial_*^2\psi_1+\omega^2\psi_1
+\mathcal A_{11}\partial_*\psi_1+\mathcal A_{12}\partial_*\psi_2
+\mathcal A_{13}\partial_*\psi_3+\mathcal A_{14}\partial_*\psi_4+\mathcal A_{15}\partial_*\psi_5
\nonumber\\*
&+\mathcal B_{11}\psi_1+\mathcal B_{12}\psi_2+\mathcal B_{13}\psi_3+\mathcal B_{14}\psi_4+\mathcal B_{15}\psi_5,
\label{eq:five_V}\\
0={}&\partial_*^2\psi_2+\omega^2\psi_2
+\mathcal A_{21}\partial_*\psi_1+\mathcal A_{22}\partial_*\psi_2
+\mathcal A_{23}\partial_*\psi_3+\mathcal A_{24}\partial_*\psi_4+\mathcal A_{25}\partial_*\psi_5
\nonumber\\*
&+\mathcal B_{21}\psi_1+\mathcal B_{22}\psi_2+\mathcal B_{23}\psi_3+\mathcal B_{24}\psi_4+\mathcal B_{25}\psi_5,
\label{eq:five_Psi}\\
0={}&\partial_*^2\psi_3+\omega^2\psi_3
+\mathcal A_{31}\partial_*\psi_1+\mathcal A_{32}\partial_*\psi_2
+\mathcal A_{33}\partial_*\psi_3+\mathcal A_{34}\partial_*\psi_4+\mathcal A_{35}\partial_*\psi_5
\nonumber\\*
&+\mathcal B_{31}\psi_1+\mathcal B_{32}\psi_2+\mathcal B_{33}\psi_3+\mathcal B_{34}\psi_4+\mathcal B_{35}\psi_5,
\label{eq:five_phi}\\
0={}&\partial_*^2\psi_4+\omega^2\psi_4
+\mathcal A_{41}\partial_*\psi_1+\mathcal A_{42}\partial_*\psi_2
+\mathcal A_{43}\partial_*\psi_3+\mathcal A_{44}\partial_*\psi_4+\mathcal A_{45}\partial_*\psi_5
\nonumber\\*
&+\mathcal B_{41}\psi_1+\mathcal B_{42}\psi_2+\mathcal B_{43}\psi_3+\mathcal B_{44}\psi_4+\mathcal B_{45}\psi_5,
\label{eq:five_chi}\\
0={}&\partial_*^2\psi_5+\omega^2\psi_5
+\mathcal A_{51}\partial_*\psi_1+\mathcal A_{52}\partial_*\psi_2
+\mathcal A_{53}\partial_*\psi_3+\mathcal A_{54}\partial_*\psi_4+\mathcal A_{55}\partial_*\psi_5
\nonumber\\*
&+\mathcal B_{51}\psi_1+\mathcal B_{52}\psi_2+\mathcal B_{53}\psi_3+\mathcal B_{54}\psi_4+\mathcal B_{55}\psi_5.
\label{eq:five_A}
\end{align}
\end{subequations}
\end{widetext}
The diagonal frequency term follows from the exact finite-$L$ identity $\widehat G=-Nh^2\widehat K$, verified directly from the quadratic action. It does not require diagonalization of the full operator. In particular, $\mathcal A$ need not vanish and $\mathcal B$ need not be symmetric in these variables. On the regular purely electric branch, $q_M=0$ implies $\phi'=0$ and $\phi=0$ under the chosen asymptotic condition. The matrix entries connecting $\{1,2\}$ to $\{3,4,5\}$ then vanish, giving the even two-field and odd three-field systems. For $q_M\ne0$, Eqs.~\eqref{eq:five_radial_equations} are solved together with the ingoing/outgoing conditions~\eqref{eq:bc}. Factoring the phase~\eqref{eq:phase} and changing to $x=r_h/r$ gives precisely Eq.~\eqref{eq:QEP}; Eqs.~\eqref{eq:radial_A} and~\eqref{eq:radial_B} are the matrices used in its implementation.

\section{Numerical validation and response accuracy}
\label{app:numerics}

This appendix validates the radial matrices and tests the
numerical accuracy of the frequencies and NED responses
in Secs.~\ref{sec:pert}--\ref{sec:ned}.

\begin{table*}[!t]
\centering
\caption{Spectral-resolution test for the five oscillatory $\ell=2$ families continued from the fundamental RN modes at $Q/M=0.8$, $\bar g=10$, $\eta=0.1$, and $\ah=\bh=0$. The frequency column gives the $N_{\rm c}=60$ result, with $\Omega_n\equiv\Omega(N_{\rm c}=n)$. This background also supports the growing mode listed in Table~\ref{tab:growing_modes}. Frequency differences are calculated before rounding.}\label{tab:spectral_resolution}
\small
\begin{tabular}{|l|c|c|c|}
\hline
Family & $\Omega\;(N_{\rm c}=60)$ & $|\Omega_{48}-\Omega_{36}|$ & $|\Omega_{60}-\Omega_{48}|$\\
\hline
even grav. & $0.4017429180-0.0926895522i$ & $3.31\times10^{-12}$ & $1.62\times10^{-11}$\\
even EM & $0.5390703862-0.1010348385i$ & $1.31\times10^{-12}$ & $9.89\times10^{-12}$\\
odd grav. & $0.1985920993-0.0748470755i$ & $6.71\times10^{-10}$ & $1.12\times10^{-11}$\\
odd EM & $0.8489962829-0.1175336138i$ & $2.04\times10^{-12}$ & $3.41\times10^{-12}$\\
scalar & $0.4226742193-0.0944325443i$ & $1.57\times10^{-12}$ & $3.39\times10^{-12}$\\
\hline
\end{tabular}
\end{table*}

\subsection{Radial matrices and RN parity-pair checks}
\label{app:operator_checks}

We check the numerical implementation of the radial matrices
$\mathcal A,\mathcal B$ in Eqs.~\eqref{eq:radial_A}
and~\eqref{eq:radial_B} against expressions derived separately
from the quadratic action.
Both calculations use the same perturbation variables,
the time normalization $N(\infty)=1$, and units $r_h=\Mp=1$.
Let $X_{\rm num}$ and $X_{\rm sym}$ denote the numerical
and symbolic expressions for either radial matrix
$X=\mathcal A,\mathcal B$.
We use $\|X\|\equiv(\sum_{i,j=1}^{5}|X_{ij}|^2)^{1/2}$,
where $X_{ij}$ are the matrix entries.
The normalized difference is
\begin{equation}
\epsilon_X =
\frac{\|X_{\rm num}-X_{\rm sym}\|}
{\max\{1,\|X_{\rm num}\|,\|X_{\rm sym}\|\}}.
\label{eq:radial_matrix_error}
\end{equation}
For five backgrounds covering positive and negative NED
coefficients, $\epsilon_X<6.5\times10^{-13}$ at 25 exterior
points in total, five on each background.

For the gravitational and EM RN pairs of
Sec.~\ref{sec:ned_composition}, each parity sector has one
independent QNM eigenvector at the reference frequency for
$Q/M=0.6$, $\ell=2$, and $N_{\rm c}=36,48,60$.
The $2\times2$ matrix $N_J$ multiplying the first-order frequency
change in each pair's equation is diagonal with nonzero entries
and hence invertible.

\subsection{Frequency and NED-shift convergence}
\label{app:resolution}

Table~\ref{tab:spectral_resolution} compares the frequencies
at different radial resolutions for the $Q/M=0.8$ benchmark
of Sec.~\ref{sec:magnetic}.
Other frequency and background-tolerance tests appear in
Sec.~\ref{sec:pert}.
A NED shift is the difference between frequencies with and
without NED terms; both must be computed at the same
resolution and background tolerance.

For every point in Figs.~\ref{fig:V_09} and~\ref{fig:V_11},
both frequencies are evaluated at $N_{\rm c}=36,48$.
At the largest plotted coefficients, $\lambda=0.05$ and
$0.16$, respectively, both are also evaluated at $N_{\rm c}=60$.
In the recorded tests across all five families, changes in
$\Delta\Omega_j$ are below
$2.0\times10^{-11}$ between 36 and 48 nodes.
At the largest-coefficient points, changes between 48 and 60
nodes are below $3.6\times10^{-11}$.
At these points, tightening the background relative tolerance from
$2\times10^{-11}$ to $2\times10^{-12}$ changes these shifts
by less than $2.1\times10^{-11}$.
Both figures use $N_{\rm c}=48$.

\subsection{Additional growing modes}
\label{app:growing_modes}

The growing modes in Sec.~\ref{sec:magnetic} have
$\Omega=i\gamma$, $\gamma>0$, and grow as $e^{\gamma t/M}$.
Table~\ref{tab:growing_modes} lists growth rates checked at
$N_{\rm c}=36,48,60$.
These backgrounds satisfy the local conditions~\eqref{eq:adc}
but support both growing and damped modes.

\begin{table}[ht]
\centering
\caption{Growing $\ell=2$ modes at $\bar g=10$, $\eta=0.1$,
and $\hat\alpha=\hat\beta=0$. The frequencies have
$\operatorname{Re}\Omega=0$; $\gamma=\operatorname{Im}\Omega>0$
is the dimensionless growth rate. Values are rounded to
five decimal places.}
\label{tab:growing_modes}
\begin{tabular}{|c|c|}
\hline
$Q/M$ & $\gamma$\\
\hline
$0.6$ & $0.06720$\\
$0.7$ & $0.13343$\\
$0.8$ & $0.17560$\\
\hline
\end{tabular}
\end{table}

For the five-component field $\bm\psi$ defined in
Eq.~\eqref{eq:tortoise_fields}, the radial equation~\eqref{eq:radial_matrix_equation}
at $\omega=i\gamma/M$ becomes
\begin{equation}
(\partial_*^2+\mathcal A\partial_*+\mathcal B)\bm\psi
=\left(\frac{\gamma}{M}\right)^2\bm\psi.
\label{eq:growing_bound_state}
\end{equation}
The QNM conditions~\eqref{eq:bc} require spatial decay here,
which we approximate by zero field values at both finite boundaries.
We use second-order finite differences with 2001, 4001, and
8001 nodes on $r_*/r_h\in[-70,130]$,
with $r_*=0$ at $r=3r_h$.
The matrices $\mathcal A,\mathcal B$ are replaced by their
zero horizon limits only for $(r-r_h)/r_h<10^{-5}$.
Doubling the radial samples for coefficient interpolation
from 700 to 1400 changes the growth rates by less than
$5\times10^{-11}$.
With 8001 nodes and this $10^{-5}$ cutoff, the growth rates
agree with the spectral results within $4.1\times10^{-7}$.
Reducing the cutoff to $10^{-6}$ changes $\gamma$ at
$Q/M=0.6$ by about $7\times10^{-7}$.
This horizon-replacement sensitivity is separate from the
interpolation test; neither is a total error bound.
For the electric limit $Q/M=0.8$, $\bar g=10$, $\eta=0$,
and $\hat\alpha=\hat\beta=0$,
a separate integration-and-matching calculation of
Eq.~\eqref{eq:RN_odd_master} also reproduces
$\Omega\simeq0.14456i,0.42539i$.
These tests confirm the listed unstable backgrounds without
determining the instability thresholds as the parameters vary.

\subsection{Accuracy of the NED expansion}
\label{app:response_accuracy}

For each mode in the NED expansion of Sec.~\ref{sec:ned_expansion},
we suppress the family index $j$ and write
$\Delta\Omega=\lambda\kappa+\lambda^2L_2+\cdots$ along each
coefficient sequence.
Here $\kappa$ and $L_2$ are the linear and quadratic coefficients.
When only $\hat\alpha$ is varied, with $\hat\beta=0$,
we denote them by $\kappa_\alpha$ and $L_{2,\alpha}$.
When only $\hat\beta$ is varied, with $\hat\alpha=0$,
we denote them by $\kappa_\beta$ and $L_{2,\beta}$.
The mixed derivative is
$H_{\alpha\beta}=(\partial_{\hat\alpha}\partial_{\hat\beta}
\Omega)_{\hat\alpha=\hat\beta=0}$.
Finite differences use steps $\delta\lambda=0.003$ for $\kappa$
and $0.015$ for $L_2$.
For the resolution test, both are recomputed from
$\lambda=0,\pm0.015,\pm0.030$.
The second-order prediction at $\lambda=0.05$, beyond these
sampled values, changes by at most $1.6\times10^{-10}$
over $N_{\rm c}=36,48,60$ for
$(\hat\alpha,\hat\beta)=(\lambda,0),(0,\lambda),(\lambda,\lambda)$.

For the even-EM example with
$(\hat\alpha,\hat\beta)=(4\lambda/7,\lambda)$ in
Fig.~\ref{fig:V_09}, the coefficients are
$\kappa=(4/7)\kappa_\alpha+\kappa_\beta$ and
$L_2=(16/49)L_{2,\alpha}+L_{2,\beta}+(4/7)H_{\alpha\beta}$.
At $\lambda=0.05$, the absolute differences between predicted
and directly computed complex shifts are $7.8\times10^{-7}$
at first order, $3.2\times10^{-9}$ at second order, and
$3.4\times10^{-7}$ without the mixed term.
These exceed the resolution changes above and support the
improvement discussed in Sec.~\ref{sec:ned_expansion}.
The remaining difference includes numerical error and
higher-order Taylor terms.

\section{Frequency-response formulas}
\label{app:response}

We give the frequency derivatives and degenerate-pair
projections used in Secs.~\ref{sec:ned_composition}
and~\ref{sec:ned_expansion}.
At fixed $(Q/M,\bar g,\eta)$, the derivatives of the
eigenfrequencies in Eq.~\eqref{eq:ned_shift} are
\begin{equation}
K_{j,p}=\left.\partial_p\Omega_j\right|_{\ah=\bh=0},
\qquad p=\hat\alpha,\hat\beta.
\label{eq:K}
\end{equation}
Derivative labels $\alpha,\beta$ below refer to the
hatted coefficients.
For an isolated simple eigenfrequency, the full NED
shift has the local expansion
\begin{align}
\Delta\Omega_j={}&\ah K_{j,\alpha}+\bh K_{j,\beta}
+\frac12\ah^2 H_{j,\alpha\alpha}\nonumber\\*
&+\ah\bh H_{j,\alpha\beta}
+\frac12\bh^2 H_{j,\beta\beta}\nonumber\\*
&+O\!\left((|\hat\alpha|+|\hat\beta|)^3\right),
\label{eq:Taylor}
\end{align}
where $H_{j,pq}=\partial_p\partial_q\Omega_j|_{\ah=\bh=0}$,
with $p,q\in\{\hat\alpha,\hat\beta\}$.
This approximates the directly computed finite-NED
frequencies; it is not a truncation imposed on the solver.
A higher-order prediction for a specific EFT would also
require the operators discussed in Appendix~\ref{app:EFT}.

Let $\mathcal P(\Omega,p)$ denote either the spectral matrix
in Eq.~\eqref{eq:QEP} or the matrix imposing the matching
conditions in the integration method of Sec.~\ref{sec:pert}.
In both cases, we express the frequency as $\omega=\Omega/M(p)$.
At a QNM frequency, nonzero right and left vectors satisfy
$\mathcal P\bm e_R=0$ and $\bm e_L^\dagger\mathcal P=0$,
where a dagger denotes complex-conjugate transpose.
The matrices $\mathcal P_{,p}$ and $\mathcal P_{,\Omega}$
are obtained by differentiating each matrix entry with respect
to $p$ at fixed $\Omega$ and to $\Omega$ at fixed $p$,
respectively.
For an isolated, nondegenerate QNM frequency,
\begin{equation}
\partial_p\Omega
=-\frac{\bm e_L^\dagger\mathcal P_{,p}\bm e_R}
{\bm e_L^\dagger\mathcal P_{,\Omega}\bm e_R},
\label{eq:simple}
\end{equation}
provided the denominator is nonzero.
The $p$ derivative includes changes in the background,
ADM mass, and boundary phases, with $(Q/M,\bar g,\eta)$
held fixed.

Along $(\ah,\bh)=\lambda(a_\alpha,a_\beta)$, the coefficients
introduced in Appendix~\ref{app:response_accuracy} are
$\kappa=\Omega'(0)$ and $L_2=\Omega''(0)/2$, with the family
index $j$ omitted on $\kappa,L_2,H_{pq}$.
Primes denote $\lambda$ derivatives.
In particular, $\kappa_\alpha=K_{j,\alpha}$ and
$\kappa_\beta=K_{j,\beta}$.
For $\Omega_n=\Omega(n\,\delta\lambda)$, fourth-order
centered approximations are
\begin{align}
\kappa(\delta\lambda)&=
\frac{\Omega_{-2}-8\Omega_{-1}+8\Omega_1-\Omega_2}
{12\,\delta\lambda},\nonumber\\*
L_2(\delta\lambda)&=
\frac{-\Omega_2+16\Omega_1-30\Omega_0+16\Omega_{-1}-\Omega_{-2}}
{24\,(\delta\lambda)^2}.
\label{eq:frequency_differences}
\end{align}
The denominator 24 accounts for $L_2=\Omega''(0)/2$.
Let $L_{2,\mathrm{eq}}$ denote the quadratic coefficient
along $(\hat\alpha,\hat\beta)=(\lambda,\lambda)$.
The second derivatives then satisfy
$H_{\alpha\alpha}=2L_{2,\alpha}$,
$H_{\beta\beta}=2L_{2,\beta}$, and
$H_{\alpha\beta}=L_{2,\mathrm{eq}}-L_{2,\alpha}-L_{2,\beta}$.
The numerical settings and resolution checks are given in
Appendix~\ref{app:response_accuracy}.

At $\eta=\bar g=\hat\alpha=\hat\beta=0$, each RN parity pair
$J=\mathrm{grav},\mathrm{EM}$ in Eq.~\eqref{eq:RN_pair_sums}
consists of independent even and odd modes with the same
frequency.
We define $V_J=(\bm e_{R,J,e},\bm e_{R,J,o})$ and
$U_J=(\bm e_{L,J,e},\bm e_{L,J,o})$,
where $s=e,o$ labels parity and the vectors satisfy
$\mathcal P\bm e_{R,J,s}=0$ and
$\bm e_{L,J,s}^\dagger\mathcal P=0$
at the common RN frequency.
Projecting onto these two modes gives the $2\times2$ matrices
\begin{align}
N_J&=U_J^\dagger\mathcal P_{,\Omega}V_J,\nonumber\\
P_J^{(p)}&=U_J^\dagger\mathcal P_{,p}V_J,
\qquad p=\hat\alpha,\hat\beta.
\label{eq:RN_projected_matrices}
\end{align}
Here $N_J$ multiplies the first-order frequency change
in the projected two-mode equation, while $P_J^{(p)}$
describes the change induced by varying $p$.

If the pair is isolated from the remaining spectrum and
$N_J$ is invertible, the two eigenvalues of
$-N_J^{-1}P_J^{(p)}$ give the first-order frequency derivatives
with respect to $p$.
Their sum is $\partial_p\Sigma_J$, so the electric-reference
response coefficients defined in
Eq.~\eqref{eq:RN_pair_derivatives} satisfy
\begin{align}
T_{J,\alpha}&=\operatorname{Tr}\!\left(-N_J^{-1}P_J^{(\hat\alpha)}\right),
\nonumber\\
T_{J,\beta}&=\operatorname{Tr}\!\left(-N_J^{-1}P_J^{(\hat\beta)}\right).
\label{eq:RN_pair_response_trace}
\end{align}
Both degenerate modes must be retained in this calculation;
the single-mode formula~\eqref{eq:simple} cannot be applied
directly to an arbitrary linear combination of them.

In this parity basis, the decoupling of the even and odd
sectors makes $N_J$ diagonal:
\begin{align}
N_J&=\operatorname{diag}(\nu_{J,e},\nu_{J,o}),\nonumber\\
\nu_{J,s}&=\bm e_{L,J,s}^\dagger
\mathcal P_{,\Omega}\bm e_{R,J,s}.
\label{eq:RN_parity_frequency_projection}
\end{align}
Nonzero $\nu_{J,e}$ and $\nu_{J,o}$ ensure that $N_J$ is
invertible; the checks at $Q/M=0.6$ are reported in
Appendix~\ref{app:operator_checks}.
If a scalar mode also shares this frequency, it must be
included in the projection unless the calculation is restricted
to the exactly decoupled gravitoelectromagnetic sector.

\section{Duality and first-order frequency relations}
\label{app:duality}

We derive the individual-branch relation
\eqref{eq:duality_individual} for equal coefficients and
the more general pair-sum relation~\eqref{eq:dualitylaw}
used in Sec.~\ref{sec:ned_composition}.

\subsection{First-order duality map}

At zero axion coupling, $g=0$, the local electric
displacement is
\begin{equation}
\bm D=\frac{\partial\cL}{\partial\bm E}
=\bm E+\alpha(E^2-B^2)\bm E
+2\beta(\bm E\cdot\bm B)\bm B.
\end{equation}
Inverting to first NED order gives the electromagnetic
Hamiltonian density
\begin{align}
\mathcal H&=\bm D\cdot\bm E-\cL\nonumber\\*
&=\frac{\mathcal S}{2}
-\frac{\alpha\mathcal U^2+\beta\mathcal V^2}{4}
+O(\alpha^2,\alpha\beta,\beta^2),
\end{align}
where $\bm E,\bm B$ are local orthonormal-frame fields,
$\mathcal S=D^2+B^2$, $\mathcal U=D^2-B^2$, and
$\mathcal V=2\bm D\cdot\bm B$.
We choose the rotation angle $\vartheta$ so that
$q_E/q_T=\cos\vartheta$ and $\eta=\sin\vartheta$.
A duality rotation of $(\bm D,\bm B)$ through $\vartheta$
rotates $(\mathcal U,\mathcal V)$ through $2\vartheta$, giving
\begin{align}
\alpha_\vartheta&=\alpha\cos^2(2\vartheta)+\beta\sin^2(2\vartheta),
\label{eq:duality_alpha_rotation}\\
\beta_\vartheta&=\alpha\sin^2(2\vartheta)+\beta\cos^2(2\vartheta).
\label{eq:duality_beta_rotation}
\end{align}
The rotation also generates a parity-odd
$\mathcal U\mathcal V$ term proportional to
$(\alpha-\beta)\sin(2\vartheta)\cos(2\vartheta)$.
It vanishes when $\alpha=\beta$.
The dimensionless rotated coefficients are
$\hat\alpha_\vartheta=\alpha_\vartheta\Mp^2/r_h^2$ and
$\hat\beta_\vartheta=\beta_\vartheta\Mp^2/r_h^2$.

Through first NED order, the rotation maps the coupled
field equations~\cite{GibbonsRasheed:1995}.
As a metric-independent canonical rotation of electric
displacement and magnetic flux densities, it preserves the
electromagnetic symplectic form and momentum density
and rotates the Gauss constraints.
Covariantly, the rotation acts on $(F,{}^\star P)$,
where ${}^\star$ denotes the Hodge dual with the
convention~\eqref{eq:dual}.
Here $F$ is the field-strength two-form and
$P=(1/2)P_{\mu\nu}\,{\rm d}x^\mu\wedge{\rm d}x^\nu$
is the constitutive two-form, with
$P^{\mu\nu}=\cL_{,\cF}F^{\mu\nu}
+\cL_{,\cG}\widetilde F^{\mu\nu}$.
The indices of $P^{\mu\nu}$ are lowered using the metric.
The closure of $F$ and ${}^\star P$ follows from the
Bianchi identity and Eq.~\eqref{eq:gauge_eom}, respectively.
The perturbation map includes
$\delta({}^\star P)={}^\star\delta P+(\delta{}^\star)\bar P$,
where $\bar P$ is the background value of $P$ and
$\delta{}^\star$ denotes the first-order change in the
Hodge-dual operator induced by the metric perturbation.

For a locally invertible constitutive relation and
regular nonextremal horizon, this finite algebraic map
preserves the physical ingoing solution space in
coordinates regular at the future horizon.
At infinity it approaches a constant Maxwell polarization
rotation and preserves the outgoing branch, understood
by analytic continuation for damped QNMs.
The geometry, ADM mass, horizon radius, and asymptotic
clock are preserved through first NED order.
The charge rotation preserves $q_E^2+q_M^2$, so the map
is compatible with fixed $Q/M$ and the hatted-coefficient
normalization.
These statements relate the original and rotated theories
through first NED order.

\subsection{Equal coefficients: individual branches}
\label{app:duality_equal}

When $\alpha=\beta$, the quartic Hamiltonian depends on
$\mathcal U$ and $\mathcal V$ only through the invariant
\begin{equation}
\mathcal U^2+\mathcal V^2
=\mathcal S^2-4|\bm D\times\bm B|^2.
\label{eq:duality_H_invariant}
\end{equation}
Both $\mathcal S$ and $\bm D\times\bm B$ are unchanged
by the canonical duality rotation.
Thus $\alpha_\vartheta=\beta_\vartheta=\alpha$, the
generated parity-odd term vanishes, and the rotated
Hamiltonian agrees with the original one through first
NED order.
The coupled gravitational--EM perturbation operators
are therefore related within the same theory, including
the boundary conditions and background normalization
described above.
The minimally coupled free scalar is decoupled at $g=0$
and does not affect this argument.

This map preserves the complete first-order frequency
corrections in each RN gravitational or EM subspace,
not only their trace.
Equivalently, the projected two-mode frequency pencils
at different $\eta$ have the same eigenvalues.
Following the branches continuously from $\eta=0$
identifies their individual shifts and gives
Eq.~\eqref{eq:duality_individual}, provided the pair
remains isolated in the gravitational--EM spectrum.
At an unresolved branch crossing the equality concerns
the unordered pair of shifts.
The two parity partners can have different linear NED
corrections; the result constrains charge composition,
not axial--polar isospectrality.
Terms of second and higher NED order in the Hamiltonian
of the quartic-truncated theory need not be duality
invariant, so finite-$\lambda$ residuals are allowed.
Adding a complete self-dual nonlinear theory would be
a different assumption from the truncation used here.

\begin{table}[t]
\centering
\caption{Integration-and-matching check of the individual-branch
relation~\eqref{eq:duality_individual} at $Q/M=0.6$,
$\bar g=0$, $\ell=2$, and $\eta=0.4$.
The entries are $|\delta\kappa_j(\delta\lambda)|$, with the two-point
estimate defined in Eq.~\eqref{eq:duality_fd_check}.
The approximately fourfold decrease on halving $\delta\lambda$
is consistent with the $O((\delta\lambda)^2)$ truncation error of
this derivative estimate.}
\label{tab:duality_individual_check}
\begin{tabular}{lcc}
\toprule
Branch & $\delta\lambda=0.003$ & $\delta\lambda=0.0015$\\
\midrule
Even-grav & $1.38\times10^{-9}$ & $3.46\times10^{-10}$\\
Odd-grav  & $5.79\times10^{-10}$ & $1.45\times10^{-10}$\\
Even-EM   & $1.45\times10^{-8}$ & $3.63\times10^{-9}$\\
Odd-EM    & $7.11\times10^{-9}$ & $1.78\times10^{-9}$\\
\bottomrule
\end{tabular}
\end{table}

For an independent integration-and-matching check of
each branch, we use the dimensionless step $\delta\lambda>0$
along $\hat\alpha=\hat\beta=\lambda$ and define
\begin{equation}
\begin{aligned}
\widehat\kappa_j(\eta;\delta\lambda)
&=\frac{\Omega_j(\eta;\delta\lambda,\delta\lambda)-\Omega_j(\eta;-\delta\lambda,-\delta\lambda)}{2\delta\lambda},\\
\delta\kappa_j(\delta\lambda)
&=\widehat\kappa_j(\eta;\delta\lambda)-\widehat\kappa_j(0;\delta\lambda).
\end{aligned}
\label{eq:duality_fd_check}
\end{equation}
The hat denotes a numerical derivative estimate.
We use the order-14 outgoing expansion with
$r_\infty/r_h=55$, $r_{\rm match}/r_h=8$, and working
precision 45, retaining each endpoint's own ADM mass.
Table~\ref{tab:duality_individual_check} shows the
$\eta=0.4$ check; the $\eta=0.2$ results have the same
step scaling.
For both compositions and all four branches, the ratio
$|\delta\kappa_j(0.003)|/|\delta\kappa_j(0.0015)|$
lies between 3.999 and 4.001.
The four-frequency derivative stencil in
Eq.~\eqref{eq:frequency_differences}, at $\delta\lambda=0.0015$,
reduces the composition differences to below
$2.3\times10^{-13}$ in complex modulus.
These small residuals are a consistency check rather
than an absolute numerical error bound.
The complete EM stencils at $\eta=0.4$ and $\delta\lambda=0.003$
are also available at $r_\infty/r_h=50$;
the resulting composition differences change by less
than $2.5\times10^{-13}$ between the two radii.
This radius comparison tests the two EM branches at
$\delta\lambda=0.003$; the other checks above test parameter-step
dependence.

\subsection{Unequal coefficients: RN pair sums}

For general coefficients, we apply the map to an RN
even--odd pair isolated from the
remaining spectrum, with invertible $N_J$, as described in
Appendix~\ref{app:response}.
At the electric reference, $N_J$ is diagonal in the parity
basis, whereas the generated parity-odd term gives an
off-diagonal perturbation matrix.
Its contribution to the trace in
Eq.~\eqref{eq:RN_pair_response_trace} therefore vanishes.
It can affect individual first-order shifts while leaving
their sum unchanged.

The NED-induced change in the pair's frequency sum is then
\begin{align}
\Delta\Sigma_J(\eta)
&\equiv\Sigma_J(\eta;\hat\alpha,\hat\beta)
-\Sigma_J(\eta;0,0)\nonumber\\
&=\hat\alpha_\vartheta T_{J,\alpha}
+\hat\beta_\vartheta T_{J,\beta}
+O(\hat\alpha^2,\hat\alpha\hat\beta,\hat\beta^2),
\label{eq:duality_pair_shift}
\end{align}
where $T_{J,\alpha}$ and $T_{J,\beta}$ are the derivatives
of the pair's frequency sum at the electric RN reference,
defined in Eq.~\eqref{eq:RN_pair_derivatives}.
Consequently,
\begin{equation}
B_J(\eta)=\Delta\Sigma_J(\eta)-\Delta\Sigma_J(0).
\label{eq:duality_pair_magnetic_difference}
\end{equation}
Using $\sin^2(2\vartheta)=4\eta^2(1-\eta^2)$
in the rotated coefficients yields Eq.~\eqref{eq:dualitylaw},
restricted to $g=0$ and first NED order.
The definition of $B_J$ remains valid for continued
families at nonzero axion coupling, but their shifts
must then be computed from the axion-coupled operator,
as in Sec.~\ref{sec:ned_composition}.

\begin{acknowledgments}
\widowpenalty=10000
\clubpenalty=10000
S.T.\ acknowledges support from JSPS KAKENHI Grant Nos.~26K07090 and 26H00847, 
and from the Waseda University Special Research Projects (No.~2026C-486). 

The authors used ChatGPT (OpenAI; GPT-6.1)
and Codex (OpenAI; GPT-6 Astra)
to assist with parts of the symbolic and numerical calculations, code development, and validation.
The authors directed this assistance, reviewed 
and revised the resulting material, and checked the reported results using the symbolic checks and independent numerical tests
described in this paper.

The authors take full responsibility for the scientific
content and the final manuscript.

\end{acknowledgments}

\section*{Data availability}
The data and numerical codes supporting the findings of this
study are available from the corresponding author upon
reasonable request.

\FloatBarrier

\bibliographystyle{mybibstyle}
\bibliography{bib}

@article{Svrcek:2006yi,
    author = "Svrcek, Peter and Witten, Edward",
    title = "{Axions In String Theory}",
    eprint = "hep-th/0605206",
    archivePrefix = "arXiv",
    reportNumber = "SLAC-PUB-11894",
    doi = "10.1088/1126-6708/2006/06/051",
    journal = "JHEP",
    volume = "06",
    pages = "051",
    year = "2006"
}

@article{Peccei:1977hh,
    author = "Peccei, R. D. and Quinn, Helen R.",
    title = "{CP Conservation in the Presence of Instantons}",
    reportNumber = "ITP-568-STANFORD",
    doi = "10.1103/PhysRevLett.38.1440",
    journal = "Phys. Rev. Lett.",
    volume = "38",
    pages = "1440--1443",
    year = "1977"
}

@article{Lee:1991jw,
    author = "Lee, Ki-Myeong and Weinberg, Erick J.",
    title = "{Charge black holes with scalar hair}",
    reportNumber = "CU-TP-515",
    doi = "10.1103/PhysRevD.44.3159",
    journal = "Phys. Rev. D",
    volume = "44",
    pages = "3159--3163",
    year = "1991"
}

@article{Filippini:2019cqk,
    author = "Filippini, Francesco and Tasinato, Gianmassimo",
    title = "{On long range axion hairs for black holes}",
    eprint = "1903.02950",
    archivePrefix = "arXiv",
    primaryClass = "gr-qc",
    doi = "10.1088/1361-6382/ab4371",
    journal = "Class. Quant. Grav.",
    volume = "36",
    number = "21",
    pages = "215015",
    year = "2019"
}

@article{Fernandes:2019kmh,
    author = "Fernandes, Pedro G. S. and Herdeiro, Carlos A. R. and Pombo, Alexandre M. and Radu, Eugen and Sanchis-Gual, Nicolas",
    title = "{Charged black holes with axionic-type couplings: Classes of solutions and dynamical scalarization}",
    eprint = "1908.00037",
    archivePrefix = "arXiv",
    primaryClass = "gr-qc",
    doi = "10.1103/PhysRevD.100.084045",
    journal = "Phys. Rev. D",
    volume = "100",
    number = "8",
    pages = "084045",
    year = "2019"
}

@article{Pereniguez:2023wxf,
    author = "Pere\~niguez, David",
    title = "{Black hole perturbations and electric-magnetic duality}",
    eprint = "2302.10942",
    archivePrefix = "arXiv",
    primaryClass = "gr-qc",
    doi = "10.1103/PhysRevD.108.084046",
    journal = "Phys. Rev. D",
    volume = "108",
    number = "8",
    pages = "084046",
    year = "2023"
}

@article{DeFelice:2023rra,
    author = "De Felice, Antonio and Tsujikawa, Shinji",
    title = "{Can we distinguish black holes with electric and magnetic charges from quasinormal modes?}",
    eprint = "2312.03191",
    archivePrefix = "arXiv",
    primaryClass = "gr-qc",
    reportNumber = "YITP-23-158, WUCG-23-12",
    doi = "10.1103/PhysRevD.109.084022",
    journal = "Phys. Rev. D",
    volume = "109",
    number = "8",
    pages = "084022",
    year = "2024"
}

@article{Nomura:2020tpc,
    author = "Nomura, Kimihiro and Yoshida, Daisuke and Soda, Jiro",
    title = "{Stability of magnetic black holes in general nonlinear electrodynamics}",
    eprint = "2004.07560",
    archivePrefix = "arXiv",
    primaryClass = "gr-qc",
    reportNumber = "KOBE-COSMO-20-08",
    doi = "10.1103/PhysRevD.101.124026",
    journal = "Phys. Rev. D",
    volume = "101",
    number = "12",
    pages = "124026",
    year = "2020"
}

@article{Nomura:2021efi,
    author = "Nomura, Kimihiro and Yoshida, Daisuke",
    title = "{Quasinormal modes of charged black holes with corrections from nonlinear electrodynamics}",
    eprint = "2111.06273",
    archivePrefix = "arXiv",
    primaryClass = "gr-qc",
    reportNumber = "KOBE-COSMO-21-16",
    doi = "10.1103/PhysRevD.105.044006",
    journal = "Phys. Rev. D",
    volume = "105",
    number = "4",
    pages = "044006",
    year = "2022"
}

@article{Berti:2003zu,
    author = "Berti, Emanuele and Kokkotas, Kostas D.",
    title = "{Asymptotic quasinormal modes of Reissner-Nordstrom and Kerr black holes}",
    eprint = "hep-th/0303029",
    archivePrefix = "arXiv",
    doi = "10.1103/PhysRevD.68.044027",
    journal = "Phys. Rev. D",
    volume = "68",
    pages = "044027",
    year = "2003"
}

@article{Onozawa:1995vu,
    author = "Onozawa, Hisashi and Mishima, Takashi and Okamura, Takashi and Ishihara, Hideki",
    title = "{Quasinormal modes of maximally charged black holes}",
    eprint = "gr-qc/9603021",
    archivePrefix = "arXiv",
    reportNumber = "TIT-HEP-311, COSMO-61",
    doi = "10.1103/PhysRevD.53.7033",
    journal = "Phys. Rev. D",
    volume = "53",
    pages = "7033--7040",
    year = "1996"
}

@article{Andersson:1996xw,
    author = "Andersson, Nils and Onozawa, Hisashi",
    title = "{Quasinormal modes of nearly extreme Reissner-Nordstrom black holes}",
    eprint = "gr-qc/9607054",
    archivePrefix = "arXiv",
    reportNumber = "TIT-HEP-339, COSMO-76, WUGRAV-96-7",
    doi = "10.1103/PhysRevD.54.7470",
    journal = "Phys. Rev. D",
    volume = "54",
    pages = "7470--7475",
    year = "1996"
}

@article{Leaver:1985ax,
    author = "Leaver, E. W.",
    title = "{An Analytic representation for the quasi normal modes of Kerr black holes}",
    doi = "10.1098/rspa.1985.0119",
    journal = "Proc. Roy. Soc. Lond. A",
    volume = "402",
    pages = "285--298",
    year = "1985"
}

@article{Leaver:1986gd,
    author = "Leaver, Edward W.",
    title = "{Spectral decomposition of the perturbation response of the Schwarzschild geometry}",
    doi = "10.1103/PhysRevD.34.384",
    journal = "Phys. Rev. D",
    volume = "34",
    pages = "384--408",
    year = "1986"
}

@article{Leaver:1990zz,
    author = "Leaver, Edward W.",
    title = "{Quasinormal modes of Reissner-Nordstrom black holes}",
    doi = "10.1103/PhysRevD.41.2986",
    journal = "Phys. Rev. D",
    volume = "41",
    pages = "2986--2997",
    year = "1990"
}

@article{Gunter:1980,
    author = "Gunter, D. L.",
    title = "{A study of the coupled gravitational and electromagnetic perturbations to 
    the Reissner-Nordstrom black hole: the scattering matrix, energy conversion, and quasi-normal modes}",
    doi = "10.1098/rsta.1980.0190",
    journal = "Phil. Trans. Roy. Soc. Lond",
    volume = "A296",
    pages = "497",
    year = "1980"
}

@article{Kokkotas:1988fm,
    author = "Kokkotas, K. D. and Schutz, Bernard F.",
    title = "{Black Hole Normal Modes: A {WKB} Approach. 3. The {Reissner-Nordstrom} Black Hole}",
    doi = "10.1103/PhysRevD.37.3378",
    journal = "Phys. Rev. D",
    volume = "37",
    pages = "3378--3387",
    year = "1988"
}

@article{Schutz:1985km,
    author = "Schutz, Bernard F. and Will, Clifford M.",
    title = "{BLACK HOLE NORMAL MODES: A SEMIANALYTIC APPROACH}",
    reportNumber = "PRINT-85-0063 (WASH.U.,ST.LOUIS)",
    doi = "10.1086/184453",
    journal = "Astrophys. J. Lett.",
    volume = "291",
    pages = "L33--L36",
    year = "1985"
}

@article{Iyer:1986np,
    author = "Iyer, Sai and Will, Clifford M.",
    title = "{Black Hole Normal Modes: A {WKB} Approach. 
    1. Foundations and Application of a Higher Order {WKB} Analysis of Potential Barrier Scattering}",
    reportNumber = "Print-86-1482 (WASH. U., ST. LOUIS)",
    doi = "10.1103/PhysRevD.35.3621",
    journal = "Phys. Rev. D",
    volume = "35",
    pages = "3621",
    year = "1987"
}

@article{Chandrasekhar:1979iz,
    author = "Chandrasekhar, Subrahmanyan",
    editor = "Wali, K. C.",
    title = "{On one-dimensional potential barriers having equal reflexion and transmission coefficients}",
    doi = "10.1098/rspa.1980.0008",
    journal = "Proc. Roy. Soc. Lond. A",
    volume = "369",
    pages = "425",
    year = "1980"
}

@article{Moncrief:1974gw,
    author = "Moncrief, Vincent",
    title = "{Odd-parity stability of a Reissner-Nordstrom black hole}",
    doi = "10.1103/PhysRevD.9.2707",
    journal = "Phys. Rev. D",
    volume = "9",
    pages = "2707--2709",
    year = "1974"
}

@article{Moncrief:1974ng,
    author = "Moncrief, Vincent",
    title = "{Stability of Reissner-Nordstrom black holes}",
    doi = "10.1103/PhysRevD.10.1057",
    journal = "Phys. Rev. D",
    volume = "10",
    pages = "1057--1059",
    year = "1974"
}

@article{Moncrief:1975sb,
    author = "Moncrief, Vincent",
    title = "{Gauge-invariant perturbations of Reissner-Nordstrom black holes}",
    doi = "10.1103/PhysRevD.12.1526",
    journal = "Phys. Rev. D",
    volume = "12",
    pages = "1526--1537",
    year = "1975"
}

@article{Zerilli:1974ai,
    author = "Zerilli, F. J.",
    title = "{Perturbation analysis for gravitational and electromagnetic radiation in a reissner-nordstroem geometry}",
    doi = "10.1103/PhysRevD.9.860",
    journal = "Phys. Rev. D",
    volume = "9",
    pages = "860--868",
    year = "1974"
}

@article{Kokkotas:1999bd,
    author = "Kokkotas, Kostas D. and Schmidt, Bernd G.",
    title = "{Quasinormal modes of stars and black holes}",
    eprint = "gr-qc/9909058",
    archivePrefix = "arXiv",
    doi = "10.12942/lrr-1999-2",
    journal = "Living Rev. Rel.",
    volume = "2",
    pages = "2",
    year = "1999"
}

@article{Nollert:1999ji,
    author = "Nollert, Hans-Peter",
    title = "{TOPICAL REVIEW: Quasinormal modes: 
    the characteristic `sound' of black holes and neutron stars}",
    doi = "10.1088/0264-9381/16/12/201",
    journal = "Class. Quant. Grav.",
    volume = "16",
    pages = "R159--R216",
    year = "1999"
}

@article{Berti:2009kk,
    author = "Berti, Emanuele and Cardoso, Vitor and Starinets, Andrei O.",
    title = "{Quasinormal modes of black holes and black branes}",
    eprint = "0905.2975",
    archivePrefix = "arXiv",
    primaryClass = "gr-qc",
    doi = "10.1088/0264-9381/26/16/163001",
    journal = "Class. Quant. Grav.",
    volume = "26",
    pages = "163001",
    year = "2009"
}

@article{Konoplya:2011qq,
    author = "Konoplya, R. A. and Zhidenko, A.",
    title = "{Quasinormal modes of black holes: From astrophysics to string theory}",
    eprint = "1102.4014",
    archivePrefix = "arXiv",
    primaryClass = "gr-qc",
    doi = "10.1103/RevModPhys.83.793",
    journal = "Rev. Mod. Phys.",
    volume = "83",
    pages = "793--836",
    year = "2011"
}

@article{Pani:2013pma,
    author = "Pani, Paolo",
    editor = "Cardoso, V. and Gualtieri, L. and Herdeiro, C. and Sperhake, U.",
    title = "{Advanced Methods in Black-Hole Perturbation Theory}",
    eprint = "1305.6759",
    archivePrefix = "arXiv",
    primaryClass = "gr-qc",
    doi = "10.1142/S0217751X13400186",
    journal = "Int. J. Mod. Phys. A",
    volume = "28",
    pages = "1340018",
    year = "2013"
}

@article{Regge:1957td,
    author = "Regge, Tullio and Wheeler, John A.",
    title = "{Stability of a Schwarzschild singularity}",
    doi = "10.1103/PhysRev.108.1063",
    journal = "Phys. Rev.",
    volume = "108",
    pages = "1063--1069",
    year = "1957"
}

@article{Chandrasekhar:1975nkd,
    author = "Chandrasekhar, Subrahmanyan",
    title = "{On the equations governing the perturbations of the Schwarzschild black hole}",
    doi = "10.1098/rspa.1975.0066",
    journal = "Proc. Roy. Soc. Lond. A",
    volume = "343",
    number = "1634",
    pages = "289--298",
    year = "1975"
}

@book{Chandrasekhar:1985kt,
    author = "Chandrasekhar, Subrahmanyan",
    title = "{The mathematical theory of black holes}",
    isbn = "978-0-19-850370-5",
    publisher = "Oxford University Press",
    address = "Oxford",
    year = "1985"
}

@article{Chandrasekhar:1975zza,
    author = "Chandrasekhar, S. and Detweiler, Steven L.",
    title = "{The quasi-normal modes of the Schwarzschild black hole}",
    doi = "10.1098/rspa.1975.0112",
    journal = "Proc. Roy. Soc. Lond. A",
    volume = "344",
    pages = "441--452",
    year = "1975"
}

@article{Zerilli:1970se,
    author = "Zerilli, Frank J.",
    title = "{Effective potential for even parity Regge-Wheeler gravitational perturbation equations}",
    doi = "10.1103/PhysRevLett.24.737",
    journal = "Phys. Rev. Lett.",
    volume = "24",
    pages = "737--738",
    year = "1970"
}

@article{Zerilli:1970wzz,
    author = "Zerilli, F. J.",
    title = "{Gravitational field of a particle falling in a schwarzschild geometry analyzed in tensor harmonics}",
    doi = "10.1103/PhysRevD.2.2141",
    journal = "Phys. Rev. D",
    volume = "2",
    pages = "2141--2160",
    year = "1970"
}

@article{LIGOScientific:2016aoc,
    author = "Abbott, B. P. and others",
    collaboration = "LIGO Scientific, Virgo",
    title = "{Observation of Gravitational Waves from a Binary Black Hole Merger}",
    eprint = "1602.03837",
    archivePrefix = "arXiv",
    primaryClass = "gr-qc",
    reportNumber = "LIGO-P150914",
    doi = "10.1103/PhysRevLett.116.061102",
    journal = "Phys. Rev. Lett.",
    volume = "116",
    number = "6",
    pages = "061102",
    year = "2016"
}

@article{Langlois:2021xzq,
    author = "Langlois, David and Noui, Karim and Roussille, Hugo",
    title = "{Asymptotics of linear differential systems and application to quasinormal modes of nonrotating black holes}",
    eprint = "2103.14744",
    archivePrefix = "arXiv",
    primaryClass = "gr-qc",
    doi = "10.1103/PhysRevD.104.124043",
    journal = "Phys. Rev. D",
    volume = "104",
    number = "12",
    pages = "124043",
    year = "2021"
}

@article{DeFelice:2024axionQNM,
  author = "De Felice, Antonio and Tsujikawa, Shinji",
  title = "{Probing the signature of axions through the quasinormal modes of black holes}",
  journal = "Phys. Lett. B", volume = "855", pages = "138808", year = "2024",
  doi = "10.1016/j.physletb.2024.138808",
  eprint = "2402.08868", archivePrefix = "arXiv", primaryClass = "gr-qc"
}

@article{DeFelice:2025NED,
  author = "De Felice, Antonio and Tsujikawa, Shinji",
  title = "{Nonsingular black holes and spherically symmetric objects in nonlinear electrodynamics with a scalar field}",
  journal = "Phys. Rev. D", volume = "111", pages = "064051", year = "2025",
  doi = "10.1103/PhysRevD.111.064051", eprint = "2412.04754", archivePrefix = "arXiv", primaryClass = "gr-qc"
}

@article{Taniguchi:2025dyonic,
  author = "Taniguchi, Kitaro and Nishimura, Shunta and Tsukamoto, Naoki and Kase, Ryotaro",
  title = "{Linear perturbations of dyonic black holes in the lowest-order $U(1)$ gauge-invariant scalar-vector-tensor theories}",
  journal = "Phys. Rev. D", volume = "112", pages = "124043", year = "2025",
  doi = "10.1103/718g-ncgn", eprint = "2504.21279", archivePrefix = "arXiv", primaryClass = "gr-qc"
}

@article{Boskovic:2018lkj,
  author = "Boskovic, Mateja and Brito, Richard and Cardoso, Vitor and Ikeda, Taishi and Witek, Helvi",
  title = "{Axionic instabilities and new black hole solutions}",
  journal = "Phys. Rev. D", volume = "99", pages = "035006", year = "2019",
  doi = "10.1103/PhysRevD.99.035006", eprint = "1811.04945", archivePrefix = "arXiv", primaryClass = "gr-qc"
}

@article{Takahashi:2025resonance,
  author = "Takahashi, Takuya and Motohashi, Hayato and Takahashi, Kazufumi",
  title = "{Resonance of black hole quasinormal modes in coupled systems}",
  journal = "Phys. Rev. D", volume = "112", pages = "064006", year = "2025",
  doi = "10.1103/59tb-2x9r", eprint = "2505.03883", archivePrefix = "arXiv", primaryClass = "gr-qc"
}

@article{Takahashi:2026ringdown,
  author = "Takahashi, Takuya and Motohashi, Hayato and Takahashi, Kazufumi",
  title = "{Avoided crossings and resonances in black hole ringdown with coupled fields: A case study of the Einstein--Maxwell--axion system}",
  year = "2026", eprint = "2609.34401", archivePrefix = "arXiv", primaryClass = "gr-qc"
}

@article{McManus:2019qnm,
  author = "McManus, Ryan and Berti, Emanuele and Macedo, Caio F. B. and Kimura, Masashi and Maselli, Andrea and Cardoso, Vitor",
  title = "{Parametrized black hole quasinormal ringdown. II. Coupled equations and quadratic corrections for nonrotating black holes}",
  journal = "Phys. Rev. D", volume = "100", pages = "044061", year = "2019",
  doi = "10.1103/PhysRevD.100.044061", eprint = "1906.05155", archivePrefix = "arXiv", primaryClass = "gr-qc"
}

@article{GibbonsRasheed:1995,
  author = "Gibbons, G. W. and Rasheed, D. A.",
  title = "{Electric-magnetic duality rotations in non-linear electrodynamics}",
  journal = "Nucl. Phys. B", volume = "454", pages = "185--206", year = "1995",
  doi = "10.1016/0550-3213(95)00409-L", eprint = "hep-th/9506035", archivePrefix = "arXiv"
}

@article{Heisenberg:1936nmg,
  author = "Heisenberg, W. and Euler, H.",
  title = "{Folgerungen aus der Diracschen Theorie des Positrons}",
  journal = "Z. Phys.", volume = "98", pages = "714--732", year = "1936",
  doi = "10.1007/BF01343663"
}

@article{Schwinger:1951nm,
  author = "Schwinger, Julian S.",
  title = "{On Gauge Invariance and Vacuum Polarization}",
  journal = "Phys. Rev.", volume = "82", pages = "664--679", year = "1951",
  doi = "10.1103/PhysRev.82.664"
}

@incollection{Dunne:2004nc,
  author = "Dunne, Gerald V.",
  title = "{Heisenberg-Euler effective Lagrangians: Basics and extensions}",
  booktitle = "From Fields to Strings: Circumnavigating Theoretical Physics",
  editor = "Shifman, M. and Vainshtein, A. and Wheater, J.",
  publisher = "World Scientific", address = "Singapore", volume = "1", pages = "445--522", year = "2005",
  doi = "10.1142/9789812775344_0014", eprint = "hep-th/0406216", archivePrefix = "arXiv"
}

@article{Ayon-Beato:1998hmi,
    author = "Ayon-Beato, Eloy and Garcia, Alberto",
    title = "{Regular black hole in general relativity coupled to nonlinear electrodynamics}",
    eprint = "gr-qc/9911046",
    archivePrefix = "arXiv",
    doi = "10.1103/PhysRevLett.80.5056",
    journal = "Phys. Rev. Lett.",
    volume = "80",
    pages = "5056--5059",
    year = "1998"
}

@article{Ayon-Beato:1999kuh,
    author = "Ayon-Beato, Eloy and Garcia, Alberto",
    title = "{New regular black hole solution from nonlinear electrodynamics}",
    eprint = "hep-th/9911174",
    archivePrefix = "arXiv",
    doi = "10.1016/S0370-2693(99)01038-2",
    journal = "Phys. Lett. B",
    volume = "464",
    pages = "25",
    year = "1999"
}

@article{Ayon-Beato:2000mjt,
    author = "Ayon-Beato, Eloy and Garcia, Alberto",
    title = "{The Bardeen model as a nonlinear magnetic monopole}",
    eprint = "gr-qc/0009077",
    archivePrefix = "arXiv",
    doi = "10.1016/S0370-2693(00)01125-4",
    journal = "Phys. Lett. B",
    volume = "493",
    pages = "149--152",
    year = "2000"
}

@article{Abbott:1982af,
    author = "Abbott, L. F. and Sikivie, P.",
    editor = "Srednicki, M. A.",
    title = "{A Cosmological Bound on the Invisible Axion}",
    reportNumber = "PRINT-82-0695 (BRANDEIS)",
    doi = "10.1016/0370-2693(83)90638-X",
    journal = "Phys. Lett. B",
    volume = "120",
    pages = "133--136",
    year = "1983"
}

@article{Acharya:2010zx,
    author = "Acharya, Bobby Samir and Bobkov, Konstantin and Kumar, Piyush",
    title = "{An M Theory Solution to the Strong CP Problem and Constraints on the Axiverse}",
    eprint = "1004.5138",
    archivePrefix = "arXiv",
    primaryClass = "hep-th",
    doi = "10.1007/JHEP11(2010)105",
    journal = "JHEP",
    volume = "11",
    pages = "105",
    year = "2010"
}

@inproceedings{Ansoldi:2008jw,
    author = "Ansoldi, Stefano",
    title = "{Spherical black holes with regular center: A Review of existing models including a recent realization with Gaussian sources}",
    booktitle = "{Conference on Black Holes and Naked Singularities}",
    eprint = "0802.0330",
    archivePrefix = "arXiv",
    primaryClass = "gr-qc",
    reportNumber = "KUNS-2108",
    month = "2",
    year = "2008"
}

@article{Arvanitaki:2009fg,
    author = "Arvanitaki, Asimina and Dimopoulos, Savas and Dubovsky, Sergei and Kaloper, Nemanja and March-Russell, John",
    title = "{String Axiverse}",
    eprint = "0905.4720",
    archivePrefix = "arXiv",
    primaryClass = "hep-th",
    doi = "10.1103/PhysRevD.81.123530",
    journal = "Phys. Rev. D",
    volume = "81",
    pages = "123530",
    year = "2010"
}

@article{Balart:2014jia,
    author = "Balart, Leonardo and Vagenas, Elias C.",
    title = "{Regular black hole metrics and the weak energy condition}",
    eprint = "1401.2136",
    archivePrefix = "arXiv",
    primaryClass = "gr-qc",
    doi = "10.1016/j.physletb.2014.01.024",
    journal = "Phys. Lett. B",
    volume = "730",
    pages = "14--17",
    year = "2014"
}

@article{Balart:2014cga,
    author = "Balart, Leonardo and Vagenas, Elias C.",
    title = "{Regular black holes with a nonlinear electrodynamics source}",
    eprint = "1408.0306",
    archivePrefix = "arXiv",
    primaryClass = "gr-qc",
    doi = "10.1103/PhysRevD.90.124045",
    journal = "Phys. Rev. D",
    volume = "90",
    number = "12",
    pages = "124045",
    year = "2014"
}

@article{Born:1934gh,
    author = "Born, M. and Infeld, L.",
    title = "{Foundations of the new field theory}",
    doi = "10.1098/rspa.1934.0059",
    journal = "Proc. Roy. Soc. Lond. A",
    volume = "144",
    number = "852",
    pages = "425--451",
    year = "1934"
}

@article{Cai:2004eh,
    author = "Cai, Rong-Gen and Pang, Da-Wei and Wang, Anzhong",
    title = "{Born-Infeld black holes in (A)dS spaces}",
    eprint = "hep-th/0410158",
    archivePrefix = "arXiv",
    doi = "10.1103/PhysRevD.70.124034",
    journal = "Phys. Rev. D",
    volume = "70",
    pages = "124034",
    year = "2004"
}

@article{Cicoli:2012sz,
    author = "Cicoli, Michele and Goodsell, Mark and Ringwald, Andreas",
    title = "{The type IIB string axiverse and its low-energy phenomenology}",
    eprint = "1206.0819",
    archivePrefix = "arXiv",
    primaryClass = "hep-th",
    reportNumber = "DESY-12-058, CERN-PH-TH-2012-153",
    doi = "10.1007/JHEP10(2012)146",
    journal = "JHEP",
    volume = "10",
    pages = "146",
    year = "2012"
}

@article{Dey:2004yt,
    author = "Dey, Tanay Kr.",
    title = "{Born-Infeld black holes in the presence of a cosmological constant}",
    eprint = "hep-th/0406169",
    archivePrefix = "arXiv",
    reportNumber = "IP-BBSR-2004-06-17",
    doi = "10.1016/j.physletb.2004.06.047",
    journal = "Phys. Lett. B",
    volume = "595",
    pages = "484--490",
    year = "2004"
}

@article{Dine:1982ah,
    author = "Dine, Michael and Fischler, Willy",
    editor = "Srednicki, M. A.",
    title = "{The Not So Harmless Axion}",
    reportNumber = "UPR-0201T",
    doi = "10.1016/0370-2693(83)90639-1",
    journal = "Phys. Lett. B",
    volume = "120",
    pages = "137--141",
    year = "1983"
}

@article{Dine:1981rt,
    author = "Dine, Michael and Fischler, Willy and Srednicki, Mark",
    title = "{A Simple Solution to the Strong CP Problem with a Harmless Axion}",
    reportNumber = "Print-81-0320 (IAS,PRINCETON)",
    doi = "10.1016/0370-2693(81)90590-6",
    journal = "Phys. Lett. B",
    volume = "104",
    pages = "199--202",
    year = "1981"
}

@article{Drummond:1979pp,
    author = "Drummond, I. T. and Hathrell, S. J.",
    title = "{QED Vacuum Polarization in a Background Gravitational Field and Its Effect on the Velocity of Photons}",
    reportNumber = "DAMTP 79/16",
    doi = "10.1103/PhysRevD.22.343",
    journal = "Phys. Rev. D",
    volume = "22",
    pages = "343",
    year = "1980"
}

@article{Dymnikova:2004zc,
    author = "Dymnikova, Irina",
    title = "{Regular electrically charged structures in nonlinear electrodynamics coupled to general relativity}",
    eprint = "gr-qc/0407072",
    archivePrefix = "arXiv",
    doi = "10.1088/0264-9381/21/18/009",
    journal = "Class. Quant. Grav.",
    volume = "21",
    pages = "4417--4429",
    year = "2004"
}

@article{Fan:2016hvf,
    author = "Fan, Zhong-Ying and Wang, Xiaobao",
    title = "{Construction of Regular Black Holes in General Relativity}",
    eprint = "1610.02636",
    archivePrefix = "arXiv",
    primaryClass = "gr-qc",
    doi = "10.1103/PhysRevD.94.124027",
    journal = "Phys. Rev. D",
    volume = "94",
    number = "12",
    pages = "124027",
    year = "2016"
}

@article{Fernando:2003tz,
    author = "Fernando, Sharmanthie and Krug, Don",
    title = "{Charged black hole solutions in Einstein-Born-Infeld gravity with a cosmological constant}",
    eprint = "hep-th/0306120",
    archivePrefix = "arXiv",
    reportNumber = "NKU-01-SF2",
    doi = "10.1023/A:1021315214180",
    journal = "Gen. Rel. Grav.",
    volume = "35",
    pages = "129--137",
    year = "2003"
}

@article{Kim:1979if,
    author = "Kim, Jihn E.",
    title = "{Weak Interaction Singlet and Strong CP Invariance}",
    reportNumber = "UPR-0120T",
    doi = "10.1103/PhysRevLett.43.103",
    journal = "Phys. Rev. Lett.",
    volume = "43",
    pages = "103",
    year = "1979"
}

@article{Kruglov:2017mpj,
    author = "Kruglov, S. I.",
    title = "{Born{\textendash}Infeld-type electrodynamics and magnetic black holes}",
    eprint = "1707.04495",
    archivePrefix = "arXiv",
    primaryClass = "gr-qc",
    doi = "10.1016/j.aop.2017.06.008",
    journal = "Annals Phys.",
    volume = "383",
    pages = "550--559",
    year = "2017"
}

@article{Maeda:2021jdc,
    author = "Maeda, Hideki",
    title = "{Quest for realistic non-singular black-hole geometries: regular-center type}",
    eprint = "2107.04791",
    archivePrefix = "arXiv",
    primaryClass = "gr-qc",
    doi = "10.1007/JHEP11(2022)108",
    journal = "JHEP",
    volume = "11",
    pages = "108",
    year = "2022"
}

@article{Preskill:1982cy,
    author = "Preskill, John and Wise, Mark B. and Wilczek, Frank",
    editor = "Srednicki, M. A.",
    title = "{Cosmology of the Invisible Axion}",
    reportNumber = "HUTP-82-A048, NSF-ITP-82-103",
    doi = "10.1016/0370-2693(83)90637-8",
    journal = "Phys. Lett. B",
    volume = "120",
    pages = "127--132",
    year = "1983"
}

@article{Preucil:2017wen,
    author = "P{\v{r}}eu{\v{c}}il, Filip and Ho{\v{r}}ej{\v{s}}{\'\i}, Ji{\v{r}}{\'\i}",
    title = "{Effective Euler{\textendash}Heisenberg Lagrangians in models of QED}",
    eprint = "1707.08106",
    archivePrefix = "arXiv",
    primaryClass = "hep-ph",
    doi = "10.1088/1361-6471/aace90",
    journal = "J. Phys. G",
    volume = "45",
    number = "8",
    pages = "085005",
    year = "2018"
}

@article{DeFelice:2024seu,
    author = "De Felice, Antonio and Tsujikawa, Shinji",
    title = "{Instability of Nonsingular Black Holes in Nonlinear Electrodynamics}",
    eprint = "2410.00314",
    archivePrefix = "arXiv",
    primaryClass = "gr-qc",
    reportNumber = "YITP-24-124, WUCG-24-09",
    doi = "10.1103/PhysRevLett.134.081401",
    journal = "Phys. Rev. Lett.",
    volume = "134",
    number = "8",
    pages = "081401",
    year = "2025"
}

@article{Rodrigues:2018bdc,
    author = "Rodrigues, Manuel E. and de Sousa Silva, Marcos V.",
    title = "{Bardeen Regular Black Hole With an Electric Source}",
    eprint = "1802.05095",
    archivePrefix = "arXiv",
    primaryClass = "gr-qc",
    doi = "10.1088/1475-7516/2018/06/025",
    journal = "JCAP",
    volume = "06",
    pages = "025",
    year = "2018"
}

@article{Shifman:1979if,
    author = "Shifman, Mikhail A. and Vainshtein, A. I. and Zakharov, Valentin I.",
    title = "{Can Confinement Ensure Natural CP Invariance of Strong Interactions?}",
    reportNumber = "ITEP-64-1979",
    doi = "10.1016/0550-3213(80)90209-6",
    journal = "Nucl. Phys. B",
    volume = "166",
    pages = "493--506",
    year = "1980"
}

@article{Chen:2025aom,
    author = "Chen, Che-Yu and De Felice, Antonio and Tsujikawa, Shinji and Sano, Taishi",
    title = "{Vector Horndeski black holes in nonlinear electrodynamics}",
    eprint = "2509.23134",
    archivePrefix = "arXiv",
    primaryClass = "gr-qc",
    reportNumber = "RIKEN-iTHEMS-Report-25, YITP-25-152, WUCG-25-11",
    doi = "10.1103/fjqh-7gb2",
    journal = "Phys. Rev. D",
    volume = "113",
    number = "2",
    pages = "024027",
    year = "2026"
}

@article{Weinberg:1977ma,
    author = "Weinberg, Steven",
    title = "{A New Light Boson?}",
    reportNumber = "HUTP-77/A074",
    doi = "10.1103/PhysRevLett.40.223",
    journal = "Phys. Rev. Lett.",
    volume = "40",
    pages = "223--226",
    year = "1978"
}

@article{Wilczek:1977pj,
    author = "Wilczek, Frank",
    title = "{Problem of Strong  $P$  and  $T$  Invariance in the Presence of Instantons}",
    reportNumber = "Print-77-0939 (COLUMBIA)",
    doi = "10.1103/PhysRevLett.40.279",
    journal = "Phys. Rev. Lett.",
    volume = "40",
    pages = "279--282",
    year = "1978"
}

@article{Witten:1984dg,
    author = "Witten, Edward",
    title = "{Some Properties of O(32) Superstrings}",
    reportNumber = "Print-84-0838 (PRINCETON)",
    doi = "10.1016/0370-2693(84)90422-2",
    journal = "Phys. Lett. B",
    volume = "149",
    pages = "351--356",
    year = "1984"
}

@article{Yajima:2000kw,
    author = "Yajima, Hiroki and Tamaki, Takashi",
    title = "{Black hole solutions in Euler-Heisenberg theory}",
    eprint = "gr-qc/0005016",
    archivePrefix = "arXiv",
    doi = "10.1103/PhysRevD.63.064007",
    journal = "Phys. Rev. D",
    volume = "63",
    pages = "064007",
    year = "2001"
}

@article{Zhitnitsky:1980tq,
    author = "Zhitnitsky, A. R.",
    title = "{On Possible Suppression of the Axion Hadron Interactions. (In Russian)}",
    journal = "Sov. J. Nucl. Phys.",
    volume = "31",
    pages = "260",
    year = "1980"
}

@article{Cho:2009cj,
    author = "Cho, H. T. and Cornell, A. S. and Doukas, Jason and Naylor, Wade",
    title = "{Black hole quasinormal modes using the asymptotic iteration method}",
    eprint = "0912.2740",
    archivePrefix = "arXiv",
    primaryClass = "gr-qc",
    reportNumber = "YITP-09-110, WITS-CTP-049",
    doi = "10.1088/0264-9381/27/15/155004",
    journal = "Class. Quant. Grav.",
    volume = "27",
    pages = "155004",
    year = "2010"
}

@article{Fortuna:2020obg,
    author = "Fortuna, Sean and Vega, Ian",
    title = "{Bernstein spectral method for quasinormal modes and other eigenvalue problems}",
    eprint = "2003.06232",
    archivePrefix = "arXiv",
    primaryClass = "gr-qc",
    doi = "10.1140/epjc/s10052-023-12350-9",
    journal = "Eur. Phys. J. C",
    volume = "83",
    number = "12",
    pages = "1170",
    year = "2023"
}

@article{Chung:2023wkd,
    author = "Chung, Adrian Ka-Wai and Wagle, Pratik and Yunes, Nicolas",
    title = "{Spectral method for metric perturbations of black holes: Kerr background case in general relativity}",
    eprint = "2312.08435",
    archivePrefix = "arXiv",
    primaryClass = "gr-qc",
    doi = "10.1103/PhysRevD.109.044072",
    journal = "Phys. Rev. D",
    volume = "109",
    number = "4",
    pages = "044072",
    year = "2024"
}

@article{Chung:2023zdq,
    author = "Chung, Adrian Ka-Wai and Wagle, Pratik and Yunes, Nicolas",
    title = "{Spectral method for the gravitational perturbations of black holes: Schwarzschild background case}",
    eprint = "2302.11624",
    archivePrefix = "arXiv",
    primaryClass = "gr-qc",
    doi = "10.1103/PhysRevD.107.124032",
    journal = "Phys. Rev. D",
    volume = "107",
    number = "12",
    pages = "124032",
    year = "2023"
}

@article{Ansorg:2016ztf,
    author = "Ansorg, Marcus and Panosso Macedo, Rodrigo",
    title = "{Spectral decomposition of black-hole perturbations on hyperboloidal slices}",
    eprint = "1604.02261",
    archivePrefix = "arXiv",
    primaryClass = "gr-qc",
    doi = "10.1103/PhysRevD.93.124016",
    journal = "Phys. Rev. D",
    volume = "93",
    number = "12",
    pages = "124016",
    year = "2016"
}

@article{Jansen:2017oag,
    author = "Jansen, Aron",
    title = "{Overdamped modes in Schwarzschild-de Sitter and a Mathematica package for the numerical computation of quasinormal modes}",
    eprint = "1709.09178",
    archivePrefix = "arXiv",
    primaryClass = "gr-qc",
    doi = "10.1140/epjp/i2017-11825-9",
    journal = "Eur. Phys. J. Plus",
    volume = "132",
    number = "12",
    pages = "546",
    year = "2017"
}

@article{Kasuya:1981tq,
    author = "Kasuya, Masahiro and Kamata, Masaru",
    title = "{AN EXACT DYON SOLUTION WITH THE REISSNER-NORDSTROM METRIC}",
    reportNumber = "INS-402",
    doi = "10.1007/BF02725482",
    journal = "Nuovo Cim. B",
    volume = "66",
    pages = "75",
    year = "1981"
}

@article{Kasuya:1981ef,
    author = "Kasuya, Masahiro",
    title = "{Exact Solution of a Rotating Dyon Black Hole}",
    reportNumber = "INS-422",
    doi = "10.1103/PhysRevD.25.995",
    journal = "Phys. Rev. D",
    volume = "25",
    pages = "995",
    year = "1982"
}

@article{Macedo:2016wgh,
    author = "Macedo, Caio F. B. and Cardoso, Vitor and Crispino, Lu{\'\i}s C. B. and Pani, Paolo",
    title = "{Quasinormal modes of relativistic stars and interacting fields}",
    eprint = "1603.02095",
    archivePrefix = "arXiv",
    primaryClass = "gr-qc",
    doi = "10.1103/PhysRevD.93.064053",
    journal = "Phys. Rev. D",
    volume = "93",
    number = "6",
    pages = "064053",
    year = "2016"
}

@article{Berti:2007dg,
    author = "Berti, Emanuele and Cardoso, Vitor and Gonzalez, Jose A. and Sperhake, Ulrich",
    title = "{Mining information from binary black hole mergers: A Comparison of estimation methods for complex exponentials in noise}",
    eprint = "gr-qc/0701086",
    archivePrefix = "arXiv",
    doi = "10.1103/PhysRevD.75.124017",
    journal = "Phys. Rev. D",
    volume = "75",
    pages = "124017",
    year = "2007"
}

@article{Ripley:2022ypi,
    author = "Ripley, Justin L.",
    title = "{Computing the quasinormal modes and eigenfunctions for the Teukolsky equation using horizon penetrating, hyperboloidally compactified coordinates}",
    eprint = "2202.03837",
    archivePrefix = "arXiv",
    primaryClass = "gr-qc",
    doi = "10.1088/1361-6382/ac776d",
    journal = "Class. Quant. Grav.",
    volume = "39",
    number = "14",
    pages = "145009",
    year = "2022"
}

@article{Gundlach:1993tp,
    author = "Gundlach, Carsten and Price, Richard H. and Pullin, Jorge",
    title = "{Late time behavior of stellar collapse and explosions: 1. Linearized perturbations}",
    eprint = "gr-qc/9307009",
    archivePrefix = "arXiv",
    reportNumber = "NSF-ITP-93-84",
    doi = "10.1103/PhysRevD.49.883",
    journal = "Phys. Rev. D",
    volume = "49",
    pages = "883--889",
    year = "1994"
}

@article{Matyjasek:2017psv,
    author = "Matyjasek, Jerzy and Opala, Micha{\l}",
    title = "{Quasinormal modes of black holes. The improved semianalytic approach}",
    eprint = "1704.00361",
    archivePrefix = "arXiv",
    primaryClass = "gr-qc",
    doi = "10.1103/PhysRevD.96.024011",
    journal = "Phys. Rev. D",
    volume = "96",
    number = "2",
    pages = "024011",
    year = "2017"
}

@article{Heisenberg:2018SVTperturbations,
  author = "Heisenberg, Lavinia and Kase, Ryotaro and Tsujikawa, Shinji",
  title = "{Odd-parity stability of hairy black holes in $U(1)$ gauge-invariant scalar-vector-tensor theories}",
  journal = "Phys. Rev. D",
  volume = "97",
  pages = "124043",
  year = "2018",
  doi = "10.1103/PhysRevD.97.124043",
  eprint = "1804.00535",
  archivePrefix = "arXiv",
  primaryClass = "gr-qc"
}

@article{Heisenberg:2018SVThairy,
  author = "Heisenberg, Lavinia and Tsujikawa, Shinji",
  title = "{Hairy black hole solutions in $U(1)$ gauge-invariant scalar-vector-tensor theories}",
  journal = "Phys. Lett. B",
  volume = "780",
  pages = "638--646",
  year = "2018",
  doi = "10.1016/j.physletb.2018.03.059",
  eprint = "1802.07035",
  archivePrefix = "arXiv",
  primaryClass = "gr-qc"
}

@article{Rout:2026RNdegeneracy,
  author = "Rout, Abhishek",
  title = "{Isospectrality breaking in the axial quasinormal modes of Reissner-Nordstr{\"o}m black holes in dynamical Chern-Simons gravity}",
  year = "2026",
  doi = "10.48550/arXiv.2610.38742",
  eprint = "2610.38742",
  archivePrefix = "arXiv",
  primaryClass = "gr-qc"
}

\end{document}